\documentclass[12pt,a4paper]{article}            
 \usepackage[skins,theorems]{tcolorbox}
\tcbset{highlight math style={enhanced,
  colframe=red,colback=white,arc=0pt,boxrule=1pt}}
  \usepackage[bookmarksopen, bookmarksnumbered, bookmarksopenlevel=2]{hyperref}
  \usepackage{tikz}
  \usepackage{tikz-3dplot}
 \usetikzlibrary{calc}
 \usetikzlibrary{decorations} %
 \usepackage[UKenglish]{babel}
 \usepackage[toc,page]{appendix}
 \usepackage{amsmath}
 \usepackage{amssymb}
 \usepackage{graphicx}
 \usepackage{hhline}
 \usepackage[bf]{caption}
\usepackage{cite}
\usepackage[vcentermath]{youngtab}
\usepackage{geometry}
\usepackage{slashed}
\usepackage{color}
\usepackage{stackrel}
\usepackage{tikz-cd} 
\usepackage{cancel} 
\usepackage{booktabs} 

\usepackage{empheq}
\usepackage{arydshln}

\newcommand{\bqa}{\begin{eqnarray}}
\newcommand{\eqa}{\end{eqnarray}}

\newcommand{\nn}{\nonumber}

\def\IH{\mathbb{H}}

\def\IC{\mathbb{C}}
\def\reff{{\rm eff}}
\def\Jac{{\rm Jac}}
\def\Jac{{J}}

\hypersetup{
    pdftitle={},
    pdfauthor={},
    pdfsubject={}
}
\numberwithin{equation}{section}
\numberwithin{table}{section}

\makeatletter

\newtheorem{Conjecture}{Conjecture}

\newcommand{\be}{\begin{equation}}
\newcommand{\ee}{\end{equation}}
\newcommand{\beq}{\begin{equation}}
\newcommand{\eeq}{\end{equation}}
\newcommand{\ba}{\begin{aligned}}
\newcommand{\ea}{\end{aligned}}

\newcommand{\bea}{\begin{eqnarray}}
\newcommand{\eea}{\end{eqnarray}}
\newcommand{\tW}{\omega}
\newcommand{\cO}{\mathcal{O}}

\newcommand{\cE}{\mathcal{E}}

\newcommand{\cL}{\mathcal{L}}

\newcommand{\cF}{\mathcal{F}}

\newcommand\bi{\begin{itemize}}
\newcommand\ei{\end{itemize}}

\newcommand{\bG}{\mathbf{G}}
\newcommand{\Cr}{C^{\rm f}_r}

\def\Tr{\mathop{\mathrm{Tr}}\nolimits}
\def\tr{\mathop{\mathrm{tr}}\nolimits}

\def\unit{{1\kern-.65ex {\rm l}}}
\def\1{{1\kern-.65ex {\rm l}}}

\def\IZ{\mathbb{Z}}
\def\IP{\mathbb{P}}

\def\IR{{\mathbb{R}}}

\def\zh{{\hat z}}

\newcount\hour \newcount\minute
\hour=\time \divide \hour by 60
\minute=\time
\def\now{%
\ifnum \hour<13
  \ifnum \hour=0 \advance \hour by 12 \number\hour:\else \number\hour:\fi%
     \ifnum \minute<10 0\fi%
     \number\minute%
\ A.M.%
\else \advance \hour by -12 \number\hour:%
  \ifnum \minute<10 0\fi%
  \number\minute%
  \ P.M.%
\fi%
}

\def\fnote#1#2{\begingroup\def\thefootnote{#1}\footnote{#2}
     \addtocounter{footnote}{-1}\endgroup}

\makeatother

\begin{document}

\begin{flushright}
{\tt\normalsize CTPU-PTC-20-10}\\
{\tt\normalsize CERN-TH-2020-077}\\
\tt\normalsize  MITP/20-026 \\
\end{flushright}

\vskip 10 pt
\begin{center}
{\large \bf
Quasi-Jacobi Forms, Elliptic Genera\\
and Strings in Four Dimensions
} 

\vskip 11 mm

Seung-Joo Lee${}^{1}$, Wolfgang Lerche${}^{2}$, Guglielmo Lockhart${}^{2}$, 
and Timo Weigand${}^{3}$

\vskip 7 mm
\small ${}^{1}${\it Center for Theoretical Physics of the Universe, \\ Institute for Basic Science, Daejeon 34051, South Korea} \\[3 mm]
\small ${}^{2}${\it CERN, Theory Department, \\ 1 Esplanade des Particules, Geneva 23, CH-1211, Switzerland} \\[3 mm]
\small ${}^{3}${\it PRISMA Cluster of Excellence and Mainz Institute for Theoretical Physics, \\
Johannes Gutenberg-Universit\"at, 55099 Mainz, Germany}

\fnote{}{seung.joo.lee, wolfgang.lerche, guglielmo.lockhart, 
timo.weigand @cern.ch}

\end{center}

\vskip 5 mm
\begin{abstract}
\vskip 5 mm
We investigate the interplay between the enumerative geometry of Calabi-Yau fourfolds with fluxes
 and the modularity of elliptic genera
in four-dimensional string theories. 
We argue that certain contributions to the elliptic genus
are given by derivatives of modular or quasi-modular forms, which may encode
BPS invariants of Calabi-Yau or non-Calabi-Yau threefolds that are embedded in the given fourfold. 
As a result, the elliptic genus is only a quasi-Jacobi form, rather than a modular or quasi-modular one in the usual sense. This manifests itself  
as a holomorphic anomaly of the spectral flow symmetry, 
and in an elliptic holomorphic anomaly equation that maps between different flux sectors. We support our general considerations by a detailed study of examples, including non-critical strings in four dimensions.

For the critical heterotic string, we explain how anomaly cancellation is restored  due to the properties of the derivative sector. Essentially, 
while the modular sector of the elliptic genus takes care of anomaly cancellation involving the universal $B$-field,
the quasi-Jacobi one accounts for additional $B$-fields that can be present.   

Thus once again, diverse mathematical ingredients, namely here the algebraic geometry of fourfolds, relative Gromow-Witten theory pertaining to flux backgrounds, and the modular properties of (quasi-)Jacobi forms, conspire in an intriguing manner precisely as required by stringy consistency.

\end{abstract}

\vfill

\thispagestyle{empty}
\setcounter{page}{0}

\setcounter{page}{1}
\newpage

\tableofcontents
\goodbreak

\section{Introduction and Overview} 

Recent developments in the context of the Weak Gravity Conjecture \cite{ArkaniHamed:2006dz}, reviewed in \cite{Brennan:2017rbf,Palti:2019pca},
have revived interest in string dualities, which underlie the emergence
of tensionless strings \cite{Lee:2018urn,Lee:2018spm,Lee:2019tst,Lee:2019xtm,Lee:2019wij,Baume:2019sry}  at infinite distance boundaries of moduli space.

Specifically, the previous work \cite{Lee:2019tst} initiated a study of the emergence of 
asymptotically tensionless heterotic strings in $N=1$ supersymmetric string
compactifications in $d=4$ dimensions. These strings arise as solitonic objects
in certain flux compactifications on Calabi-Yau fourfolds in F-theory.
In suitable limits, they furnish the dominant degrees of freedom, and
become weakly coupled when described in the proper duality eigenframe.
The main purpose of that work was to test various Quantum Gravity Conjectures in a 
controlled four-dimensional setting.\footnote{See \cite{Grimm:2018ohb,Blumenhagen:2018nts,Grimm:2018cpv,Corvilain:2018lgw,Joshi:2019nzi,Erkinger:2019umg,Gendler:2020dfp} for a small sample of complementary, quantitative geometrical tests especially of the Swampland Distance Conjecture \cite{Ooguri:2006in,Klaewer:2016kiy}.}

The confirmation of the Weak Gravity Conjecture in \cite{Lee:2018urn,Lee:2018spm,Lee:2019tst} crucially hinged on the modular properties of a certain index-like partition function, the elliptic genus, of the asymptotically tensionless strings. Related aspects of modularity in this context have been discussed in \cite{ArkaniHamed:2006dz,Heidenreich:2015nta,Montero:2016tif,Heidenreich:2016aqi,Cottrell:2016bty}.
As a somewhat surprising outcome of the analysis in \cite{Lee:2019tst}, the elliptic genera of four-dimensional strings turn out not to be modular or even quasi-modular.

The goal of the present paper
is to study the interplay between geometry and modular properties of elliptic genera in
much greater depth.
We will observe that the deviation from the expected (quasi-)modularity of the elliptic genus in four dimensions is due to the appearance of derivatives of (quasi-)modular Jacobi forms.
These derivatives yield so-called {\it quasi-Jacobi forms} in the sense of~\cite{KanekoZagier,libgober2009elliptic}, in agreement with general conjectures made in~\cite{Oberdieck:2017pqm}.
As we will see, they originate from underlying, formally six-dimensional sectors of the theory. 
This fact manifests itself in an intriguing way in the geometry of the Calabi-Yau fourfolds and certain threefolds embedded therein.
These structures will likely be of use for
further applications  to non-perturbative, especially non-critical, strings in four dimensions.

In the next subsection we will review known results in order to set the stage,  followed by
 a summary of our findings as a road map. The rest of the text is then
devoted to a detailed analysis. In Section \ref{sec_geom} we will set up the geometry underlying the fourfold and flux configurations that we consider. The geometric objects which we will study are relative\footnote{The term relative invariants here refers to invariants for curves of form $C_{\rm b} + C_{\rm fib}$, where $C_{\rm b}$ lies in the base and  $C_{\rm fib}$ in the fiber of the elliptic fibration.} genus-zero Gromov-Witten, or {equivalently} BPS invariants on elliptically fibered fourfolds.
 In Section \ref{sec_geomelliptic} we  relate these invariants to a geometric, generally non-perturbative definition of elliptic genera of four-dimensional strings, with focus on the relationship between their modular properties and the underlying flux configurations.
 In this context we also observe an intriguing relation between partition functions associated with transversal and non-transversal fluxes, and this turns out to be a manifestation of the elliptic holomorphic anomaly equation of~\cite{Oberdieck:2017pqm}.
 Our analysis, condensed into Conjecture \ref{Conj2}, applies not only to the elliptic genus of solitonic heterotic strings, but also
 more generally to those of non-critical strings such as four-dimensional versions of E-strings. Section \ref{sec_anomalies} is devoted to the
 interplay between modularity, elliptic genera and anomaly cancellation, for the special case of a heterotic solitonic string. Specifically,
in Section \ref{subsec_effanomaly} we  provide a match between the Green-Schwarz terms in the effective action, and the various modular sectors of the elliptic genus.   In Sections \ref{sec_examples} and \ref{sec_non-criticalstrings} we present a detailed technical analysis of several examples for heterotic and non-critical strings.
Further mathematical facts are relegated to the Appendices.

\subsection{Review of Known Results}\label{subsec_recap}

An important quantity that captures certain robust features of string theories is the elliptic
genus \cite{Schellekens:1986yi,Schellekens:1986xh,Witten:1986bf,Alvarez:1987wg},
which serves as a loop space extension of the ordinary chiral index in quantum
field theories. By turning on background fields, a wealth of exact information about the chiral spectrum and its charges
can be extracted. In this paper we will mainly consider elliptic genera for four-dimensional string theories in
a $U(1)$ gauge background.  More concretely, what we will consider are the
partition functions in the
Ramond-Ramond sector of  superconformal worldsheet theories with right-moving supersymmetry, {which in $d=2n+2$ dimensions are defined as}
\be\label{Zellgendef}
Z(q,\xi)\ =\ \tr_{RR}\Big[(-1)^{F_R} {F_R^n} \,q^{H_L}\bar q^{H_R}\xi^J\Big]\,,\qquad
\qquad  q\equiv e^{2\pi i\tau}, \, \xi\equiv e^{2\pi i z}\,.
\ee
Here $\tau$ denotes the modular parameter of the toroidal  worldsheet, and
$z$ represents the background gauge field strength, or fugacity, which couples to a left-moving, holomorphic  $U(1)$ current,~$J$. 
{In order to obtain a non-vanishing result, the zero modes are saturated by inserting an appropriate power $F_R^n$ of the right-moving fermion number operator. In the present, four-dimensional, context, we have $n=1$ and the worldsheet theory possesses $N=(0,2)$ supersymmetry.}

The elliptic genus (\ref{Zellgendef})  should be contrasted with the familiar
one of $N=(2,2)$ superconformal theories. For these one can refine the elliptic genus in a canonical way as to keep track of the left-moving $U(1)$ superconformal R-symmetry. On the other hand, the current $J$ in the present context is just the worldsheet incarnation of some four-dimensional gauge symmetry, which for concreteness we have taken to be $U(1)$.\footnote{In Section \ref{sec_Estring}, we will also discuss a non-abelian extension.} This is a generic, model-dependent symmetry which does not pertain to any left-moving $N=2$ superconformal algebra.
 
It is familiar from the earliest works \cite{Schellekens:1986yi,Schellekens:1986xh,Witten:1986bf,Alvarez:1987wg} that the elliptic genus (\ref{Zellgendef}), defined as the RR partition function of a weakly coupled,
toroidal worldsheet theory, enjoys distinguished transformation properties under the modular group, $SL(2,\IZ)$:
For a string in $d=2n+2$ dimensions, it transforms with modular weight $w=-n$.
As we will recall later in Section~\ref{sec_anomalies}, for the special case of a critical heterotic string this has important implications for
anomaly cancellation \cite{Schellekens:1986yi,Schellekens:1986xh,Lerche:1987sg,Lerche:1987qk}  via the Green-Schwarz \cite{Green:1984sg} mechanism.

When the elliptic genus is refined by an extra $U(1)$ gauge background, as considered here, one might expect that 
it will be promoted to a weak Jacobi form \cite{Kawai:1993jk,Gritsenko:1999fk,Gritsenko:1999nm}. 
This means that   $Z(q,\xi)=\Phi_{-n,m}(\tau,z)$, where $\Phi_{w,m}$ denotes
a generic weak Jacobi form of modular weight $w$ and index $m$ (the index $m$ is model dependent and specifies the level of the underlying $U(1)$ Kac-Moody algebra, or equivalently, the spacing of the charge lattice).
As summarized in Appendix~\ref{app_jacobi}, such Jacobi forms enjoy distinguished modular and shift
transformation properties, which play an important r\^ole for elliptic genera in general (for a review, see e.g.~\cite{Dabholkar:2012nd}).

{While this expectation is indeed realised in six dimensions, we find}
 that the elliptic genus in four-dimensional string theories is not necessarily a modular or quasi-modular weak Jacobi form, but rather what is known as a quasi-Jacobi form (see again Appendix~\ref{app_jacobi}).  This is not in conflict with the arguments of \cite{Kawai:1993jk} which are based on spectral flow \cite{Schwimmer:1986mf}, as these arguments do not apply to generic 
$U(1)$ currents in $(0,2)$ models. Indeed it is known \cite{Lerche:1987ca} that (left-right asymmetric) spectral flow is not necessarily a symmetry of the theory. In fact the situation is not that bad, in that 
the elliptic genus will still be closely related to weak Jacobi forms, albeit in a more intricate way: namely, at least in special situations,  as a collection of formally {\it six-dimensional} elliptic genera in disguise. We will explain these matters, which are among our main findings, in detail in the next subsection.

\goodbreak
{Historically the elliptic genus of critical strings as written in (\ref{Zellgendef}) refers to a weakly coupled,
 conformal worldsheet theory and 
as such it is an intrinsically perturbative, one-loop quantity. 
However, it was later understood how elliptic genera for critical as well as for non-critical strings can also be defined and computed in non-perturbative settings, by resorting to a variety of methods such as mirror symmetry, the topological vertex, localization, and 2d CFT technology \cite{Klemm:1996hh,Minahan:1998vr,Haghighat:2013gba,Huang:2013yta,Haghighat:2013tka,Hohenegger:2013ala,Haghighat:2014pva,Hosomichi:2014rqa,Cai:2014vka,Kim:2014dza,Haghighat:2014vxa,Huang:2015sta,Haghighat:2015coa,Hohenegger:2015cba,Kim:2015gha,Honda:2015yha,Gadde:2015tra,Haghighat:2015ega,Hayashi:2015zka,Kim:2015fxa,Kim:2016foj,DelZotto:2016pvm,Gu:2017ccq,Hayashi:2017jze,Kim:2017jqn,Bastian:2017ing,DelZotto:2017mee,DelZotto:2017mee,Zhu:2017ysu,Kim:2018gak,Kim:2018gjo,DelZotto:2018tcj,Lee:2018urn,Duan:2018sqe,Gu:2018gmy,Lee:2019tst,Gu:2019dan,Gu:2019pqj}. This has the advantage of being far more general than for perturbative strings based on weakly coupled worldsheet theories, and applies also to non-perturbative heterotic as well as to non-critical strings.

 In this paper we will exploit the fact that elliptic genera can be computed geometrically in terms of Gromov-Witten invariants  arising in dual string compactifications in M- or F-theory.} Most of the work has been done, so far, for six-dimensional theories. Essentially, the idea is to consider solitonic strings that arise in F-theory from D3-branes wrapping some curve, $C_{\rm b}$,
which lies in the base of an elliptic threefold, $Y_3$.  In the dual M-theory formulation, the charged excitations of the string wrapped on an extra $S^1$ correspond to M2-branes on $\mathbf{C} = C_{\rm b} + n  \mathbb E_\tau + C^{\rm f}_r$.
Here $\mathbb E_\tau$ is the elliptic fiber of $Y_3$ and the other fibral curves $C^{\rm f}_r$ are associated with the gauge symmetry. 
The degeneracies that are encoded  in the elliptic genus (\ref{Zellgendef})
then have an interpretation as the 
genus-zero BPS invariants, $N_{C_{\rm b}}(n,r)$, associated with $\mathbf C$. These invariants can be assembled into the following free energy, which is defined relative to~$C_{\rm b}$:
\be\label{Fnr}
\cF_{C_{\rm b}}(q,\xi)\ =\ \sum N_{C_{\rm b}}(n,r)q^n\xi^r\,.
\ee
Here we assumed just one extra $U(1)$ gauge symmetry.\footnote{The generalisation to multiple $U(1)$ factors should be straightforward, 
in terms of (quasi-)Jacobi forms with multiple elliptic variables, along the lines of~\cite{Lee:2018spm}.} Physically the M2 brane wrapping numbers $n$ and $r$ correspond to
levels and charges of excitations of the solitonic string.

Via duality the free energy $\cF_{C_{\rm b}}(q,\xi)$  can be argued to coincide with the elliptic genus\footnote{Throughout this work we are considering genus-zero BPS invariants. In six dimensions, with a suitable omega background turned on, one can define an elliptic genus that also captures higher genus BPS invariants of Calabi-Yau threefolds as in \cite{Haghighat:2013gba}.
Note however that for compact Calabi-Yau fourfolds only genus zero and genus one invariants are relevant~\cite{Cox:2000vi}.} (\ref{Zellgendef}) of the solitonic string,
upon identifying the modulus of the elliptic fiber with the modulus of the toroidal worldsheet 
(and similarly for the $U(1)$ fugacity):
\be
Z_{C_{\rm b}}(q,\xi)\ =\ -q^{E_0}  \cF_{C_{\rm b}}(q,\xi)
\,,
\ee
where $E_0$ is the ground state energy of the Ramond sector of the string worldsheet theory.

In \cite{Lee:2019tst} we have addressed the analogous situation for four-dimensional F-theory compactifications on fourfolds, $Y_4$, focussing on geometries that lead to dual heterotic strings.
This is much more involved not the least because BPS  invariants on fourfolds, $N_{\alpha;C_{\rm b}}$,
depend on extra data. Namely they need to be
defined \cite{Cox:2000vi,Klemm:2007in,Haghighat:2015qdq,Cota:2017aal} with respect to some basis of cohomology classes,
$\omega^\alpha\in H^{2,2}(Y_4,\IR)$. In physics terms these data correspond to choices for the background four-flux, 
$\bG= c_\alpha \omega^\alpha$.
Thus for a given fourfold $Y_4$,
we have in general a collection of independent elliptic genera labelled by background fluxes,
\be
Z_{\alpha;C_{\rm b}}(q,\xi)\
=\ - q^{E_0} \, \sum_{n,r} N_{\alpha;C_{\rm b}}(n,r)q^n\xi^r\,,
\ee
so that the full elliptic genus for a given flux background $\bG$ is given by a linear combination
\be
Z_{\bG;C_{\rm b}}(q,\xi) = \sum_{\alpha=1}^{\dim H^{2,2}(Y_4,\IR)}c_\alpha\, Z_{\alpha;C_{\rm b}}(q,\xi)\,.
\ee
As far as the modular properties are concerned, it was found in \cite{Lee:2019tst} that depending on the flux sectors labelled by $\alpha$, the various building blocks $Z_{\alpha;C_{\rm b}}(q,\xi)$ behave very differently. To be more
specific, let us introduce the following symbolic notation (now labelling by modular weight and index rather than by flux and curve):
\be\label{Zdecomp}
\big[ Z_{-1,m}^* \big]\ =\ \big[Z_{-1,m}^M\big] \oplus \big[Z_{-1,m}^{QM}\big] \oplus \big[Z_{-1,m}^{QJ}\big]\,,
\ee
where the superscripts refer to\footnote{With ``modular'' (and similarly with ``quasi-modular'') we refer 
in this context to the transformation properties of weak Jacobi forms, which comprise besides (\ref{jacobitrApp})  also the double quasi-periodicity (\ref{periodicity}). } ``modular'', ``quasi-modular'' and ``quasi-Jacobi'', respectively. We will explain these terms in due course.
Note that at this point there is an ambiguity in that any $ Z_{-1,m}^{QM}$ is a priori defined only up to a modular piece,
and $Z_{-1,m}^{QJ}$ up to modular and quasi-modular pieces.
The precise alignment between the modular properties and fluxes in $H^{2,2}(Y_4)$, in relation to the overall geometry of the fourfold $Y_4$, will be a main issue in the present paper and will be discussed later in detail.

Let us go through the various components of $[Z_{-1,m}^*]$ and briefly characterize their modular properties.
The fully modular piece, $Z_{-1,m}^M$ in (\ref{Zdecomp}) is, by definition, given by some weak Jacobi form \cite{Kawai:1993jk,Gritsenko:1999fk,Gritsenko:1999nm} (see Appendix~\ref{app_jacobi}).
The quasi-modular piece, $Z_{-1,m}^{QM}$, is a benign modification, the only difference being that it is a quasi-modular and not fully modular Jacobi form. By this we mean that besides the ordinary modular Eisenstein series $E_4$ and $E_6$, also the quasi-modular series $E_2$ appears. 
As is familiar, this mild violation of modularity can be repaired by replacing $E_2$ by its
modular, but non-holomorphic cousin
 \be\label{Etwohat}
 \hat E_2=E_2-\frac3{\pi{\rm Im}\tau}\,,
 \ee  
 which transforms uniformly with modular weight two. 
 In field theoretic terms, this reflects a regularization ambiguity 
in the zero mode sector, which is resolved by imposing modular invariance
at the expense of holomorphicity.

This is just a manifestation of the celebrated holomorphic anomaly \cite{Bershadsky:1993ta}, which has many manifestations in physics.
 In the present context (and for genus-zero invariants) it is well-known
 \cite{Minahan:1997if, Minahan:1997ct,Minahan:1998vr,Hosono:1999qc,Klemm:2012sx,Alim:2012ss,Huang:2015sta} 
to mean that   the base curve $C_{\rm b} = C^0$, which corresponds to the heterotic string, splits over certain subloci:
$C^0= C^1_E+C^2_E$.
The curves $C^i_E$ in turn are associated with non-critical E-strings, and the split just {reflects the fact that
the heterotic string can fractionate into two E-strings \cite{Haghighat:2014pva}}.
In the dual heterotic language these correspond to having extra 5-branes in the geometry, which
means that the quasi-modular sector of the theory is intrinsically non-perturbative as seen from
the heterotic perspective.
As we will discuss later in Section~\ref{sec_anomalies}, this will have also a non-trivial bearing on
anomaly cancellation, which is closely tied to the modular properties of the elliptic genus.

Finally, most peculiar and thus most interesting is the last
component of the elliptic genus,  $Z_{-1,m}^{QJ}$, in (\ref{Zdecomp}).
It was found in \cite{Lee:2019tst} that it cannot be an ordinary modular or quasi-modular Jacobi form, since 
it does not obey the characteristic
transformation properties (\ref{jacobitrApp}) and (\ref{periodicity}). 
However it was observed that the 
coefficients of an expansion in powers of $z$ are quasi-modular forms term by term, so that one can at least
assign a uniform overall modular weight, $w=-1$, to it. 

\goodbreak
\subsection{Summary of Present Work} \label{sec_summary}

The main new result of the present paper is that the component $Z^{QJ}_{-1,m}$ of the ellitpic genus in (\ref{Zdecomp}) is actually 
also expressible in terms of the more familiar (quasi)-modular Jacobi forms, though in an intriguing way. Namely, it is given by a derivative
 \beq \label{Zm1QJ}
 Z_{-1,m}^{QJ}(q,\xi)=
\xi\frac\partial{\partial\xi}  Z_{-2,m}(q,\xi)\,
\eeq
of a partition function $Z_{-2,m}(q,\xi)$ of modular weight $-2$ and index $m$. Depending on the geometry it can be either a modular or quasi-modular weak Jacobi form. Thus, just like for the (quasi-)modular sector, the extra sector splits into two, namely into a perturbative and a non-perturbative piece, and we can refine the symbolic decomposition (\ref{Zdecomp}) as follows:
\be\label{betterZdecomp}
\big[Z_{-1,m}^* \big]\ =\ \Big(\big[Z_{-1,m}^M\big] \oplus \big[Z_{-1,m}^{QM}\big]\Big) \oplus \xi\frac\partial{\partial\xi}\Big(\big[Z_{-2,m}^{M}\big]\oplus \big[Z_{-2,m}^{QM}\big]\Big).
\ee
Accordingly, from now  on we will refer to these extra components as the ``derivative sector'', 
which by itself can come in a modular and quasi-modular version.

One main concern in the present paper will be to understand the mathematical origin and physical interpretation of this sector in terms of the underlying fourfold geometry and flux background. 
More concretely we aim to understand how the set of possible four-form fluxes
maps into the space  (\ref{betterZdecomp}) of elliptic genera, i.e.,
\be\label{thebasicmap}
 \big[\bG\big] \ \longrightarrow \big[Z_{-1,m}^* \big]\,.
 \ee
This question will be addressed in Section~\ref{sec_geomelliptic}, with special emphasis on geometries that
are dual to heterotic strings in Section~\ref{sec_geommod}.
In this concrete setting we can explicitly match the geometrical data to the 
decomposition (\ref{betterZdecomp}) of the elliptic genus in terms of modular objects. 
Specifically, we anticipate, as described in our key equation~(\ref{Zhetstr1}), that
\be \label{Zhetstr1first}
Z_{-1,m} = {g}^{0}  \, Z_{-1,m}^{0} +  {g}^{E}  \, Z_{-1,m}^{E} + \sum_i   {g}^{i}  \frac{1}{2m} \xi \partial_\xi \, Z^i_{-2,m} \,,
\ee
where the index $m$ is determined by a certain topological 
intersection product to be explained later. The flux-dependent coefficients, $g^*$,  are intersection products as well and are given in eqs.~(\ref{fluxcoeffs}). The first term in~\eqref{Zhetstr1first} represents the fully modular piece 
of the elliptic genus, while the second constitutes a possible non-perturbative, quasi-modular piece. 
It originates from a possible blowup of the base threefold, $B_3$,  which introduces an exceptional divisor, $E$,
and  is also associated with non-critical E-strings. As mentioned
in the previous section,
this generalizes well-known results in six dimensions  \cite{Klemm:2012sx,Alim:2012ss,Huang:2015sta}. 
The novel piece in four dimensions is then the derivative piece, which  is in general given by a sum of terms, as labelled by $i$ in~\eqref{Zhetstr1first}.

In fact, the derivative structure nicely ties in with statements given in the mathematical literature
~\cite{libgober2009elliptic,Oberdieck:2017pqm}. In particular it was proven in 
\cite{libgober2009elliptic} that elliptic genera  will in general  lie in the ring
of quasi-Jacobi forms,
which is a broader notion than just Jacobi or quasi-modular Jacobi forms.
It is easy to see from the definitions given in Appendix~\ref{app_modular},
that $Z_{-1,m}^{QJ}$ in (\ref{Zm1QJ})
yields a simple and concrete realization of such quasi-Jacobi forms, which explains the superscript. 
The paper \cite{Oberdieck:2017pqm} furthermore conjectured that the generating functions of relative BPS invariants in any elliptically fibered variety can generally be expressed in terms of quasi-Jacobi forms.
Our results for Calabi-Yau fourfolds were found in an independent manner, and thus provide a nontrivial check of these  conjectures.

Given that the derivative terms do not behave well under either $SL(2,\IZ)$  transformations
(\ref{jacobitrApp}) or under shifts (\ref{periodicity}), one might raise
the point of consistency of the physical theories. Specifically the shift
$z\rightarrow z+\lambda\tau$, $\lambda\in\IZ$, which
is a manifestation of spectral flow in the
$U(1)$ sector of the theory, would seem to be broken
for flux  backgrounds that lead to derivative contributions to the elliptic genus.

This is analogous to the problem discussed in the
previous section, where the quasi-modular 
Eisenstein series $E_2$ appears in  the component $Z^{QM}_{*,*}$ of the elliptic 
genus. In that case, the
cure for restoring modular invariance was to add a non-holomorphic piece to $E_2$, replacing it by the modular, non-holomorphic
weight-two Eisenstein series $ \hat E_2$ as in (\ref{Etwohat}).
 In the present context of quasi-Jacobi forms, we have a similar remedy: we can augment the derivative piece
by introducing a non-holomorphic term, and declare the following to be
the (``derivative'' part of the) physical elliptic genus:\footnote{This does not change the counting of states, since
the $q$-expansion is defined in the regime ${\rm Im}\tau\rightarrow\infty$.}
\be \label{pznonhol}
\hat Z_{-1,m}^{QJ}(\tau,z) \ = 
\nabla_{z,m}Z_{-2,m}^*(\tau,z) \ \equiv
\left( \partial_z + 
4\pi i m\frac{{\rm Im}z}{{\rm Im}\tau}  \right) Z_{-2,m}^*(\tau,z)\,.
\ee
This restores not only the modular $SL(2,\IZ)$ symmetry  
(\ref{jacobitrApp}), but also invariance under the shifts 
$z\rightarrow z+\lambda\tau$, $\lambda\in\IZ$, ie., spectral flow. 
In other words, what we encounter here as a novel phenomenon, on top of the known modular anomaly,
is an anomaly of the spectral flow which can be cancelled upon sacrificing holomorphicity.

\begin{figure}[t!]
\centering
\includegraphics[width=11cm]{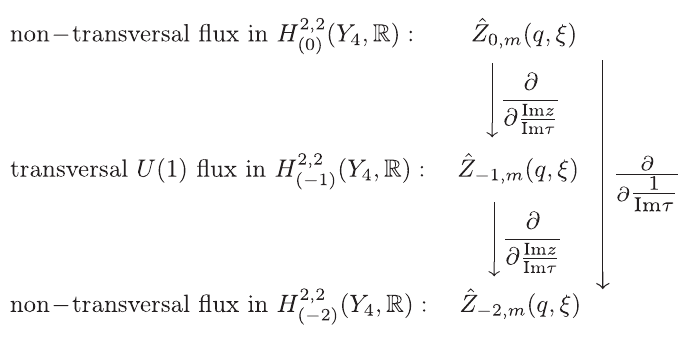}
\caption{
A sketch of the interplay between elliptic holomorphic
anomalies and (almost holomorphic) partition functions, $\hat
Z_{w,m}(q,\xi)$, pertaining to the various fluxes as classified
in (\ref{H22vertdecomp}).  Here $\hat Z_{-1,m}(q,\xi)$  is  (the
non-holomorphic cousin of) the elliptic genus in four dimensions.
Moreover $\hat Z_{-2,m}(q,\xi)$ encodes relative invariants of certain
embedded threefolds $\mathbb Y_3^i$, which can, at least formally,
be associated to elliptic genera in six dimensions.}
\label{anomalyfluxes}
\end{figure}

As we will explain in Section \ref{sec_m2flux}, eq.~(\ref{pznonhol}) has an interpretation in
terms of an elliptic generalization of  the holomorphic anomaly equation \cite{Bershadsky:1993ta,Minahan:1997if, Minahan:1997ct,Minahan:1998vr,Hosono:1999qc,Klemm:2012sx,Alim:2012ss,Huang:2015sta}.
 It is analogous to the familiar holomorphic anomaly equation,
which in essence states that  given an almost holomorphic, modular function which depends on $\hat E_2$, there is 
a functional identity between the non-holomorphic sector and a derivative with respect to $E_2$. More specifically,
we trivially infer from (\ref{pznonhol})  that
\be
\frac{\partial} {\partial\frac{{\rm Im}z}{{\rm Im}\tau}}
 \hat Z_{-1,m}^{QJ}(\tau,z)\ =\ 4 \pi i m Z_{-2,m}^*(\tau,z)\,.
 \label{eq:ell_anomaly}
\ee
The surprising point  is that the image of the derivative yields the generating function of BPS invariants related to a {\it different} flux sector.
Indeed we will find in Section \ref{sec_m2flux} that the $Z_{-2,m}^i(\tau,z)$ coincide with invariants related to certain non-transversal fluxes, 
even though these have no interpretation in terms of gauge fluxes in four-dimensional F-theory!
See Figure~\ref{anomalyfluxes} for a sketch of the relations between the various flux sectors.

This makes contact with the work of \cite{Haghighat:2015qdq,Cota:2017aal},
where the  BPS invariants induced by non-transversal  fluxes have been observed to organise into quasi-modular partition functions. 
More-over in ref.~\cite{Cota:2017aal} a holomorphic anomaly equation
was found that relates flux sectors associated with modular weights $w=0$ and $w=-2$; the rightmost arrow in Figure~\ref{anomalyfluxes} refers to this.
Our result (\ref{eq:ell_anomaly}) relates the elliptic genus of weight $w=-1$ to
a flux sector  associated with modular weight $w=-2$. 
In fact, after translating our setup to the formalism \cite{Oberdieck:2017pqm}, equation (\ref{eq:ell_anomaly}) turns out to be a manifestation of the elliptic holomorphic anomaly equation that was introduced in a more abstract form in that reference.

\begin{figure}[t!]
\centering
\includegraphics[width=10cm]{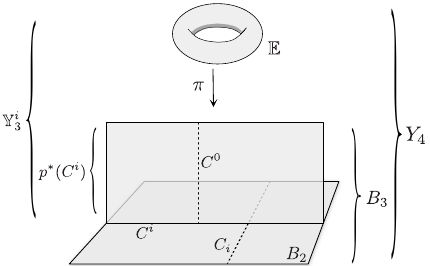}
\caption{
The derivative contribution to the elliptic genus of a four-dimensional string is captured, in suitable situations, by the relative BPS invariants of certain threefolds $\mathbb Y_3^i$ inside the given elliptic fourfold $Y_4$. Here we present a sketch of how these threefolds are embedded for the example of a heterotic string.
The heterotic string arises in this F-theory geometry from a D3-brane that wraps the rational fiber $C^0$ in the base threefold, $B_3\subset Y_4$. The extra 
two-form fields $B_i$ that are needed for reinstating Green-Schwarz anomaly cancellation arise by expanding the ten-dimensional four-form, ${\mathbf C}_4$, with respect to the divisor classes $p^\ast(C^i)$ for  $i=1,\dots h^{1,1}(B_2)$.}
\label{fig_3fold}
\end{figure}

Moreover, we have found another, related
interpretation of the $Z_{-2,m}^i(\tau,z)$:
  We will see that for certain geometries, the $Z^i_{-2,m}(\tau,z)$ are literally the elliptic genera of certain six-dimensional string theories. For example, in the context of solitonic heterotic strings, where the base threefold $B_3$ of the elliptic fourfold $Y_4$ is itself fibered over some surface, $B_2$,
the $Z^i_{-2,m}$ are labelled by curve classes $C_i$ in $B_2$. 
To each of these $C_i$ we can 
associate a certain specific threefold, $\mathbb Y_3^i$, which is defined by the restriction of the fourfold to the pullback divisor $p^\ast(C^i)$
as follows:
\be\label{Yidef}
\mathbb Y_3^i\ :=\ Y_4|_{p^\ast(C^i)} \,.
\ee
Here $\{C^j\}$ denotes the basis of curves dual to the basis $\{C_k\}$ on $B_2$. 
This geometric setup is schematically depicted in Figure \ref{fig_3fold}.
As we will argue in Section \ref{sec_geommod}, the $Z^i_{-2,m}$ just encode
the relative BPS invariants of these auxiliary threefolds.
This alternative interpretation of the $Z^i_{-2,m}$ then provides an intriguing
relationship between background fluxes in $H^{2,2}_{(-2)}$ and the enumerative
geometry of the threefolds $\mathbb Y_3^i$. It also gives independent 
support to some of the
conjectures of ref.~\cite{Oberdieck:2017pqm}.

In many cases, the embedded threefolds, $\mathbb Y_3^i$, can be Calabi-Yau by themselves for an appropriately chosen basis $\{C^i\}$.
 Since they are elliptically fibered as well, one can then associate to each of them a chiral, 
 six-dimensional F-theory compactification.  We will show that in this situation, the  collection of the $Z^i_{-2,m}$ that feature in the sum (\ref{Zhetstr1first})  are nothing but the
 elliptic genera associated with these threefolds  $\mathbb Y_3^i$. This phenomenon  generalises also to non-critical strings which will be the subject of section \ref{sec_non-criticalstrings}.
 
More generally, however, it turns out that the embedded threefolds $\mathbb Y_3^i$ are not necessarily Calabi-Yau. 
We nonetheless conjecture, and support with some arguments,
that the $Z^i_{-2,m}$ continue to encode BPS invariants of the embedded
threefolds,  $\mathbb Y_3^i$. For these cases, an interpretation 
 in terms of elliptic genera
 likely persists only as a formal analogy.

To support our considerations, we will present 
a detailed analysis of several examples. 
In Section~\ref{sec_Ex1het} we will discuss an example where the derivative sector comprises two embedded 
threefolds both of which
are Calabi-Yau. On the other hand, Section~\ref{sec_Ex2het} shows an example with a single embedded threefold, which has negative
anti-canonical bundle; the derivative piece of the elliptic genus  can then be associated via duality, 
in a formal sense, to a certain elliptic surface with 36 singular fibers.
Furthermore we see in examples that the derivative structure appears even more broadly, and also applies to  elliptic genera of non-critical strings.  This suggests,
as stated in Conjecture~\ref{Conj2}, that it is a general feature of elliptic genera in four dimensions.
In Section \ref{sec_non-criticalstrings} we will see how it applies to what we will call \emph{four-dimensional E-strings},\footnote{{It would be interesting to make explicit the relation of these four-dimensional theories to the compactifications of 6d E-string theories with flux, whose study was initiated in \cite{Kim:2017toz}.}} as well as to a non-critical string arising from a D3-brane wrapping a curve on $B_3 = \mathbb P^3$.

From a physics perspective, one may wonder about the Green-Schwarz cancellation of the $U(1)$ gauge anomaly, which is known to be closely tied to the modular properties of the elliptic genus.
For the example of a flux background that is dual to
 the heterotic string, we will show in Section~\ref{sec_anomalies} that anomaly cancellation persists also when derivative pieces are present,  albeit in a modified way.
 
 As we will recall in Section~\ref{subsec_Anomaliesmod},  the modular properties of the elliptic genus underlie the
standard Green-Schwarz mechanism which involves the universal $B$-field. The derivative terms  in the elliptic genus, $Z_{-2}^i$, appear precisely when, depending on the geometry and flux, further 2-form fields $B_i$ contribute to the Green-Schwarz mechanism\cite{Blumenhagen:2005ga}.\footnote{For clarity we neglect here the quasi-modular sector, which brings in its own $B$-fields, as explained in Section \ref{subsec_Anomaliesmod}.}
The additional 2-form fields arise from the curve classes $C_i$ in the base threefold $B_2$, which also determine, as per (\ref{Yidef}), the corresponding elliptic threefolds $\mathbb Y_3^i$. To close the circle of ideas, the threefolds in turn encode the BPS invariants pertaining to the $i^{\rm th}$ derivative sector of the elliptic genus, and altogether everything conspires such that anomalies are cancelled.

\section{Geometric Foundations}\label{sec_geom}

In Section \ref{subsec_BPS} we briefly review the geometric definition of BPS invariants for curves on Calabi-Yau fourfolds and their computation via mirror symmetry. In Section \ref{subsec_relBPS} we then specialise  to the relative BPS invariants on elliptic fibrations in the presence of fluxes. These concepts become particularly important in light of F-theory/heterotic duality, whose salient geometric manifestation we recall in Section~\ref{subsec_P1fibered}.

\subsection{BPS Invariants on Calabi-Yau Fourfolds}
\label{subsec_BPS}
In this work we investigate the structure of certain integral BPS invariants for Calabi-Yau fourfolds, $Y_4$,
which are analogous to the familiar integral BPS invariants on Calabi-Yau three-folds.
There are two ways to approach the definition of the invariants, either on purely geometric grounds or via mirror symmetry, and we will briefly review both. For more details we refer e.g. to \cite{Cox:2000vi,Klemm:2007in,Haghighat:2015qdq,Cota:2017aal} and references therein.

We begin with the geometric approach by first defining the (in general rational) Gromov-Witten invariants of a Calabi-Yau fourfold. 
Consider a curve class $C \in H_2(Y_4,\mathbb Z)$ 
and the moduli space of stable holomorphic maps from a Riemann surface of genus $g$ to $C$ with $s$ points fixed, denoted as $\overline{{\cal M}}_{g,s}$.
This moduli space has  expected or virtual (complex) dimension ${\rm dim}_{\rm vir}(\overline{{\cal M}}_{g,s}) = 1-g+s$.
For genus $g=0$ and $s=1$,  the virtual dimension of  $\overline{{\cal M}}_{g,s}$ is two and thus one can define a topological invariant  by, loosley speaking, integrating a suitably quantized element $\bG \in H^4(Y_4,\mathbb R)$ over the moduli space.
More precisely, following \cite{Cox:2000vi} we denote by $\mu$  the virtual fundamental class of $\overline{{\cal M}}_{0,1}$ associated with a curve $C$ (with one point fixed). Then this defines the genus-zero Gromov-Witten invariants $n_{\bG}(C)$ of $C$ with respect to $\bG$ as 
\be \label{GWdef}
n_{\bG}(C) = \int_\mu {\rm ev}^\ast(\bG) \,,
\ee
where ${\rm ev}^\ast$ is the evaluation map applied to $\bG$. By holomorphicity, this integral is non-zero only if $\bG \in H^{2,2}(Y_4,\mathbb R)$.
While Gromov-Witten invariants are in general not integral, there exist related
integral BPS invariants for fourfolds which are 
analogues of the integral BPS invariants of threefolds \cite{Gopakumar:1998ii,Gopakumar:1998jq,Katz:1999xq}.
This was first conjectured in \cite{Klemm:2007in} and proven for $g=0$ in \cite{ionel2013gopakumarvafa}. We will denote these
integral BPS invariants by $N_{\bG}(C)$.
 At genus zero,  and as long as we do not consider multiples of curve classes, the two notions of invariants are equivalent;  throughout this work we will be in this situation and can hence use both notions of invariants interchangeably.

The BPS invariants $N_{\bG}(C)$ can be computed by mirror symmetry \cite{Greene:1993vm,Mayr:1996sh}, by interpreting the Calabi-Yau fourfold $Y_4$ as the compactification space of Type IIA string theory to two dimensions, and the element $\bG \in H^{2,2}(Y_4,\mathbb R)$
as a four-form background flux. 
The space $H^{2,2}(Y_4,\mathbb R)$ of supersymmetric flux backgrounds admits a decomposition \cite{Greene:1993vm,Braun:2014xka}
\be \label{H22decomp}
H^{2,2}(Y_4,\mathbb R) = H^{2,2}_{\rm vert}(Y_4,\mathbb R) \oplus H^{2,2}_{\rm hor}(Y_4,\mathbb R)  \oplus H^{2,2}_{\rm rest}(Y_4,\mathbb R)  \,,
\ee
where the vertical subspace $H^{2,2}_{\rm vert}(Y_4,\mathbb R)$ is generated by all products of two elements in $H^{1,1}(Y_4)$, while the horizontal subspace $H^{2,2}_{\rm hor}(Y_4,\mathbb R)$
is obtained by the variation of Hodge structure from the unique $(4,0)$-form on $Y_4$.
In a  flux background given by $\bG \in H^{2,2}_{\rm vert}(Y_4,\mathbb R)$, the $N=(2,2)$ supersymmetric compactification of Type IIA string theory on $Y_4$ is partially determined by the free energy, ${\cal F}_{\bG}(t)$, which depends holomorphically on the 
 K\"ahler moduli $t^a$, $a = 1, \ldots, h^{1,1}(Y_4)$. 
The two-point functions for the chiral fields in the effective action associated with the K\"ahler moduli are then given by \cite{Mayr:1996sh,Klemm:1996ts}
\be
C_{ab,\bG} = \partial_a \partial_b {\cal F}_\bG(t) \,.
\ee
The free energy ${\cal F}_\bG(t)$, which plays the r\^ole of a superpotential in two dimensions,
encodes the genus-zero invariants as follows:  Define first the variables 
\be
q_a = e^{2\pi i t_a} \,, \qquad a = 1,\ldots, h^{1,1}(Y_4) \,,
\ee
and expand  a given curve class $C_\beta$ in terms of the basis $C_a$ of $H_2(Y_4)$ with complexified volumes $t^a$,
\be
C_\beta = \beta^a C_a \,.
\ee
Hence to each curve $C_\beta$ one can associate the product 
\be
q_{{\beta}} = q_1^{\beta^1} \ldots  q_{h^{1,1}(Y_4)}^{\beta^{h^{1,1}(Y_4)}}  \,.
\ee
The free energy ${\cal F}_G(t)$ then enjoys a worldsheet instanton expansion of the form
\be \label{freeenergy-def}
{\cal F}_{\bG}(t) = \sum_{{\beta}} N_\bG(C_\beta)  \, {\rm Li}_2(q^{{\beta}}) \,, \qquad N_\bG(C_\beta)  \in \mathbb Z \,,
\ee
where we have suppressed possible classical pieces which are polynomial in $t_a$.
The function ${\cal F}_\bG(t)$, and hence the invariants $N_\bG(C_\beta)$,  can in turn be computed by mirror symmetry: Type IIA string theory on $Y_4$ with flux $\bG \in H^{2,2}_{\rm vert}(Y_4,\mathbb R)$ is dual to Type IIA string theory on the mirror $\tilde Y_4$ with a dual flux $\tilde \bG \in H^{2,2}_{\rm hor}(Y_4,\mathbb R)$. Under the mirror map the free energy ${\cal F}_{\bG}(t)$ maps to 
\be
\cF_{\tilde \bG} = \int_{\tilde Y_4} \tilde \bG \wedge \tilde \Omega^{4,0} \,,
\ee 
which is a holomorphic function of the complex structure moduli of $\tilde Y_4$.
It is, in  principle, exactly computable as a period integral, which eventually determines the invariants $N_\bG(C_\beta)$. For more details see \cite{Mayr:1996sh,Klemm:1996ts,Lerche:1997zb,Gukov:1999ya}.

 \subsection{Relative BPS Invariants on Elliptic Calabi-Yau Fourfolds} \label{subsec_relBPS}
 
We now focus on invariants  $N_\bG(C_\beta)$ of those Calabi-Yau fourfolds $Y_4$ 
which admit an elliptic fibration of the form
\bea \label{piproj}
\pi :\quad \mathbb{E}_\tau \ \rightarrow & \  \ Y_{4} \cr 
& \ \ \downarrow \cr 
& \ \  B_3  
\eea
The base  $B_3$ of the fibration is a K\"ahler threefold, which, in order for such a fibration to exist, must have an effective anti-canonical divisor, $\bar K_{B_3}$. 

As we will see, for certain choices of curve class $C_\beta$ and flux $\bG$, the genus-zero invariants $N_\bG(C_\beta)$ admit yet another interpretation in terms of an elliptic genus of a string.
To arrive at this interpretation, we view the K\"ahler threefold $B_3$ as the compactification space of F-theory \cite{Vafa:1996xn} to four dimensions. Compactification of this theory on a further circle, $S^1_F$, gives rise to a theory in three dimensions, which coincides with M-theory compactified on $Y_4$.\footnote{The following well-known elements of F-theory are reviewed for example in \cite{Weigand:2018rez,Cvetic:2018bni}, to which we refer for details and original references.}
 
We will furthermore assume that the gauge group of the four-dimensional F-theory is non-trivial.
For any non-abelian factor $G$ of the gauge group, there must be a divisor $b_G$ on $B_3$ which is wrapped by a stack of 7-branes.
For the geometry of $Y_4$, this implies that the generic elliptic fiber $\mathbb{E}_\tau$ splits into several holomorphic curves over  $b_G$. In the dual M-theory picture, M2-branes wrapping these fibral curves give rise to the non-abelian gauge bosons that are not in the Cartan subalgebra of $G$.
The fibral curves may split further over curves on $b_G$,  in which case additional matter fields charged under $G$ appear.
Even for $G=U(1)$,  massless charged matter exists only if the elliptic fiber splits over certain curves on $B_3$.

For definiteness we now focus on gauge group $G= U(1)$.
Geometrically, in such situation the fourfold $Y_4$ then exhibits an extra rational section, $S$, in addition to the zero-section, $S_0$.
Associated with $S$ is the divisor class
\be
\sigma \equiv \sigma(S) \in H^{1,1}(Y_4) \,,
\ee
which is the image of the Shioda map. It has the defining properties that
\be
\sigma \circ \pi^\ast w_6 = 0 \,, \qquad \sigma \circ S_0 \circ \pi^\ast w_4 = 0 \qquad \forall \, w_6 \in H^6(B_3) \,, \quad w_4 \in H^4(B_3) \,.
\ee
Here and in what follows we use the notation $\circ$ for the intersection product on $Y_4$, i.e.,   
\be
w_a \circ w_b \circ \ldots \circ w_{8-a-b-\ldots}   = \int_{Y_4} w_a \wedge w_b \wedge \ldots \wedge w_{8-a-b-\ldots} \,.
\ee
Given $\sigma$ we can expand the M-theory 3-form as $C_3  = A \wedge \sigma + \ldots$, where the 1-form field $A$ becomes the $G=U(1)$ gauge potential in the dual M-theory.

In the language of Type IIB/F-theory, the abelian gauge group is associated with a linear combination of 7-branes, each wrapping a 4-cycle on $B_3$.
The linear combination of four-cycles associated with the $U(1)$ in this way can be identified with the 
 the so-called height-pairing
\be \label{bU1def}
b_{U(1)} \equiv b = - \pi_\ast (\sigma \circ \sigma) \,.
\ee

As mentioned before, in addition to the gauge potential there will in general be a collection of
massless charged matter fields.
In the Type IIB/F-theory picture, massless $N=1$ chiral multiplets with 
\be
\text{$U(1)$ charge} \, \,  Q = r
\ee
 arise from open strings stretched between the 7-branes. The open strings give rise to massless states of charge $r$ which
 are localized on certain (self-)intersecting curves of the 7-branes on $B_3$. We will call these  ``matter curves'' and denote them by $\Sigma_r$.
In the M-theory picture, the charged matter fields are obtained by wrapping M2-branes on curves $\Cr$ which sit in the fiber of $Y_4$ over $\Sigma_r$.
Their charges are determined by the intersection product with the Shioda map:
\be
r = \sigma \circ \Cr \,.
\ee

Apart from the geometry intrinsic to the fourfold $Y_4$, the effective theory also depends on the background flux, which, via the duality to M-theory, is encoded in a flux $\bG \in H^4(Y_4,\mathbb R)$ in M-theory. It is quantized such that 
$\bG + \frac{1}{2} c_2(Y_4) \in  H^4(Y_4,\mathbb Z)$. Importantly for us,
the primary vertical subspace $H^{2,2}_{{\rm vert}}(Y_4,\mathbb R)$, as sketched in (\ref{H22decomp}), receives additional structure 
if $Y_4$ is elliptically fibered. 
In this case, $H^{2,2}_{{\rm vert}}(Y_4,\mathbb R)$ is spanned by three different types of 4-forms which can be characterised as follows: 
\be
H^{2,2}_{{\rm vert}}(Y_4,\mathbb R) = H^{2,2}_{(0)}(Y_4,\mathbb R)  \cup H^{2,2}_{(-1)}(Y_4,\mathbb R)  \cup H^{2,2}_{(-2)}(Y_4,\mathbb R)  
\ee
with 
\be
\begin{aligned} \label{H22vertdecomp}
 H^{2,2}_{(0)}(Y_4,\mathbb R)  &= \langle   (S_0  + \frac{1}{2} \pi^\ast(\bar K_{B_3}))  \wedge \pi^\ast(w_i) \rangle    \\
 H^{2,2}_{(-1)}(Y_4,\mathbb R)  &= \langle \sigma(S) \wedge \pi^\ast(w_i) \rangle    \\
H^{2,2}_{(-2)}(Y_4,\mathbb R)  &= \langle \pi^\ast(w_i) \wedge \pi^\ast(w_j) \rangle  \,.
\end{aligned}
\ee
Here $\{w_i\}$ is a basis of $H^{1,1}(B_3)$, $S_0$  denotes the zero-section and $\sigma(S)$ the Shioda map image associated with the additional independent section $S$, as before.
Note that not all elements in the set $\{\pi^\ast(w_i) \wedge \pi^\ast(w_j)\}$ are linearly independent within  $H^{2,2}_{{\rm vert}}(Y_4,\mathbb R)$.
As will become clear later,
the subscript in $H^{2,2}_{(w)}(Y_4,\mathbb R)$ refers to the modular weight $w$
of the partition function, $Z_{w,m}$, that is associated with the given flux.

Of special importance for us is the so-called transversal subspace $H^{2,2}_{(-1)}(Y_4,\mathbb R)$ of $H^{2,2}_{{\rm vert}}(Y_4)$.  It is orthogonal to the other two subspaces in  (\ref{H22vertdecomp}), i.e., a flux $\bG \in H^{2,2}_{(-1)}(Y_4,\mathbb R)$ by definition satisfies the two conditions 
\be \label{vertcond}
\bG \circ  \pi^\ast(w_i)   \circ  \pi^\ast(w_j)  = 0 \,, \qquad \bG \circ S_0 \circ  \pi^\ast(w_i) = 0    \quad \forall w_i \in H^{1,1}(B_3) \,.
\ee
The transversality conditions (\ref{vertcond}) ensure that the flux $\bG$, which is a priori defined as background in the M-theory compactification on $Y_4$, is compatible with the duality to F-theory on $B_3$, in the sense of giving rise to a four-dimensional effective theory with full Poincar\'e invariance in $\mathbb R^{1,3}$. 
Such transversal fluxes related to the $U(1)$ symmetry will be denoted by 
\be \label{GU1def}
\bG_{U(1)} = \sigma \wedge \pi^\ast(F) \,, \qquad \quad F \in H^{1,1}(B_3) \,.
\ee
All other elements in $H^{2,2}_{\rm vert}(Y_4)$, while corresponding to valid flux backgrounds in M-theory or Type IIA string theory,
 are not liftable to F-theory.
In the more general context of M-theory/Type IIA string theory on $Y_4$, one can in any case analyse the BPS invariants of an elliptic fibration in a non-transversal flux background, as pioneered in 
 \cite{Haghighat:2015qdq,Cota:2017aal}.
  
 Let us now recall how the transversal fluxes determine the chiral index of massless charged matter in the context of four-dimensional F-theory compactifications \cite{Weigand:2018rez}.
As noted above, massless matter fields with $U(1)$ charge $Q=r$ are localised on a curve $\Sigma_r$ on the base $B_3$. The fiber over $\Sigma_r$ contains the curve 
$\Cr$, and an M2-brane wrapping $\Cr$ gives rise to a BPS particle in the dual M-theory picture.
In fact, the fibration of $\Cr$ defines a surface  $\hat \Sigma_r$,
\bea \label{rproj}
\pi_r :\quad   \Cr \ \rightarrow & \  \ \hat \Sigma_r \cr 
& \ \ \downarrow \cr 
& \ \  \Sigma_r  
\eea
in terms of which the chiral index of massless matter of charge $Q=r$ is computed as
\be \label{chiQ=r}
\chi_{\bG_{U(1)}, Q=r} = n^{+}_r - n^-_r = \int_ {\hat \Sigma_r} \bG_{U(1)} = r \int_{\Sigma_r} F =r  \, (\Sigma_r \cdot F)  \,.
\ee
The third equality is a consequence of the factorized form (\ref{GU1def}). 
Furthermore we have introduced, after the last equality, the intersection product on $B_3$ that we will henceforth denote by a dot.

In fact, the integral invariant $\chi_{\bG_{U(1)},Q=r}$ is exactly the genus-zero Gromov-Witten invariant for the fibral curve $\Cr$ with respect to $\bG_{U(1)}$ \cite{Lee:2019tst}, 
\be \label{chi=GW}
\chi_{\bG_{U(1)},Q=r} = n_{\bG_{U(1)}}(\Cr)  = N_{\bG_{U(1)}}(\Cr) \,.
\ee
The first equation follows from the geometric definition (\ref{GWdef}) because the moduli space of $\Cr$ with one point fixed coincides with the surface $\hat \Sigma_r$.
The second equation holds because for the rational curves in the fiber, $\Cr = r C^{\rm f}_{r=1}$ in cohomology; if $n_{\bG_{U(1)}}(\Cr) \neq 0$, there must exist an actual curve in this class in the fiber.
Hence the non-vanishing invariants  $n_{\bG_{U(1)}}(\Cr) \neq 0$ do not involve multiple wrappings and therefore agree with the BPS invariants $N_{\bG_{U(1)}}(\Cr)$. 

More generally, we are interested in the structure of genus-zero integral BPS invariants for curves of the form
\be\label{Cdef}
\mathbf{C} = C_{\rm b} + n \, \mathbb E_\tau +  C^{\rm f}_r \,,
\ee
with respect to fluxes in $H^{2,2}_{\rm vert}(Y_4)$ that satisfy (\ref{vertcond}).
We denote these invariants as
\be
N_{\bG}[C_{\rm b} + n \, \mathbb E_\tau + \Cr] =: N_{\bG; C_{\rm b}}(n, r) \,.
\ee
As long as $C_{\rm b}$ is not a multiple of an integral curve class on $B_3$, these integral invariants coincide with the Gromov-Witten invariants for the same curve.
They are called the {\it relative Gromov-Witten} invariants with respect to the elliptic fibration $\pi$.  

These integral invariants can be packaged into a generating function
\be \label{genfun1} {\cal F}_{\bG; C_{\rm b}} = \sum_{n,Q} N_{\bG; C_{\rm b}}(n, r) q^n \, \xi^r \,.
\ee
Here we have defined the variables
\be
q = e^{2 \pi i \tau} \,, \qquad  \quad \xi = e^{2\pi i z} \,,
\ee
where $\tau$ is the K\"ahler parameter of the generic elliptic fiber $\mathbb E_\tau$ and $z$ is the K\"ahler parameter of the fibral curve $C^{\rm f}_{r=1}$. 
From the perspective of Type IIA string theory on $Y_4$, ${\cal F}_{\bG, C_{\rm b}}$ contributes to the two-dimensional 
superpotential ${\cal F}(t)$, as defined in (\ref{freeenergy-def}). 

As stressed above, in the context of F-theory we must insist that $\bG$ is a transversal flux. Under this proviso, (\ref{genfun1})  coincides with the elliptic genus of a four-dimensional solitonic string (up to a prefactor), as will be explained in Section~\ref{subsec_solstrings}. 
On the other hand, for Type IIA compactifications on fourfolds,
there is no restriction to transversal flux backgrounds, 
and one can consider generating functions (\ref{genfun1}) for the other,
non-transversal types of flux as well.
As exemplified in \cite{Haghighat:2015qdq,Cota:2017aal},
the partition functions for fluxes in $H^{2,2}_{(-2)}(Y_4)$ and $H^{2,2}_{(0)}(Y_4)$   are meromorphic (quasi-)modular forms of weight $-2$ and $0$,  respectively.

\subsection{$\mathbb P^1$-Fibered Base Spaces and F-Theory/Heterotic Duality}  \label{subsec_P1fibered}

As a special case of the structure outlined in the previous section, we now consider the situation where the base space $B_3$ 
by itself admits a further, 
rational fibration with section $S_-$, possibly blown up along one or several curves on the section.
The projection of the rational fibration will be denoted by
\bea \label{pproj}
p :\quad C^0 \ \rightarrow & \  \ B_{3} \cr 
& \ \ \downarrow \cr 
& \ \  B_2  
\eea
The generic fiber is some rational curve $C^0$.
Prior to performing any blowup, the fibration can be understood as the projectivised bundle $\mathbb P({\cal O} \oplus {\cal L})$, where ${\cal L}$ is a line bundle on $B_2$.
This means that the section $S_-$, which is oftentimes referred to as an exceptional section, has self-intersection $S_- \cdot S_- = -  S_- \cdot p^\ast(c_1({\cal L}))$.
One can therefore define another section
$S_+: = S_- + p^\ast c_1({\cal L})$
such that 
\be
S_- \cdot S_+ = S_- \cdot (S_- + p^\ast c_1({\cal L})) = 0 \,.
\ee


We can also perform an optional blowup along some curve $\Gamma_a$ in the base $B_2$.
After the blow-up, the  rational fiber $C^0$ over the curve $\Gamma_a$
splits into two rational curves,
\be \label{CEdef1}
C^0 = C_{E_a}^1 + C_{E_a}^2 \,.
\ee
The blowup introduces an exceptional divisor $E_a$, which is itself a $\mathbb P^1$-fibration over $\Gamma_a$. 
We label the curves $C_{E_a}^1$ and $C_{E_a}^2$ such that $C_{E_a}^2$ 
is the fiber of the divisor $E_a$.
With this convention the intersection numbers of the exceptional curves $C_{E_a}^{1,2}$ with the sections $S_\pm$ and with $E_a$ become
\begin{equation}
\begin{aligned} 
S_\pm \cdot C^0 &= 1 \,, \qquad S_\pm \cdot C^1_{E_a} &= 1 \qquad  S_\pm \cdot C^2_{E_a} &= 0    \label{E1E2int} \cr
E_a     \cdot C^0 &= 0 \,, \qquad E_a \cdot C^1_{E_a} &= 1 \qquad  E_a \cdot C^2_{E_a} &= -1 \,. 
\end{aligned}
\end{equation}

This process can of course be repeated for several different curves $\Gamma_a$ and  followed up by successive blowups in the fiber.
For simplicity of presentation, however, we assume only one such blow-up locus and hence drop the label $a$.

Whenever the base $B_3$ is endowed with such a $\mathbb P^1$-fibration, F-theory on $B_3$ has a clearly identifiable
 heterotic dual \cite{Friedman:1997yq}. Viewed from the dual, weakly coupled eigenframe, the heterotic string theory appears as a four-dimensional compactification
 on a certain Calabi-Yau 3-fold ${\cal Z}_3$. The latter is elliptically fibered over the same base $B_2$ as before,
\bea \label{phet}
\rho :\quad \mathbb E_{\rm het} \ \rightarrow & \  \ {\cal Z}_{3} \cr 
& \ \ \downarrow \cr 
& \ \  B_2  
\eea
Apart from the geometry of ${\cal Z}_3$,  the dual heterotic theory is determined by a gauge background in form of some polystable $E_8^{1} \times E_8^{2}$ bundle $W = V_1 \oplus V_2$.
The particular choice of background depends both on the details of the elliptic fibration $Y_4$ and on the original
F-theory background flux,~$\bG_{U(1)}$. 

Moreover, the optional blow-up along the curve $\Gamma$ on $B_2$ on the F-theory side translates to 
heterotic 5-brane that is wrapped on the same curve $\Gamma$ in $B_2$, now viewed as the base of 
the heterotic 3-fold ${\cal Z}_3$. Such compactifications are inherently non-perturbative from the heterotic perspective.

\section{Elliptic Genera and the Geometry of Modularity}\label{sec_geomelliptic}

We are now in a position to discuss the identification of the relative BPS invariants, defined in the previous section, with the degeneracies of states contributing to the elliptic genus of four-dimensional solitonic strings.
We state this connection in Conjecture \ref{Conjecture1} of Section \ref{subsec_solstrings}, which is a four-dimensional version of the correspondence between BPS invariants and elliptic genera in six dimensions \cite{Klemm:1996hh,Minahan:1998vr,Haghighat:2013gba,Huang:2013yta,Haghighat:2013tka,Hohenegger:2013ala,Haghighat:2014vxa,Huang:2015sta,Hayashi:2015zka,Gu:2017ccq,Hayashi:2017jze,Kim:2017jqn,DelZotto:2017mee,DelZotto:2017mee,Lee:2018urn,Duan:2018sqe,Gu:2018gmy,Gu:2019dan,Gu:2019pqj}.
In Conjecture~\ref{Conj2} of Section \ref{sec_geommod} we present the modular properties of the four-dimensional elliptic genus.
In Section~\ref{subsec_Y3} we point out an intriguing relation between the derivative sector of the elliptic genus and
the BPS invariants of certain threefolds embedded in the Calabi-Yau fourfold.
In Section \ref{sec_m2flux} we explain how these threefold invariants can alternatively be computed from the non-transversal
 $(-2)$-fluxes, even though these do not have a direct interpretation in F-theory.  This leads to an elliptic holomorphic anomaly equation.

\subsection{The Elliptic Genus of Solitonic Strings in Four Dimensions} \label{subsec_solstrings}

From a physical point of view, the main objective of this paper is to obtain a better understanding of four-dimensional critical and non-critical strings. This crucially rests on the observation, which was already put to use in \cite{Lee:2019tst}, that the generating function (\ref{genfun1}) for the relative genus-zero Gromov-Witten invariants coincides, up to a factor, with the elliptic genus of a solitonic string. The aim of this section is to spell out this relationship in greater detail and formulate it as a general conjecture that, supposedly,
applies to all four-dimensional solitonic strings.

Let us first start by discussing how the solitonic strings arise in our context. Consider an F-theory compactification with base space $B_3$. A D3-brane wrapped on a curve $C_{\rm b}$ 
in the base $B_3$ gives rise to a string in the four-dimensional extended spacetime. The worldsheet theory of this string is an $N=(0,2)$ supersymmetric field theory \cite{Lawrie:2016axq}. 
One can now define the elliptic genus, as in (\ref{Zellgendef}), as a trace in the Ramond-Ramond sector
of the $N=(0,2)$ superconformal worldsheet theory of the solitonic string, or equivalently as a partition function on a torus with modular parameter $\tau$.
As before, we consider  a configuration with four-dimensional gauge group $G=U(1)$, 
associated with charge operator $J$.
Then the elliptic genus takes the form (\ref{Zellgendef}):
\be \label{Zellgenusdef1}
Z_{\bG_{U(1)}; C_{\rm b}}(q,\xi) = {\rm tr}_{RR} \left[(-1)^{F_R} F_R  q^{H_L}  \bar q^{H_R}  \xi^J \right] \,,\quad q = e^{2 \pi i \tau},\ \ \xi= e^{2 \pi i z}\,,
 \ee
where $F_R$ is the right-moving fermion number. The extra insertion of $F_R$ is needed to saturate the fermionic
zero modes in the right-moving sector.
The elliptic genus does not only depend on the choice of curve wrapped by the D3-brane, but also on the background flux $\bG_{U(1)}$ of the parent F-theory compactification.
In order for an F-theory interpretation to exist, this flux must satisfy the transversality conditions (\ref{vertcond}).

As a consequence of the supersymmetry in the right-moving sector, the trace (\ref{Zellgenusdef1}) is a meromorphic function of $q$ and $\xi$. In can be expanded as
\be \label{ZexpNdef}
Z_{\bG_{U(1)}; C_{\rm b}} =   - q^{E_0} \sum_{n\geq 0, r} {\cal N}_{\bG_{U(1)};C_{\rm b}}(n,r) \, q^n \, \xi^r \,,
\ee
where 
\be \label{E0def}
E_0 = - \frac{1}{2} \bar K_{B_3} \cdot C_{\rm b}
\ee
 is the zero point energy of the string on $\mathbb T^2$. As we will discuss in a few moments,
  the degeneracies ${\cal N}_{\bG_{U(1)},C_{\rm b}}(n,r)$ at level $n$ and charge $r$ for the flux ackground $\bG_{U(1)}$, are conjectured to 
 agree with the relative BPS invariants  ${N}_{\bG_{U(1)},C_{\rm b}}(n,r)$   that we have
 defined in the previous section.

For a general curve $C_{\rm b}$, the solitonic string that arises from a wrapped D3-brane
 is generically some strongly coupled,  non-critical string in four dimensions  \cite{Mayr:1996sh}. 
We can distinguish three  possible types of strings. 
First, if $C_{\rm b}$ is a shrinkable curve, we can decouple the dynamics of the string from the fields in the bulk of the base $B_3$ by taking the volume of $B_3$ to infinity.
In this case we arrive at a four-dimensional superconformal field theory in the limit of decoupled gravity.
An example of such a string would be a D3-brane wrapped on an exceptional curve. For instance, this can be of the form $C_{\rm b} = C^{1,2}_{E_a}$ as defined in (\ref{CEdef1}), with normal bundle $N_{C_{\rm b}/B_3} = {\cal O}_{C_{\rm b}}(-1) \oplus {\cal O}_{C_{\rm b}}$.
Such strings could be viewed as four-dimensional analogs of the familiar E-strings in six dimensions \cite{Seiberg:1996vs,Ganor:1996mu,Witten:1996qb} and will be discussed in Section \ref{sec_Estring}.

There are also non-critical strings associated with curves whose volume cannot be taken to zero without shrinking $B_3$. 
Such non-critical strings cannot be decoupled from gravity. An example would be for instance a curve $C_{\rm b} = H \cdot H$,
where $H$ is the hyperplane class on $B_3 = \mathbb P^3$. For this curve the normal bundle is $N_{C_{\rm b}/B_3} = {\cal O}_{{C_{\rm b}}}(1) \oplus {\cal O}_{C_{\rm b}}(1)$. This example will be investigated in Section \ref{sec_P3string}.

\pagebreak

Finally, the case where $N_{C_{\rm b}/B_3} = {\cal O}_{{C_{\rm b}}} \oplus {\cal O}_{C_{\rm b}}$ is special: The curve $C_{\rm b}$ must be the fiber, $C^0$, of either a rational fibration of the form (\ref{pproj}) or of an elliptic fibration.
In the first case, a D3-brane wrapped on $C^0$ gives rise to a solitonic, critical heterotic string.\footnote{If $C^0$ is the fiber of an elliptically fibered base $B_3$, we expect instead a Type II string dual in a non-geometric background. The six-dimensional version has been discussed in \cite{Lee:2019wij}. We will not investigate this type of strings further as their elliptic genus vanishes in four dimensions.}
 In its proper duality eigenframe,
this string becomes precisely the fundamental heterotic string compactified on the threefold ${\cal Z}_3$ as given in (\ref{phet}),
additionally equipped with some gauge bundle $W$. Moreover the elliptic genus (\ref{Zellgenusdef1}), as defined via the $\bG$-flux background in F-theory, turns into the (not necessarily perturbative)
chiral partition function of that heterotic string compactification.
More precisely, the degeneracy ${\cal N}_{\bG;C^0}(n,r)$ counts (with signs) the excitations in the Ramond sector
 at excitation level $n$ and charge $r$.
 
For the critical heterotic string, the vacuum energy in (\ref{ZexpNdef}) is
$E_0 = - \frac{1}{2} C^0 \cdot \bar K_{B_3}  = -1$.
The mass of the physical states at excitation level $n$ thus is 
\be
M^2 = 8 \pi T (n-1) \,,
\ee
where $T$ is the tension of the heterotic string. This identifies ${\cal N}_{\bG_{U(1)};C^0}(1,r)$ as
the chiral index over the massless states of charge $r$.
By duality with F-theory, this spectrum must coincide with the physical massless spectrum in the original
 F-theory compactification on $Y_4$, and
therefore the index ${\cal N}_{\bG_{U(1)};C^0}(1,r)$ must agree with the chiral index (\ref{chiQ=r}):
\be \label{chi=calN}
\chi_{\bG_{U(1)},r} = {\cal N}_{\bG_{U(1)};C^0}(1,r) \,.
\ee
Furthermore recall from (\ref{chi=GW}) that $\chi_{\bG_{U(1)},r}$ coincides with the genus-zero Gromov-Witten invariant for the fibral curve $C^{\rm f}_{Q=r}$, which in turn is the same as the BPS invariant $N_{\bG_{U(1)}}(C^{\rm f}_r)$.

This demonstrates that the degeneracies ${\cal N}_{\bG_{U(1)},C^0}(n=1,r)$ are computable from certain BPS invariants of $Y_4$
for the special case of $C_{\rm b} = C^0$ for which ${\cal N}_{\bG_{U(1)};C^0}(1,r) =N_{\bG_{U(1)}}(C^{\rm f}_r)$.
With the situation in six dimensions serving as inspiration, it is natural to conjecture a far more general connection. More precisely, we conjecture that up to an overall factor of $q^{E_0}$, 
the elliptic genus (\ref{Zellgenusdef1}) for {\it any} kind of solitonic string agrees with the generating function of relative BPS invariants (\ref{genfun1}) at genus zero: \\

\noindent\fbox{%
    \parbox{\textwidth}{%

\begin{Conjecture} \label{Conjecture1}
{The generating function ${\cal F}_{\bG_{U(1)}, C_{\rm b}}$ for the relative BPS invariants at genus zero associated with the base curve  $C_{\rm b}$,  for any four-flux background $\bG_{U(1)}$ that satisfies the transversality conditions (\ref{vertcond}),
is proportional to the elliptic genus (\ref{Zellgenusdef1}) for the solitonic string obtained by wrapping a D3-brane on $C_{\rm b}$:}
\be\label{conone}
Z_{\bG_{U(1)}; C_{\rm b}} =  - q^{-\frac{1}{2} C_{\rm b} \cdot \bar K_{B_3}} \,  {\cal F}_{\bG_{U(1)}; C_{\rm b}} \,.
\ee
{\it In particular the relative BPS invariants for $C_{\rm b}$, ${N}_{\bG_{U(1)};C_{\rm b}}(n,r)$, agree with the index-like degeneracies ${\cal N}_{\bG_{U(1)};C_{\rm b}}(n,r)$ of the excitations of the solitonic string at level $n$ and charge $r$:}
\be
{\cal N}_{\bG_{U(1)};C_{\rm b}}(n,r)  = {N}_{\bG_{U(1)};C_{\rm b}}(n,r)  \,.
\ee 
\end{Conjecture}

 }%
}
\vspace{4mm}

\pagebreak

This statement is the analogue of the well-tested duality between certain free energies of elliptic Calabi-Yau threefolds
and the elliptic genera of solitonic heterotic and non-critical strings in six dimensions. The new ingredient in four dimensions, of course, is the dependence on
the F-theory four-form flux $\bG_{U(1)}$ and its respective manifestation in the dual solitonic string.

As a corollary of this proposed general duality and eq.~(\ref{chi=calN}), the relative BPS invariants  $N_{\bG_{U(1)};C^0}(1,r)$, where $C^0$ is the fiber of the $\mathbb P^1$-fibration $B_3$, must agree with the chiral index of states of the F-theory compactification, and thus
\be \label{NGC0id}
N_{\bG_{U(1)};C^0}(1,r) = {N}_{\bG_{U(1)}}(\Cr) = \chi_{\bG_{U(1)},r} \,.
\ee
We will demonstrate this identity in the examples of Section \ref{sec_examples}, via explicit computations in mirror symmetry.

 \subsection{Modular Properties of Four-Dimensional Elliptic Genera} \label{sec_geommod}

The elliptic genus has supposedly  distinguished modular properties, which reflect its definition as a chiral partition function
of  a string wrapped on a torus $\mathbb T^2$ with modular parameter $\tau$. 
For example, 
the elliptic genus of a perturbative heterotic string in $d=6$ dimensions with a single $U(1)$ gauge group factor is a meromorphic Jacobi form of weight $w= - (d-2)/2 = -2$ \cite{Schellekens:1986yi}, as recalled in Section \ref{subsec_recap}.
For more general solitonic strings  in six dimensions the elliptic genus is a meromorphic {quasi}-modular Jacobi form of weight $w=-2$ (see Appendix~\ref{app_modular}).  This applies in particular to solitonic strings that are dual to fundamental heterotic strings in the presence of 5-branes \cite{Lee:2018urn}.
Such modular behaviour is in general agreement with the relation between the elliptic genus and the BPS invariants on elliptic Calabi-Yau threefolds, whose modular properties have been analyzed in \cite{Klemm:2012sx,Alim:2012ss,Oberdieck:2017pqm,Haghighat:2017bep,Schimannek:2019ijf,Cota:2019cjx}.

One  might  expect that  this simple pattern carries over to the strings obtained by wrapped D3-branes in
F-theory compactifications to $d=4$ dimensions that we consider here. 
The expectation would be that the elliptic genus should be a {meromorphic quasi-modular Jacobi form} of weight $w=-(d-2)/{2} = -1$.

 As noticed 
  in \cite{Lee:2019tst} for the special case of a heterotic string, this is not necessarily the case.
 Rather, for some explicit examples studied in that work, it was found that the elliptic genus (\ref{Zellgenusdef1}) can
   in general also {receive contributions which are not given by
 modular or quasi-modular forms.}
 One of the main observations of the present work is that these contributions, while not modular by themselves, can actually be written as derivatives of  modular or quasi-modular Jacobi forms; recall the symbolic representation  (\ref{betterZdecomp}) given in the Introduction. Such objects are special examples of  so-called {\it quasi-Jacobi} forms as defined in Appendix~\ref{app_modular}.  What we encounter is in fact 
a concrete realisation of the mathematical conjecture of \cite{Oberdieck:2017pqm} that relative GW invariants of elliptic fibrations generally assemble into generating functions with values in the ring of  {quasi-Jacobi} forms. In Section \ref{subsec_Y3} we will in addition assign a specific geometrical meaning to the (quasi)-modular Jacobi forms whose derivative appears in the elliptic genera, namely in terms of BPS invariants of certain embedded threefolds.

Extrapolating from these observations we make the following general proposal: \\

\noindent\fbox{%
    \parbox{\textwidth}{%
        \begin{Conjecture} \label{Conj2}
{The four-dimensional, $U(1)$ refined elliptic genus (\ref{Zellgenusdef1}) can 
be written as a sum of meromorphic modular and quasi-modular Jacobi forms of weight $w=-1$ and fugacity index $m= \frac{1}{2} b \cdot C_{\rm b}$, where $b$ is the height-pairing divisor~\eqref{bU1def}, plus derivatives of
modular and quasi-modular Jacobi forms of weight $w=-2$ and the same fugacity index.
More precisely,
\be \label{Zconf2}
Z_{\bG_{U(1)}; C_{\rm b}}(q,\xi) =  {g}^M Z^{M}_{-1,m}(q,\xi) +  { g}^{QM}  Z^{QM}_{-1,m}(q,\xi) +    {g}_\partial^{M}  \xi \partial_\xi Z^{M}_{-2,m}(q,\xi) + {g}_\partial^{QM} \xi \partial_\xi Z^{QM}_{-2,m}(q,\xi) \,,
\ee
where ${g}^{M}$, ${g}^{QM}$, ${ g}_\partial^{M}$, and ${ g}_\partial^{QM}$ are flux-dependent coefficients and 
\bea
Z^{M/QM}_{-1,m}(q,\xi)  = \frac{\Phi^{M/QM}_{w,m}(q,\xi) }{\eta(q)^{12 C_{\rm b} \cdot \bar K_{B_3}}} \,,  \qquad 
Z^{M/QM}_{-2,m}(q,\xi)= \frac{\Phi^{M/QM}_{w-1,m}(q,\xi) }{\eta(q)^{12 C_{\rm b} \cdot \bar K_{B_3}}} 
\,
\eea
with $q=e^{2\pi i \tau}$ and $\xi=e^{2\pi i z}$. Here, the numerators 
$\Phi^{M/QM}_{w,m}(\tau,z)$ denote generic (quasi-)Jacobi-forms of indicated weight and (integer) index
given by
\be
 \quad w= -1 + 6 (C_{\rm b} \cdot \bar K_{B_3}) \,, \qquad m = \frac{1}{2} (b \cdot C_{\rm b})  \,.
\ee
} 
\end{Conjecture}
 }%
}
\vspace{3mm}

As mentioned in the Introduction, the novel (quasi-)modular partition functions $Z^{M/QM}_{-2,m}$ of weight $w=-2$
have the properties characteristic of elliptic genera of chiral,  six-dimensional theories. Indeed,
we will argue that 
they encode the relative BPS invariants 
of certain elliptic three-dimensional sub-manifolds, $\mathbb Y_3$, of $Y_4$.\footnote{As will be clarified in Section~\ref{subsec_Y3}, such a connection to the BPS invariants of three-folds in $Y_4$, as defined by~\eqref{Y3idef}, necessarily arises when the base curve $C_{\rm b}$ is a fiber over either a surface or a curve. Additional supporting evidence for the latter situation is provided in Section~\ref{sec_Estring}. See also~\cite{toappear} for the argument in a most general setup.}
In special cases, when the elliptic threefolds $\mathbb Y_3$ are themselves Calabi-Yau spaces, 
this assertion can be verified explicitly by mirror symmetry.  More generally, we will provide  various general consistency checks which support this claim also when 
$\mathbb Y_3$ is not a Calabi-Yau space. 
As we will demonstrate in Sections \ref{sec_examples} and \ref{sec_non-criticalstrings}, the  general structure we propose
can be verified for a variety of examples, in particular
for all the three different basic types of base curves, $C_{\rm b}$, as characterised in Section \ref{subsec_solstrings}. 

Let us  illustrate at this point the structure for
the important example of the solitonic heterotic string. 
To this end consider a base $B_3$ of $Y_4$ which is the blow-up of a 
rational fibration (\ref{pproj}), and for simplicity assume that the blow-up has been performed only over a single curve $\Gamma$ in 
the base, $B_2$, of this rational fibration. Our notation for this type of geometries has been introduced in Section 
\ref{subsec_P1fibered}.
Again we take the gauge group to be $G = U(1)$,
 so that the only fluxes satisfying the transversality condition (\ref{vertcond}) are the $U(1)$ fluxes~\eqref{GU1def} given by
 \beq
 \bG = \bG_{U(1)} = \sigma \wedge \pi^\ast(F) \,.
 \eeq

In this concrete situation, as exemplified in Section \ref{sec_examples}, one can write the elliptic genus of the heterotic string in the following geometric, closed form:
\be \label{Zhetstr1}
Z_{\bG_{U(1)};C^0} = {g}^{0}  \, Z_{-1,m}^{0} +  {g}^{E}  \, Z_{-1,m}^{E} + \sum_i   {g}^{i}  \frac{1}{2m} \xi \partial_\xi \, Z^i_{-2,m} \,,
\ee
with flux-dependent coefficients explicitly given by
\be\label{fluxcoeffs}
{g}^{0} =  F \cdot C^0\,, \qquad 
{g}^{E}  =    F \cdot C^1_E  \,, \qquad 
{g}^{i}  =   F \cdot b \cdot p^\ast(C_i)\,.
\ee
Here $C^0$ and  $C^1_E$ denote the curves defined around equation (\ref{CEdef1}),  the divisor $b$ is the height-pairing as given in (\ref{bU1def}), and the set $\{C_i\}$ is a suitable basis of curves on the base $B_2$ of $B_3$. The latter point 
will be explained in more detail in Section \ref{subsec_Y3}. Mathematical and physical constraints restrict 
the (quasi-) modular forms in (\ref{Zhetstr1}) as follows:
\bea\label{generalZstructure}
Z_{-1,m}^{0}(q,\xi) &=&\frac1{\eta^{24}(q)} \,\  \phi_{-1,2}(q,\xi)\,\Phi^{0,M}_{12,m-2}(q,\xi)\nn
\,, \\
 Z_{-1,m}^{E}(q,\xi)& =& \frac1{\eta^{24}(q)} \,\ \phi_{-1,2}(q,\xi)\,\Big[ \Phi^{E,M}_{12,m-2}(q,\xi) + \phi_{-2,1}(q,\xi)E_2(q)\Phi^{E,QM}_{12,m-3}(q,\xi)\Big]
 \label{Zstruc}
  \,,\\
Z^i_{-2,m}(q,\xi)  &=& \frac1{\eta^{24}(q)} \,\ \Big[  \Phi^{i,M}_{10,m}(q,\xi) + E_2(q) \Phi^{i,QM}_{8,m}(q,\xi)  \Big]\nn
\,.
\eea
As always, $\Phi^{*,(Q)M}_{w,m}$ denotes a generic weak Jacobi form of specified weight and index, which can be written as a polynomial in the generators of the ring $\Jac^{*,(Q)M}_{w,m}$ of (quasi-)modular Jacobi forms (see Appendix~\ref{app_jacobi}).

Note that since the weight of the $Z_{-1,m}^*$ is odd, these objects are necessarily proportional to the
unique odd-weight generator $\phi_{-1,2}$ of the ring of Jacobi forms. Furthermore, since $\phi_{-1,2}\sim z$,  
they vanish identically unless we refine the elliptic genus with regard to at least one $U(1)$ factor.
Note also that there cannot be a $1/q$ pole in  $ Z_{-1,m}^*$ because the left-moving
ground state is uncharged and so cannot be multiplied by $z$. Thus the  $Z_{-1,m}^*$ are actually holomorphic in $q$
and take a very restricted form, as indicated.

In summary, the relationship (\ref{Zhetstr1})--(\ref{Zstruc}) between flux-dependent geometric intersection 
data on the one hand, and modular, quasi-modular and derivative sectors on the other, is one of the main results of the present work, and is a concrete manifestation of the map (\ref{thebasicmap}) mentioned in the Introduction.

\subsection{Geometric Interpretation of the Derivative Sector of $Z_{-1,m}(q,\xi)$} \label{subsec_Y3}

We now point out an intriguing interpretation of the derivative contributions, $Z^i_{-2,m}(q,\xi)$, to the elliptic genus,
namely in terms of BPS invariants of certain 
threefold geometries, $\mathbb Y_3$, which are embedded in the given elliptic fourfold, $Y_4$.
For the example of the heterotic string, we will be able to explain this interpretation based on our understanding of 
 the moduli space of at least some of the curves whose BPS invariants enter the elliptic genus. 
 The relation between the derivative contributions and certain threefold invariants is, however, not restricted to heterotic strings, as we will show explicitly in Section~\ref{sec_non-criticalstrings}.

To understand the heterotic setup,  let us first assume that the Mori cone of 
effective curves on $B_2$ is simplicial and identify its generators with the basis $\{C_i\}$ on which the coefficients in the sum (\ref{Zhetstr1}) depends
via (\ref{fluxcoeffs}). We will drop the assumption of a simplicial Mori cone at the end of this section. 
The curve classes $b \cdot p^\ast(C_i)$ in   (\ref{Zhetstr1}) can be written as 
\be \label{bpCiform}
b \cdot p^\ast(C_i) = 2m \, C_i +   ({\rm curve \, \, in \,\, the \,\, fiber \,\, of } \,  B_3) \,.
\ee
To see this, note that we can parametrise the height pairing divisor as $b = 2m S_- + b_E E +  \sum_i b_i \, p^\ast(C_i)$, in agreement with $b \cdot C^0 = 2m$.
Then (\ref{bpCiform}) follows from the fact that $S_- \cdot p^\ast(C_i)$ gives back the curve class $C_i$ on $B_2$, while the intersection of $p^\ast(C_i)$ with the divisors $E$ and $p^\ast(C_j)$
lie entirely in the fiber of the rational fibration $B_3$.

Since the curves $C_i$ are the generators of the Mori cone of $B_2$, the dual curves $C^i$ on $B_2$ are generators of the K\"ahler cone of $B_2$, whose closure is contained in the closure of the cone of effective curves.
We assume for simplicity that the $C^i$ are integral, leaving the discussion of a much more general setting to the end of this section.
They are given by
\be \label{etaij}
C^i =  \eta^{ij} C_j     \,,  \qquad {\rm where} \,\,  \,\eta_{ij} = C_i \cdot_{B_2} C_j \,,
\ee
and have the important property, characteristic of generators of the K\"ahler cone, that 
\be \label{Cipos}
C^i \cdot_{B_2} C^j \geq 0  \,.
\ee
In particular, since $C^i \cdot_{B_2} C^i   \geq 0 $, each curve $C^i$ moves in a family on $B_2$.

Now, from (\ref{bpCiform}) we infer that the dual of the component of $b \cdot p^\ast(C_i)$ on $B_2$ is the curve ${C^i}/{2m}$.
{To arrive at integral classes}, let us factor out ${1}/{2m}$ and consider the
 pullback $p^\ast(C^i)$ as a divisor on $B_3$. 
 Since each $C^i$ moves in a family on $B_2$, so does the divisor $p^\ast(C^i)$ on $B_3$.
We can then define a collection of elliptic threefolds,  $\mathbb Y_3^i$, by restricting the elliptic fibration of $Y_4$ to the a {\it generic} member of this family of divisors:
\be \label{Y3idef}
\mathbb Y_3^i = Y_4|_{p^\ast(C^i)} \,.
\ee
Any single such threefold is by construction an elliptic fibration with projection
\be
 \pi_i: \mathbb Y^i_3 \rightarrow \mathbb B_2^i \,, \qquad  \mathbb B_2^i  = p^\ast(C^i)  \,.
 \ee
In fact, the base $p^\ast(C^i)$ is a $\mathbb P^1$-fibration over $C^i$ with generic fiber $C^0$ (blown up at the intersection of $\Gamma$ with $C^i$).  See Figure~\ref{fig_3fold} in the Introduction for an illustration.

Depending on whether $C^i \cdot_{B_2} C^i = 0 $ or $C^i \cdot_{B_2} C^i > 0$, 
the anti-canonical bundle of 
$\mathbb Y_3^i$ is either trivial or negative.
This follows from the adjunction formula:
\be \label{adjunction_sec3}
c_1(\bar K_{\mathbb Y^i_3}) = c_1(\bar K_{Y_4}|_{\mathbb Y^i_3}) - c_1(N_{{\mathbb Y^i_3}/Y_4}) =  - \pi_i^\ast (p^\ast(C^i)|_{p^\ast(C^i)})    \leq  0 \,.
\ee
Even if  $\mathbb Y_3^i$ is by itself not Calabi-Yau,
we can consider the relative BPS invariants with respect to $C^0$ on $\mathbb Y_3^i$, denoted by $N^i_{C^0}(n,r)$,
and package them into a generating function,
${\cal F}^i_{C^0}(\tau,z)$.
We propose that these invariants determine the derivative pieces in the elliptic genus (\ref{Zhetstr1}):
\be \label{ZiFidef}
Z^i_{-2,m}(\tau,z) = - \frac{1}{q} {\cal F}^i_{C^0}(\tau,z) = - \frac{1}{q} \sum_{n,r} N^i_{C^0}(n,r) q^n \xi^r \,.
\ee
The $N^i_{C^0}$ are not to be confused with the invariants $N_{\bG;C^0}$ of the fourfold $Y_4$ discussed before, which were defined relative
to a transversal four-form flux $\bG_{U(1)}$. Rather, we claim that they are relative BPS  invariants of a generic member, $\mathbb Y_3^i$, of the family of elliptic threefolds embedded inside $Y_4$
as  in (\ref{Y3idef}).

If we assume this, the general considerations of \cite{Oberdieck:2017pqm} imply that the generating function $Z^i_{-2,m}(\tau,z)$ 
should be a (quasi-)modular form of weight $w=-2$, 
regardless of whether $\mathbb Y_3^i$ is itself Calabi-Yau or not.
In physics terms this can  be equivalently understood  by first assuming  that $\mathbb Y_3^i$ {\it is} Calabi-Yau and considering F-theory on $\mathbb Y_3^i$ as an auxiliary, chiral theory in six dimensions.
With our assumption on the nature of the $N^i_{C^0}(n,r)$, the object $Z^i_{-2,m}(\tau,z)$ is then simply the elliptic  genus of the $N=(0,4)$ supersymmetric string obtained by wrapping a D3-brane on $C^0$ within $\mathbb Y_3^i$.
As such, $Z^i_{-2,m}(\tau,z)$ is (quasi-)modular of weight $w=-(d-2)/2=-2$.
In fact, if $c_1(\mathbb Y_3^i) = 0$,
 the fourfold $Y_4$ is expected to admit a fibration over the curve $C_i$ whose fiber is a generic member of the family $\mathbb Y_3^i$.\footnote{This natural expectation is the direct analogue of the existence of a K3/$T^4$-fibration for Calabi-Yau threefolds according to Ooguiso's criteria \cite{Ooguiso}. Consistently, note that since $\mathbb Y_3^i$ is defined as the restriction of $Y_4$ to $p^\ast(C^i)$, where $C^i$ is the curve dual to $C_i$ on $B_2$, the intersection product yields $C_i \circ  \mathbb Y_3^i =1$, in agreement with the fibration structure
(\ref{pi3fold}).} That is, there exists a projection

\pagebreak

\bea \label{pi3fold}
\rho_i :\quad \mathbb Y_3^i \ \rightarrow & \  \ Y_4 \cr 
& \ \ \downarrow \cr 
& \ \  C_i  
\eea
The underlying six-dimensional theory naturally arises in the decompactification of the base curve $C_i$ and so is well-defined by itself.

By contrast, if $C^i \cdot C^i > 0$, which implies $c_1(\bar K_{\mathbb Y^i_3}) <  0$, we can view $Y_4$ only locally as a fibration with fiber $\mathbb Y_3^i$ over the normal direction to the divisor $p^\ast(C^i)$ in $B_3$.
Since the normal bundle to $p^\ast(C^i)$  is in this case non-trivial, even in the decompactification limit we cannot define a bona-fide six-dimensional theory by restricting F-theory to ${\mathbb Y^i_3}$. It is even more intruiging to find that nevertheless
$Z^i_{-2,m}(\tau,z)$ behaves in many respects  like the elliptic genus of a six-dimensional string even in this case, as will be discussed further in Section \ref{sec_anomalies}. In particular, one  can formally
associate  $Z^i_{-2,m}(\tau,z)$ to an elliptic surface with $(24  + 12 n)$ singular fibers, where $n = C^i \cdot_{B_2} C^i$. An example will be presented in Section \ref{sec_Ex2het}.
In the sequel we will formally relate the $Z^i_{-2,m}(\tau,z)$ to such six-dimensional sectors, with the understanding that a direct interpretation as elliptic genera in six dimensions is possible only if $\mathbb Y_3^i$ is Calabi-Yau.

Let us now turn to the crucial claim that $Z^i_{-2,m}(\tau,z)$ encodes the BPS invariants $N^i_{C^0}(n,r)$ pertaining to the
threefolds $\mathbb Y_3^i$  as shown in (\ref{ZiFidef}).
In the remainder of this section we will prove this assertion at level $n=1$.
To this end we interpret the expression (\ref{Zhetstr1}) for the elliptic genus as a statement about the decomposition of the moduli spaces of curve classes ${\mathbf C} = C^0 + n \mathbb E_\tau +  \Cr$  into various components.
For the special case where $n=1$ this will allow us to deduce that the multiplicities   $N^i_{C^0}(n=1,r)$ are indeed BPS invariants on $\mathbb Y_3^i$. Extrapolating this observation to all $n$ then leads to (\ref{ZiFidef}).

To arrive at this picture, we first define the following generating functions associated with the non-derivative contributions to the elliptic genus (\ref{Zhetstr1}):
\bea \label{NC0Edef}
{\cal F}^0_{C^0}(\tau,z) &=& - q Z_{-1,m}^{0} = \sum_{n,r} N^0_{C^0}(n,r) q^n \xi^r \,,\\
{\cal F}^E_{C^0}(\tau,z) &=& - q Z_{-1,m}^{E} = \sum_{n,r} N^E_{C^0}(n,r) q^n \xi^r \,.  
\eea
This is analogous to (\ref{ZiFidef}), but with the important difference that, unlike the invariants $N^i_{C^0}(n,r)$ which appear in (\ref{ZiFidef}),
the invariants $N^0_{C^0}(n,r)$ and $N^E_{C^0}(n,r)$ do not correspond to BPS invariants of some auxiliary elliptic {\it three}folds.  Expanding both sides of~(\ref{Zhetstr1}) order by order in $q$ and $\xi$
then gives a relation between the BPS invariants $N_{\bG;C^0}(n,r)$ in $Z_{\bG;C^0}$ on the left hand side, and the invariants $N^0_{C^0}(n,r)$, $N^E_{C^0}(n,r)$ and $N^i_{C^0}(n,r)$ on the right.

Recall next that the quantities on the left hand side,
$N_{\bG;C^0}(n,r)$, are the flux dependent,
 relative Gromov-Witten invariants, as defined in (\ref{GWdef}), for the curve ${\mathbf C} = C^0 + n \mathbb E_\tau + \Cr$ .
 According to (\ref{GWdef}), 
 \be
 N_{\bG_{U(1)};C^0}(n,r) = \int_{\hat \Sigma_{n,r}} {\rm ev}^\ast\bG_{U(1)}  \,,
 \ee
 where we denote by $\hat \Sigma_{n,r}$ the class of the moduli space of stable holomorphic maps at genus $g=0$ and with one point marked associated with ${\mathbf C}= C^0 + n \mathbb E_\tau + \Cr$.
When both sides are expanded in $q^n \xi^r$, 
the relation (\ref{Zhetstr1})  therefore boils down to the statement that
\bea \label{Sigmanrequ}
\int_{\hat \Sigma_{n,r}} {\rm ev}^\ast\bG_{U(1)} =  (F \cdot C^0) N^0_{C^0}(n,r) + (F \cdot C_E^1)N^E_{C^0}(n,r) + 
\sum_i (F \cdot b \cdot p^\ast(C_i))  \frac{1}{2m}  \, r \, N^i_{C^0}(n,r) \,, \nn \\
\eea
where the extra factor of $r$ in front of the $N^i_{C^0}(n,r)$  is caused by the derivative in (\ref{Zhetstr1}).

To interpret (\ref{Sigmanrequ}) further, we note that while for general $n$ the moduli spaces $\hat \Sigma_{n,r}$ are difficult to construct,
we are in a comfortable situation when $n=1$ and  $C^0$ is the rational fiber of $B_3$:
The moduli space $\hat \Sigma_{1,r}$ is equal to the moduli space of stable holomorphic maps of genus zero for the purely fibral curve $\Cr$, up to terms orthogonal to any transversal flux satisfying (\ref{vertcond}).
This follows from the identity (\ref{NGC0id}), which in turn is a consequence of the duality between F-theory and the heterotic string.
Since the moduli space for $\Cr$ is the surface $\hat \Sigma_r$ given in (\ref{rproj}) with base $\Sigma_r$, we know that for transversal flux
\be
\int_{\hat \Sigma_{1,r}} {\rm ev}^\ast\bG_{U(1)} = \int_{\hat \Sigma_r} \bG_{U(1)}  = r \, (F \cdot \Sigma_r) \,.
\ee
Plugging this into (\ref{Sigmanrequ}), which must hold for any element $F \in H^{1,1}(B_3)$, yields 
\be \label{Sigmar-relation}
\Sigma_r = \frac{1}{r}   N^0_{C^0}(1,r) \, C^0  + \frac{1}{r} N^E_{C^0}(1,r)\,  C_E^1  + \sum_i \frac{1}{2m}  \, N^i_{C^0}(1,r) \,  (b \cdot p^\ast(C_i)) \,.
\ee

\begin{figure}[t!]
\centering
\includegraphics[width=6cm]{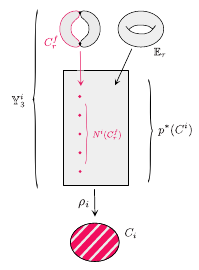}
\caption{This figure shows the component of the matter curve $\Sigma_r$ (drawn in red) along a base curve $C_i$.
It refers to
a geometry where the fourfold $Y_4$ is fibered over  $C_i$ with generic fiber $\mathbb Y_3^i$.
The issue is to determine the multiplicity $N^i(\Cr)$ of $C_i$  
in the decomposition (\ref{Sigmar-relation-2}), and from the picture we see that it coincides with the number of points on $\mathbb Y_3^i$ over which the elliptic fiber $\mathbb E_\tau$ degenerates.
The remaining components of $\Sigma_r$ in the decomposition (\ref{Sigmar-relation-2}) lie within special fibers of the fibration $\rho_i$
(\ref{pi3fold}) and are not depicted.
}
\label{fig_fibr-degen}
\end{figure}

The point is now that, at least for $n=1$, we can understand 
why the $N^i_{C^0}(1,r)$ are invariants of the embedded threefolds $\mathbb Y_3^i$: Suppose first that a given  threefold $\mathbb Y_3^i$ is by itself a Calabi-Yau space. In this case, $Y_4$ admits a fibration with fiber $\mathbb Y_3^i$ over the curve $C_i$ of the form (\ref{pi3fold}), see Figure \ref{fig_fibr-degen}.
The invariant $N^i_{C^0}(1,r)$ appearing in (\ref{Sigmar-relation}) can thus be interpreted as the multiplicity of the base curve $C_i$ of this fibration as a component of the matter curve $\Sigma_r$.\footnote{To see this,
note that $N^i_{C^0}(1,r)$ is the multiplicity of the component $\frac{1}{2m} b \cdot p^\ast(C_i) = C_i + \ldots$ within $\Sigma_r$, where we used (\ref{bpCiform}).} 
In other words
\be \label{Sigmar-relation-2}
\Sigma_r = 
 \sum_i   \, N^i_{C^0}(1,r) \,  C_i  + \text{(curve classes in the fiber of $B_3$)}\,.
\ee
The multiplicity  $N^i_{C^0}(1,r)$ in the above decomposition is simply given by the number of points on $\mathbb Y_3^i$ over which the elliptic fiber degenerates such as to support a state of charge $r$.
Indeed, 
\be
N^i_{C^0}(1,r)  = \Sigma_r   \cdot p^\ast(C^i) \,,
\ee
where we used that $p^\ast(C^i)$ does not intersect with fibral curves and that $p^\ast(C^i) \cdot C_j = \delta^i_j$, which follows from the fact that $C^i$ is the curve class dual to $C_i$ on $B_2$.
But the intersection of  $\Sigma_r $ with the base   $\mathbb B_2^i = p^\ast(C^i)$ of $\mathbb Y_3^i$ gives exactly the locus on  $\mathbb B_2^i$ where the fiber of  $\mathbb Y_3^i$ supports the fibral curve $\Cr$.
Under the present assumption that $C_i$ are the generators of the simplicial Mori cone, the intersection number $\Sigma_r \cdot p^\ast(C^i)$ is non-negative and counts the number of points on $\mathbb B_2^i$ where the degeneration occurs, in codimension-two on $\mathbb B_2^i$.
 This identifies  $N^i_{C^0}(1,r)$ with the BPS invariant for the curve $\Cr$ viewed as a curve inside the threefold $\mathbb Y_3^i$. The same argument leading to (\ref{NGC0id}), now applied to threefolds, shows that this in turn agrees with the BPS invariant of the curve $C^0 + \mathbb E_\tau + \Cr$ as a curve inside $\mathbb Y_3^i$. Hence we have shown that the $N^i_{C^0}(1,r)$ are indeed
  relative BPS invariants of $\mathbb Y_3^i$.

{These considerations continue to hold if $\mathbb Y^i_3$ is not a Calabi-Yau space itself, as the argument was purely intersection theoretic. 
Indeed, the general formalism of \cite{GathmannHabil} implies that the virtual class of the moduli space of curves $C^0 + n \mathbb E_\tau + \Cr$ on $\mathbb Y^i_3$ is related to the class of the moduli space on $Y_4$ by restriction.
Our elementary considerations for the special curves above are a manifestation of this.  }

While it is harder to make this explicit for general wrapping numbers~$n$ from first principles,
we can at least translate the relation  (\ref{Sigmanrequ}) into the following geometric statement:
Up to terms orthogonal to the transversal subspace of $H^{2,2}_{\rm vert}(Y_4)$ defined in (\ref{vertcond}),
$\hat \Sigma_{n,r}$ is equivalent to the class of a fibration of $\Cr$ over a curve $\Sigma_{C^0,n,r}$ which is given by
\be
\Sigma_{C^0,n,r} = 
\frac{1}{r} N^{0}_{C^0}(n,r) \, C^0 + \frac{1}{r} N^{E_1}_{C^0}(n,r)  \, C^1_E  +  \sum_i \frac{1}{2m}  N^i_{C^0}(n,r) \,  (b \cdot p^\ast( C_i))  \,.
\ee
Indeed, integrating $\bG_{U(1)} =  \sigma  \wedge \pi^\ast(F)$ over such a fibration would give a factor of $r$ from the fibral piece along with the class $F$, precisely as reflected in (\ref{Sigmanrequ}).

The crucial claim here,  proven above only for $n=1$, is that the integral numerical coefficients $N^i_{C^0}(n,r)$ are by themselves BPS invariants of the threefolds $\mathbb Y_3^i$ embedded in $Y_4$. 
This implies that  the objects $Z^i_{-2,m}(\tau,z)$ in  (\ref{ZiFidef}) are modular or quasi-modular Jacobi forms of weight $w=-2$, while
the extra factor of $r$ obtained by integration of $\bG_{U(1)}$  explains why the  $Z^i_{-2,m}(\tau,z)$ appear with a derivative in (\ref{Zhetstr1}).
It would be extremely interesting to establish a geometric proof of this interpretation of the $N^i_{C^0}(n,r)$ also for $n>1$.

We have been assuming that the basis $\{C_i\}$ corresponds to the generators of the simplicial Mori cone of $B_2$, so that they are effective and their dual curves satisfy (\ref{Cipos}).  However the results do not depend on this restriction.
More generally, it suffices to pick a basis of effective curves $C^i$ defining the threefold geometries $\mathbb Y^i_3$ as in (\ref{Y3idef}), and to take $C_i$ as the dual basis of 2-cycle classes on $B_2$.
In general this may imply, first, that  not all $C_i$ are effective, and second that $C^i \cdot C^i$ may be  negative. The intersection theoretic argument why $N^i_{C^0}(1,r) $ describes BPS invariants on $\mathbb Y^i_3$ does not hinge upon effectiveness of $C_i$, however, since we can view $N^i_{C^0}(1,r) $ simply as the coefficient in the expansion (\ref{Sigmar-relation-2})
in terms of curve classes. In this more general situation, $N^i_{C^0}(1,r) $ may in particular be negative. This is to be interpreted in such a way that the degeneration locus in $\mathbb Y^i_3$ where the fiber contains the curve $\Cr$ occurs in codimension-one on $\mathbb B_2^i$. Note also that if $C^i \cdot C^i < 0$, the curve $C^i$ is rigid and hence also $\mathbb Y^i_3$ is rigid as a divisor in $Y_4$, but this does not invalidate our arguments either.

The discussion in this section was tailor-made for the special case of heterotic elliptic genera. For non-critical strings the derivative terms in the elliptic genus encode the relative BPS numbers of embedded threefolds sometimes constructed in a slightly different way.  For this we refer to Section \ref{sec_non-criticalstrings}.

\subsection{Elliptic Holomorphic Anomaly Equation   
} \label{sec_m2flux}

So far  we have focused in this section on the properties of the relative BPS invariants in transversal flux backgrounds, which were defined via eq.~(\ref{vertcond}). 
This is required for having a four-dimensional interpretation within F-theory on $Y_4$, and a meaningful elliptic genus in the first place.
We have seen that in general derivative contributions to the elliptic genus appear, such as indicated in (\ref{Zhetstr1}). These take the form of
derivatives of Jacobi forms, $Z^i_{-2,m}(q,\xi)$,
 which encode relative BPS invariants associated with the embedded threefolds, $\mathbb Y_3^i$, of $Y_4$.
 
 As previewed in  Section~\ref{sec_summary}, this results not only in a violation of modular invariance (\ref{jacobitrApp}), but also
 of the invariance under  the transformations $z\rightarrow z+\lambda\tau$ in  (\ref{periodicity}), which 
 can be interpreted as spectral flow in the $U(1)$ subsector.\footnote{Originally spectral flow was understood \cite{Schwimmer:1986mf}
as a property of an $N=2$ superconformal symmetry on the worldsheet, hence the name, but in essence
this notion applies to any $U(1)$ current algebra associated with a free compact boson. We reiterate again that
it is not an automatic symmetry of the theory.} From the point of view of the four dimensional elliptic genus, the derivative terms inflict
an anomaly on these symmetries, and turn it into a quasi-Jacobi form. However, in analogy to  the treatment of the quasi-modular form $E_2$,
one can cancel these anomalies by trading the derivative terms in $Z_{\bG_{U(1)}; C_{\rm b}}(q,\xi)$
 against non-holomorphic derivatives as in (\ref{pznonhol}). This means that we pass from the holomorphic, but non-modular elliptic genus $Z_{\bG_{U(1)}; C_{\rm b}}(q,\xi)$  to the almost holomorphic, but modular quantity 
 $\hat Z_{\bG_{U(1)}; C_{\rm b}}(q,\xi)$, for which we replace the derivatives
   $ \xi \partial_\xi \equiv \frac1{2 \pi i }\partial_z$ as follows (see Appendix (\ref{app_jacobi})):
 \be
 \xi \partial_\xi     \rightarrow \frac1{2 \pi i }\,  \nabla_{z,m}:= \frac1{2 \pi i }\left(\partial_z + 4\pi i m  \frac{{\rm Im}z}{{\rm Im}\tau}  \right) \,.
 \ee
The resulting non-holomorphicity of $\hat Z_{\bG_{U(1)}; C_{\rm b}}(q,\xi)$ is characterised by the equation
\be \label{holellanom}
\frac{d}{d \alpha} \hat Z_{\bG_{U(1)}; C_{\rm b}}(q,\xi) = ( 2m)  {g}_\partial  Z_{-2,m}(q,\xi)  \,,
\quad  \qquad  \alpha  \equiv \frac{{\rm Im}z}{{\rm Im}\tau}\,,
\ee
where we  parametrise the elliptic genus as in (\ref{Zconf2}) and have abbreviated
 ${g}_\partial^{M}  Z^{M}_{-2,m}(q,\xi) + {g}_\partial^{QM}  Z^{QM}_{-2,m}(q,\xi)  =:  {g}_\partial  Z_{-2,m}(q,\xi)$.
This equation 
is a version of the {\it elliptic holomorphic anomaly equation} which was introduced in the context of generating functions for relative BPS invariants of elliptic fibrations  in \cite{Oberdieck:2017pqm}.
Compared to elliptic genera in six dimensions, which are lacking a derivative contribution, we see that the appearance of such an elliptic holomorphic anomaly equation is a genuinely new feature in four dimensions. 
 
 The elliptic holomorphic anomaly equation admits a beautiful geometric interpretation. Recall that the right-hand side of (\ref{holellanom}) consists of the (quasi-)modular objects, $Z_{-2,m}(q,\xi)$,  that encode the relative BPS invariants of the threefolds $\mathbb Y_3$.
 Remarkably, as we explain now, the same threefold invariants appear also 
 as fourfold invariants for a suitable choice of non-transversal background  flux $\bG_{-2} \in H^{2,2}_{(-2)}(Y_4,\mathbb R)$, as defined in (\ref{H22vertdecomp}).
 We will refer to such non-transversal fluxes as ``$(-2)$-fluxes'' for brevity.\footnote{We reiterate from the discussion in Section~\ref{subsec_relBPS}   that such fluxes do not admit a lift to F-theory, but define bona fide Type IIA/M-theory backgrounds.} 
 
  In fact, it has already been observed in  \cite{Haghighat:2015qdq,Cota:2017aal}
 that the generating function for certain relative genus-zero BPS invariants in a $(-2)$-flux background
 is a meromorphic (quasi-)modular form of weight $w=-2$. Such invariants are non-vanishing even in absence of a refinement by an extra gauge symmetry (in F-theory language). This is to be contrasted with the relative BPS invariants for the transversal flux backgrounds that were considered in the previous sections. 
 The important point is that  the elliptic holomorphic anomaly equation in its form (\ref{holellanom}) admits a representation in terms of the generating function for a specific type of $(-2)$-flux. 
 More precisely we have
 \be \label{holellanom-2}
\frac{d}{d \alpha} \hat Z_{\bG_{U(1)}; C_{\rm b}}(q,\xi) = Z_{\bG_{-2}; C_{\rm b}}(q,\xi)   \,,
\ee
where the $(-2)$-flux associated with $\bG_{U(1)} = \sigma \wedge \pi^\ast(F)$ is given by 
\be   \label{Gm2U1}
\bG_{-2} =     \pi^\ast(F) \wedge \pi^\ast(b)  \,.
\ee
Here $b$ is the height pairing associated with the $U(1)$, and, as always, we have defined
\be \label{G-2gendef}
Z_{\bG_{-2}; C_{\rm b}}(q,\xi)  = - q^{-\frac{1}{2} C_{\rm b} \cdot \bar K_{B_3}} {\cal F}_{\bG_{-2}; C_{\rm b}}(q,\xi)    \,.
\ee

The specific form of $\bG_{-2}$ that appears in (\ref{holellanom-2}) can be deduced from the general considerations of  \cite{Oberdieck:2017pqm} applied to our situation.
Alternatively, we can arrive at the same conclusion via more elementary geometric observations based on the results of Section \ref{subsec_Y3}, which are  supported by our detailed analysis of examples in Sections~\ref{sec_examples} and~\ref{sec_non-criticalstrings}.
 
  For illustration, consider the relative BPS invariants associated with a  curve $C^0$ with zero self-intersection, 
which figures as the fiber of some rational fibration, $B_3\subset Y_4$, see (\ref{pproj}); recall that this is the situation where a solitonic heterotic string appears.
In this case the elliptic holomorphic anomaly equation for the elliptic genus (\ref{Zhetstr1}) takes the form
  \be \label{holanomhet1}
  \frac{d}{d \alpha} \hat Z_{\bG_{U(1)}; C^0}(q,\xi)  =  \sum_i F \cdot b \cdot p^\ast(C_i)  \, Z^i_{-2,m}   (q,\xi)  =   Z_{\bG_{-2}; C^0}(q,\xi)    \,,
   \ee
  for $Z_{\bG_{-2}; C_{\rm b}}(q,\xi)$  as in (\ref{G-2gendef}).
  While the first equality follows immediately from (\ref{Zhetstr1}), the second equality is non-trivial and 
  rests on the following geometric considerations:
  First introduce a convenient basis for the space of non-transveral $(-2)$-fluxes in $H^{2,2}_{(-2)}(Y_4,\mathbb R)$:
  \bea
 \bG^i &=&  \pi^\ast(S_-) \wedge   \pi^\ast(p^\ast(C^i))  \,,  \nn \\ 
  \bG^{E} &=&  \pi^\ast(E) \wedge   \pi^\ast(p^\ast(C^\Gamma))  \,,  \label{basism2gen}      \\
  \bG^{0} &=&  \pi^\ast(p^\ast(C^{i_0}))  \wedge \pi^\ast(p^\ast(C_{i_0}))   \nn  \,.
 \eea  
In the notation of Section \ref{subsec_P1fibered}, $C^\Gamma$ denotes a curve class on $B_2$ with $C^\Gamma \cdot \Gamma =1$, and we have picked a pair of dual curves $C_{i_0}$ and $C^{i_0}$ on $B_2$.
The classes $\bG^{E}$ and $\bG^{0}$ are hence dual to the curves $C_E^2$ and $C^0$ in the fiber of $B_3$, while $\bG^i$ is dual to the curve $C^i$ on $B_2$.
 
The important claim is that  
  the contributions $Z^i_{-2,m}(q,\xi)$ to the elliptic genus (\ref{Zhetstr1})  can be computed as
 \be \label{Zifromm21}
 Z^i_{-2,m}(q,\xi) = - \frac{1}{q} {\cal F}_{\bG^i}(q, \xi)        \,,
 \ee
while the remaining two basis elements lead to the following vanishing BPS invariants,
\be \label{Z0ZEvanishing}
 {\cal F}_{\bG^E}(q, \xi)  = 0 \,, \qquad  {\cal F}_{\bG^0}(q, \xi)  = 0    \,.
\ee
We will provide arguments for these assertions below. Assuming  (\ref{Zifromm21}) and (\ref{Z0ZEvanishing}) for now, we proceed by
expanding (\ref{G-2gendef}), viewed as class of  the $(-2)$ flux,  in the above basis as
\be
\bG_{-2} =  \sum_i ( F \cdot b \cdot p^\ast(C_i)) \, \bG^i   - ( F \cdot b \cdot E) \, \bG^{E} +  ( F \cdot b \cdot S_+) \, \bG^0   \,.
\ee
Hence by linearity of the BPS invariants, together with  (\ref{Zifromm21}) and (\ref{Z0ZEvanishing}),  it is obvious that 
\bea
Z_{\bG_{-2}; C^0}(q,\xi) 
 &=&  \sum_i ( F \cdot b \cdot p^\ast(C_i)) \, Z^i_{-2,m}(q,\xi) \, .
\eea 
This explains also the second equality in  the elliptic holomorphic anomaly equation  (\ref{holanomhet1}).

It remains to justify  (\ref{Zifromm21}) and (\ref{Z0ZEvanishing}). While we cannot give formal proofs beyond the non-trivial checks in the examples of Section \ref{sec_examples}, we provide instead 
some intuition why (\ref{Zifromm21}) should hold. To this end,
 consider the relative BPS invariants $N_{\bG^i; C^0}(n,r)$ for $n=0$ and $r=0$.
According to our claim, these invariants should agree with the threefold invariants $N^i_{C^0}(n,r)$  for $n=0$ and $r=0$ in the expansion (\ref{ZiFidef}) of $Z^i_{-2,m}(q,\xi)$.
Our starting point to verify this is the definition of $N_{\bG^i; C^0}(n,r)$ as the overlap of the flux $\bG^i$ with the virtual fundamental
class of the moduli space of the curve $C^0$ in $Y_4$ with one point fixed.
Since $C^0$ is fibered over $B_2$, this class can be identified with the class of the surface obtained by fibering $C^0$ over the canonical divisor $K_{B_2}$;\footnote{
To understand this claim in physics terms, recall, for instance,
how in F-theory on elliptic fourfolds one computes the BPS invariants for the rational fibers of the exceptional divisors appearing in codimension-one: These are obtained by integrating
 the flux over
the restriction of the rational fiber to the canonical class, $K_D$. Here $D$ is the divisor over which the rational curve is fibered. 
See for example eq.~(9.43) in \cite{Weigand:2018rez}.}
that is, with the class inside $Y_4$ given by
\be \label{S0KB2}
\mu = S_0 \wedge \pi^\ast(p^\ast(K_{B_2}))  \,.
\ee
Here $S_0$ is the zero-section of the fourfold $Y_4$.
This allows us to compute $N_{\bG^i; C^0}(0,0)$ as 
\bea \label{NiGicomp1}
N_{\bG^i;C^0}(0,0) &=&  \int_\mu \bG^i = S_0 \circ \pi^\ast(p^\ast(K_{B_2})) \circ \pi^\ast(S_-) \circ   \pi^\ast(p^\ast(C^i))  = p^\ast(K_{B_2}) \cdot S_- \cdot  p^\ast(C^i) \nn \\
&=&C^i \cdot_{B_2} K_{B_2} \,,
\eea
where we have used the fact that $S_0$ is a section on $Y_4$ and $S_-$ is a section on the $\mathbb P^1$-fibration $B_3$.

The expression for $N_{\bG^i;C^0}(0,0)$  in (\ref{NiGicomp1}), in fact, can be seen to exactly agree with the invariants $N^i_{C^0}(n,r)$ for $n=r=0$.
To see this, suppose first that the curve $C^i$   is a rational curve with $C^i \cdot_{B_2} C^i =0$. According to the discussion in Section \ref{subsec_Y3}, in this case the threefold $\mathbb Y_3^i$ is a Calabi-Yau space which is K3-fibered over $C^i$. 
The invariant  $N^i_{C^0}(n,r)$ for $n=r=0$ is then simply the BPS invariant for $C^0$ within the Calabi-Yau threefold $\mathbb Y_3^i$, and hence $N^i_{C^0}(n,r) =-2$. This is because  within $\mathbb Y_3^i$,  $C^0$  is fibered over the rational base curve  $C^i$ and the (signed) Euler character of its moduli space is the integral $\int_{C^i} K_{C^i} $. By the adjunction formula
\be
\int_{C^i} (K_{C^i}- C^i)  =  C^i \cdot_{B_2} K_{B_2}  \,,
\ee
we see that $\int_{C^i} K_{C^i} $ agrees with  $C^i \cdot_{B_2} K_{B_2}$ for the rational curve $C^i$ with $C^i \cdot C^i =0$. 

{More generally, even if $\mathbb Y_3^i$ is not Calabi-Yau, $N^i_{C^0}(0,0) = C^i \cdot_{B_2} K_{B_2}$ nonetheless holds. As noted already before, this reflects the fact that, by the results of \cite{GathmannHabil}, the virtual class for the moduli space of the curve $C^0$ on $\mathbb Y^i_3$ is related to the class of its moduli space on $Y_4$ by restriction.}

In the examples of Sections~\ref{sec_examples} and~\ref{sec_non-criticalstrings}, we will indeed observe
a precise match between the $N_{\bG^i; C^0}(n,r)$ and the invariants $N^i_{C^0}(n,r)$ for zero, but also for non-zero, values of $n$ and $r$. 
Beyond such explicit examples it is much harder to make a direct argument for general values, based on the moduli space of curves.
At any rate, we will explain the analogue of (\ref{NiGicomp1}) for the elliptic genera also for other types than heterotic strings in four dimensions. 
Furthermore, it is clear that $N_{\bG^0;C^0}(0,0)=0$ and $N_{\bG^E;C^0}(0,0)=0$, because the overlaps in (\ref{NiGicomp1}) vanish geometrically. The vanishing of the invariants at all levels, as claimed by (\ref{Z0ZEvanishing}), 
will be explicitly verified for the examples further below.

\section{Elliptic Genera,  Anomalies and Modularity}\label{sec_anomalies}

The observation of the previous sections was that the four-dimensional elliptic genus need not be modular or quasi-modular in the usual sense.
Once applied to heterotic strings, this raises the question how this phenomenon is compatible with the structure of anomaly cancellation.
In this section we first review the well-known interplay \cite{Schellekens:1986yi,Schellekens:1986xh} of the elliptic genus of the heterotic string with the structure of 1-loop anomalies and their cancellation by the Green-Schwarz mechanism. 
We then explain how four-dimensional anomaly cancellation works
even when the elliptic genus is not modular but rather has a derivative component, and also discuss the situation when it is quasi-modular rather than fully modular.

In $d=2n +2$ dimensions, the 1-loop gauge and gravitational anomalies are characterized by the anomaly polynomial
\be \label{Idp2def}
I_{d+2} = \sum_{\bf R,s} n_s({\bf R}) I_s({\bf R})\Big|_{d+2} \,,
\ee
where we sum over all massless particle species of multiplicity $n_s({\bf R})$  in representation ${\bf R}$ of the gauge group and with spin $s$. The $(d+2)$-form $I_s({\bf R})|_{d+2}$ is 
formed by products of the gauge field strength $F$ and the curvature 2-form $R$.
For example, a complex chiral Weyl fermion contributes to (\ref{Idp2def}) with
\be
I_{1/2}({\bf R}) = {\rm tr}_{\bf R} \,  e^{F}  \, \hat A(T) \,,
\ee
where  $\hat A(T)$ is the A-roof genus.

In the following we will focus  on the gauge anomalies associated with a single $U(1)$ gauge group and define
\be
I_{d+2, U(1)}  = A^{(d)} \, F^\frac{d+2}{2} \,.
\ee
Based on the above expressions, the anomaly coefficients $A^{(d)}$ in $d=6$ and $d=4$ dimensions are given by: 
\bea \label{A1loopnorm}
A^{(6)} &=& \frac{1}{4!} \sum_r  n_{1/2}(r) \,  r^4  =  \frac{1}{4!} \sum_r   n_{\rm half-hyper}(r) \, r^4 \,,      \\
A^{(4)} &=& \frac{1}{3!} \sum_r  \left(n_{1/2}(r) -n_{-1/2}(r)\right)  \, r^3 = \frac{1}{3!} \sum_r \chi_r \, r^3  \,.
\eea
Here we sum over the Weyl fermions of $U(1)$ charge $r$. In $d=6$ dimensions, if we consider a theory with minimal $N=(1,0)$ supersymmetry, the number of charged Weyl fermions agrees with the number of half-hypermultiplets of corresponding charge. In $d=4$ dimensions, the anomaly coefficient involves the chiral index $\chi_r$, i.e., the number of chiral minus anti-chiral Weyl fermions of charge $r$.

\subsection{Modular Elliptic Genera} \label{subsec_Anomaliesmod}

We will first review anomaly cancellation in the more familiar case of a modular invariant heterotic string elliptic genus.
In the present context, this primarily concerns flux compactifications that are dual to perturbative heterotic strings.

Let us recall that 
 the $U(1)$-refined elliptic genus $Z_{-n,m}(q,z)$ of a perturbative, $d=2n+2$ dimensional heterotic string is expected to be
 a weak Jacobi form  \cite{Kawai:1993jk} of modular weight $w=-n$ and some index $m$ (in this work, $n=1,2$ will be relevant).
 As we have seen, this is not necessarily true in $d=4$ and so the statements that follow will eventually be adapted to the more general situation. Let us however for the moment assume that $Z_{-n,m}(q,z)$   is a  weak Jacobi form, and turn later to the required modifications. See also the remarks in the Introduction and the defining modular transformation
properties of Jacobi forms given in (\ref{jacobitrApp}) and~(\ref{periodicity}).

From a weak Jacobi form one can always strip off the quasi-modular Eisenstein series $E_2$ by writing
\beq\label{jacm}
Z^{(d)}_{-n,m}(q,\xi)=  e^{\frac m{12}E_2\zh^2}\check Z^{(d)}_{-n,m}(q,\zh)\,,    \ \ \  {\rm where }\ \ 
 \xi=e^{2\pi i z} \equiv e^\zh
 \,,
\eeq
so that the remainder,
\beq
\check Z^{(d)}_{-n,m}(q,\zh) =\zh^{n ({\rm mod\ }2)} \!\sum_{k\geq0}^\infty  M_{2(k- [\frac{n}{2}] )}(q)\zh^{2k},
\eeq
is modular invariant term by term. It involves meromorphic modular forms $M_{2l}$ of 
weight $2l$ which lie in the ring generated by $E_4$ and $E_6$, divided by  $1/\eta^{24}$. 
The idea of how this modular structure implies the Green-Schwarz anomaly cancellation goes back to 
\cite{Schellekens:1986yi,Schellekens:1986xh} and rests on two properties of the elliptic genus:

1) The anomaly coefficient (\ref{A1loopnorm}) is the coefficient of $\zh^{n+2}$ of the elliptic genus at $q^0$. More precisely:
\bea\label{Anom}
A^{(d)}&=& {-}\frac{1}{2}  \frac1{(n+2)!} \Big[ {\partial_\zh}^{n+2}Z^{(d)}_{-n,m}(q,\zh) \Big]_{\zh^0q^0}\ \\
 &=&  -\frac{1}{2} \, \left[ M_2(q) +  \frac m{12}E_2M_0(q)+ \frac12 (\frac m{12}E_2)^2M_{-2}(q)+ ....\right]_{q^0}\,.\nn
\eea
Key is the observation that due to well-known properties of modular forms, $M_2(q)$ cannot have a constant piece and therefore does
not contribute. Thus all contributing terms must involve $E_2$'s from the exponential, which brings down powers of $\zh^2\sim \Tr F^2$
(recall that $F=\zh J$, where $J$ is the charge generator).
This is tantamount to saying that the anomaly polynomial must necessarily factorize. This in turn implies that the anomaly can be cancelled; that is, in familiar terms:   $I_{2n+4}(F)\sim {\rm Tr} F^2 \wedge X_{2n}(F)$.

2) The Green-Schwarz anomaly cancelling term is given by $S_{GS}\sim\int B\wedge X_{2n}(F)$, and its numerical coefficient,
\be
A^{(d)}_{GS}\equiv X_{2n}(F)\big\vert_{F^n} \,,
\ee
 is computed as a one-loop amplitude in the heterotic string
 \cite{Lerche:1987sg,Lerche:1987qk}.  
The integrand is given by the modular invariant coefficient of $\zh^n$ in the
elliptic genus. More precisely, what enters is the modified, almost holomorphic elliptic genus, $\hat Z_{-n,m}(q,\zh)$,  
for which all occurrences of $E_2$'s are replaced by their modular invariant, but non-holomorphic version, 
$\hat E_2=E_2-\frac3{\pi{\rm Im}\tau}$. Explicitly:
\bea\label{ACT} \nn
A^{(d)}_{GS} \!&=&\!   - \frac{1 }{n!} \Big[ {\partial_\zh}^{n}\frac1{16\pi}\int_{{\cal F}_\tau}\! d\tau\, \hat Z^{(d)}_{-n,m}(q,\zh) \Big]_{\zh^0} \\ 
 \!&=&\!
- \frac{1}{16\pi}\int_{{\cal F}_\tau}\! d\tau  \left[ M_0(q)+  \frac m{12}\hat E_2M_{-2}(q)+  ...\right] \\ \nn
\! &=&\! 
- \frac{1}{24}   \Big[ \hat E_2 M_0(q) + \frac12 \frac m{12}\hat {E_2}^2M_{-2}(q)+  ...\Big]_{q^0} \,,\nn
\eea
where the integration over the worldsheet $\mathcal F_\tau$ is performed via the formula  \cite{Lerche:1987qk}
\beq
\int_{{\cal F_\tau}} d\tau {\hat E_2(q)}^k M_{-2k}(q)=\frac{2\pi}3\frac1{k+1}E_2(q)^{k+1} M_{-2k}(q)\big\vert_{q^0} \,.
\eeq
Therefore one finds
\be \label{anomalycancell}
m A^{(d)}_{GS}\ =\ A^{(d)}   \,,
\ee
which expresses that the Green-Schwarz term precisely cancels the anomaly. The extra factor of $m$ arises from
the Chern-Simons term to which the other leg of the $B$-field couples, as we will show later.
 
We can be more explicit if we specify the dimension. In $d=6$, the scarcity of independent modular forms, plus the requirement that the
left-moving ground state be uncharged, implies that the elliptic genus up to order $\zh^4$
 is fixed up to one model dependent parameter besides $m$, which we denote by $c$:
 \be
Z_{-2,m}^{(6)}(q,\zh) =  \frac1{ \eta^{24}}
e^{\frac m{12}E_2\zh^2}\Big[ 2 E_{4} E_{6}-\frac{m \zh^2}{12}  \left(2 E_{6}^2+c \left(E_{6}^2-E_{4}^3\right)\right)
+\left(\frac{m \zh^2}{12}\right)^2 E_{4}^2   E_{6}
+\cO(\zh^6)
   \Big]\,.
 \ee
 This then leads to
 \be
 A^{(6)}\ =\ m A^{(6)}_{GS} \ =\  - 6(1+c)m^2\,.
 \ee
For reference,  examples of elliptic genera that we will meet again further below
are given by
 \be\label{ZE4E6}
 Z^{(6)}_{-2,m}(q,z) \ =\  \frac{1}{12}    \sum_k  \frac{n_k}{\eta^{24}}   E_{4,\mu^k_1} E_{6,\mu^k_2}\,, \  \ \  m=\mu^k_1+\mu^k_2\,,     \qquad \sum_k n_k = 24  \,,
 \ee
 where $E_{w,m}=E_{w,m}(q,z)$ are the Eisenstein-Jacobi forms\footnote{
 Strictly speaking we use  for higher levels $m$ here and in the following
 the integral expansions~(\ref{integralE4}-\ref{integralE6}). However their
parametric ambiguities do not project down to order $\zh^4$, 
 and therefore do not contribute to the anomaly.}
 defined in Appendix \ref{app_modular}. This bilinear form of elliptic genera naturally appears
in perturbative heterotic strings compactified on $K3$, with bundles switched on such as to
leave a single $U(1)$ unbroken. In this case $c$ is fixed such that
  \be\label{A6gen}
 A^{(6)}\ =\ m A^{(6)}_{GS}  \ =\   -  \sum_k   \frac{n^k}{4}    ((\mu^k_1)^2-(\mu^k_2)^2)\,.  
 \ee
 
Analogous statements apply to perturbative heterotic strings in $d=4$. Given the various constraints, we find
for the modular part of the elliptic genus:
 \be\label{UniqueZ}
 Z^{(4),M}_{-1,m}(q,\zh) =    \frac 1{\eta^{24}(q)} \,\phi_{-1,2}(q,\zh)\,\Phi^{M}_{12,m-2}(q,\zh)
\simeq   \zh e^{\frac m{12}E_2\zh^2}\Big[1+\cO(\zh^4)\Big]\,,
 \ee
up to an overall numerical factor $c$.
The actual elliptic genus is then obtained by multiplying this with $g^M$ as in (\ref{Zconf2}), which
depends on the four-form flux that needs to be switched on.
The two overall factors fix the elliptic genus up to order $\zh^3$ and we thus have
 \be\label{4dcancell}
 A^{(4)}\ =\   m A^{(4)}_{GS} \ =\ { -}  \frac c{24}  g^M m \,.
 \ee
A four-dimensional analog of (\ref{ZE4E6}) in terms of natural building blocks of the heterotic string can be written as
\be\label{antiDZ}
Z_{-1,m}^{(4),M}(q,\zh) \ {=} \ \frac1{12}\frac {1}{\eta^{24}}  \Big[\frac1{\mu_1}(\partial_\zh E_{4,\mu_1}) E_{6,\mu_2}-\frac1{\mu_2} E_{4,\mu_1} (\partial_\zh E_{6,\mu_2})\Big]
  \,, \  \ \  m=\mu_1+\mu_2\,,
 \ee
 which by virtue of the relations (\ref{nablarelations})  is a Jacobi form proportional to $\phi_{-1,2}(q,\zh)$ and thus a special case of
(\ref{UniqueZ}). Including the flux dependent pre-factor, it leads to  $A^{(4)}=mA^{(4)}_{GS} \, {=} \, { -}  g^M m$.
Obviously, if several such terms appear, they can simply be summed over.

The structure of derivatives as exhibited in (\ref{antiDZ}) points to a deeper r\^{o}le derivatives play for four-dimensional elliptic genera.
One can view (\ref{antiDZ}) as a special linear combination that happens to be modular, while
the other natural combination is precisely of the form of the non-modular, derivative piece of the elliptic genus:
\bea\label{DzZ}
 Z^{(4),\partial}_{-1,m}(q,\zh) &{ =}&  \frac1{\eta^{24}}\partial_\zh ( E_{4,\mu_1} E_{6,\mu_2}(q,\zh)) \nn\\
  &{=} & \partial_\zh    Z^{(6)}_{-2,m}(q,\zh)  \,.
 \eea
Of course,
 in the physical partition function one needs to equip this with an additional flux-dependent prefactor, which we denote by~$g^{M,\partial}$.
 
 This leads us to discuss the derivative elliptic genera $Z^{(4),\partial}_{-1,m}(q,\zh)$ more generally, and in particular to
the question how anomalies can be cancelled in view of the fact that the $Z^{(4),\partial}_{-1,m}(q,\zh)$ are
 not modular invariant, while modular invariance was instrumental in proving anomaly cancellation in the first place.
 
Note, however, that because of the derivative relationships between the elliptic genus and the anomaly polynomial (\ref{Anom}) on the one hand, and the anomaly cancelling term (\ref{ACT}) on the other, 
it follows that the factorization of the anomaly in 4d is inherited from the 
six-dimensional one, even though the 4d elliptic genus $Z^{(4),\partial}_{-1,m}(q,\zh)$ does not transform as a Jacobi form. 
Neverthless, as already noticed in \cite{Lee:2019tst},  it has good quasi-modular
properties at any given fixed order in the $\zh$-expansion so that a well-defined modular integration, as required in (\ref{ACT}),
can be performed. 
  
More precisely, for a four-dimensional elliptic genus of the derivative form (\ref{DzZ}), expressions (\ref{Anom}) and (\ref{ACT}) evaluate to
\bea
A^{(4),\partial} &=&   {-} {\partial_\zh}^4 \frac1{12} Z^{(6)}_{-2,m}(q,\zh)\big\vert_{\zh^0q^0} = 4 A^{(6)} = 4 m A_{GS}^{(6)} \,,  \label{4d6d1}\\
A_{GS}^{(4),\partial} &=& { -}{\partial_\zh}^2 \frac1{16\pi } \int_{\cal F} d\tau \hat Z^{(6)}_{-2,m}(q,\zh) \big\vert_{\zh^0}=  2  A_{GS}^{(6)} \,. \label{4d6d2}
 \label{AGS 54d6d}
\eea
We can understand the relative factor of $4$ between $A^{(4),\partial}$  and $A^{(6)}$ diagrammatically, in that it corresponds to
 the $4$ choices of one of the $4$ external legs to be given a VEV in the transition from the quartic six-dimensional anomaly to the cubic four-dimensional anomaly. Similarly the factor of $2$ relating  $A_{GS}^{(4),\partial}$ and $A_{\rm}^{(6)}$ originates in the choice of two external legs in the six-dimensional Green-Schwarz terms, as compared to the single external leg in its four-dimensional analogue.

The mismatch by a factor of two in the derivative sector,
\be\label{deridef}
A^{(4),\partial}/(mA_{GS}^{(4),\partial}) = 2,
\ee
indicates that the anomaly is not cancelled by the standard Green-Schwarz term involving the universal $B$-field.
Rather, as we will argue later in Section~\ref{subsec_effanomaly}, the hidden six-dimensional geometry of the 
derivative sector will always correlate with the correct number of additional $B$-fields 
and their couplings, such that  in the end all such anomalies will be cancelled.

Before discussing this point, however, we turn to the other subsector of the $d=4$ elliptic genus, namely the one for which
the elliptic genus is quasi-modular rather than modular.

\subsection{Quasi-Modular Elliptic Genera} \label{subsec_AnomaliesQuasimod}   

We have indicated before that for certain fluxes ``non-perturbative'' elliptic genera can arise that are only quasi-modular, which means that
$E_2$ pieces can appear in $\check Z_{-n, m}^{(d)} (q,\hat z)$ even after stripping off the exponential prefactor as in~(\ref{jacm}). For these fluxes
the arguments about factorization of the anomaly polynomial do not hold. Moreover the 
one-loop computation of the standard Green-Schwarz term that involves the universal $B$-field 
will in general not be applicable, as it can possibly capture only the perturbative piece of the anomaly, and so
further Green-Schwarz terms that involve extra $B$-fields will necessarily come into play.

Let us be more specific and consider first six-dimensional theories for which the
non-pertur\-bative, quasi-modular piece of the heterotic elliptic genus has the form:
\be\label{QMin6d}
Z_{-2,m}^{QM}(q,\zh) = \frac1{12}E_2 (q) Z^E_{-2,m_1}(q,\zh)Z^E_{-2,m_2}(q,\zh)\,,\ \ \ \qquad  m=m_1+m_2\,.
\ee
It arises whenever the curve $C^0$ associated with the heterotic string splits into to two curves
$C^i_E$, each associated to an E-string. 
In this case, the $U(1)$ indices $m_1$ and $m_2$ are determined geometrically as
\be\label{m1m2}
m_1 = \frac{1}{2} b \cdot C_E^1 \,, \qquad    m_2 = \frac{1}{2} b \cdot C_E^2    \,.
\ee
This is precisely what is reflected by the two  factors $Z^E$,
each of which can be associated to the partition function of a non-critical E-string \cite{Klemm:1996hh}.
In fact, its form (up to order $\zh^6$) is completely fixed by the decomposition (\ref{jacm}) and by
requiring that the ground state is uncharged (no $\zh$-dependence of  the $1/q$ term):
\be\label{singleZE}
Z^E_{-2,m}(q,{\zh}) =
\frac1{\eta^{12}} e^{\frac m{12}E_2{\zh}^2}\left[E_4(q) -\frac m{12}{\zh}^2 E_6+ \frac 12 \left(\frac m{12}{\zh}^2\right)^2E_4^2
- \frac 16 \left(\frac m{12}{\zh}^2\right)^3 E_4E_6+\cO({\zh}^8)\right]\,.
\ee
As a consequence we can compute the anomaly (\ref{Anom}) and the putative Green-Schwarz term 
(\ref{ACT}) in closed form:
\bea \label{A6QMellgen}
A^{(6),QM}&=& -\frac14  ({m_1}^2+{m_2}^2)\,,\\
m A^{(6),QM}_{GS}&=&  -\frac 1{12}(m_1+m_2)^2 \,. \label{AGS6QMellgen}
\eea
 Note that the anomaly nicely separates
 into independent pieces related to the two individual E-strings, and 
in particular that cross terms of the form 
 $({\partial_\zh}^{4-k}Z^E_{4,m_1})({\partial_\zh}^k Z^E_{4,m_2})$, {$k=1,2,3$}, vanish. 
 This fits the Ho\v{r}ava-Witten picture where the gauge symmetry is localized on two end-of-the-world branes. 
 
More importantly, note that the putative Green-Schwarz term (\ref{AGS6QMellgen}) 
does not cancel the anomaly (\ref{A6QMellgen}).
Rather, this anomaly is supposedly cancelled by the Chern-Simons terms localized on the heterotic M5-branes that are necessarily present in this situation. 
Clearly this is a non-perturbative sector of the theory that cannot be captured by 
any perturbative calculation of the Green-Schwarz term, so that this is entirely as expected.

We now turn to the four-dimensional version of the story.  
From general properties (such as having no charged ground state) we can infer {\it a priori} that the quasi-modular piece of the elliptic genus and must be of the highly restricted form
 \be\label{generalQM}
 Z_{-1,m}^{QM}(q,\zh)\ =\ 
\phi_{-1,2} \phi_{-2,1}E_2\,\Phi_{0,m-3}\big[E_4, E_6, \phi_{0,1},\phi_{-2,1}\big] \simeq    \zh^3 m\, e^{\frac m{12}E_2\zh^2}\Big[1+\cO(\zh^4)\Big]\,,
\ee
up to an overall numerical factor, $c$. Again we should here keep in mind  that for the actual elliptic genus, this expression needs to be
 multiplied with a flux factor, $g^{QM}$.

The overall factor of $\zh^3$ arises from the Jacobi forms of negative weight.
An immediate consequence of this is that the naive, perturbative Green-Schwarz term vanishes identically, because 
according to (\ref{ACT}) it is determined by a single derivative with respect to $\zh$. On the other hand, the cubic anomaly
will in general be non-zero:
\bea\label{A4QM}
A^{(4),QM}&=& - \frac{ c}{12}m  \, g^{QM},\\
A^{(4),QM}_{GS}&=&0\,.
\eea
This shows even more clearly than in six dimensions that the quasi-modular part of the anomaly must be cancelled by other, non-perturbative contributions.

{ In a spirit similar  to eq.~(\ref{antiDZ}) for fully modular elliptic genera,} we can write $ Z_{-1,m}^{QM}$ in a suggestive
form which naturally makes  contact to the underlying heterotic/E-string geometry:
\be\label{antiDZQM}
Z_{-1,m}^{(4),QM}(q,\zh) \ =\ \frac1{12}\frac {1}{\eta^{24}} E_2 \Big[\frac1{m_1}(\partial_\zh E_{4,m_1}) E_{4,m_2}-\frac1{m_2} E_{4,m_1} (\partial_\zh E_{4,m_2})\Big]\,,
 \ee
where $m=m_1+m_2$. 
This being proportional to $\phi_{-1,2} \phi_{-2,1}$ (as per (\ref{nablarelations}))  is a special case of (\ref{generalQM}) and leads to 
\be \label{A4QMm1-m2}
A^{(4),QM} =(m_2-m_1) g^{QM} \,.
\ee
Recall that the $m_i$ are determined in terms of intersection numbers of the E-string geometry as shown
in~\eqref{m1m2}, which we will derive in the next section. 

By considering the other natural combination of the derivatives, we can capture the quasi-modular,
derivative sector as well:
\be\label{DZPQM}
Z_{-1,m}^{(4),\partial,QM}(q,\zh) \ =\ \partial_\zh Z_{-2,m}^{(6),QM}(q,\zh) \ =\ \frac1{12}\frac {1}{\eta^{24}} E_2\,\partial_\zh( E_{4,m_1} E_{4,m_2})\,,\ \ \qquad m=m_1+m_2\,.
 \ee
Here we get
 \bea\label{A4DQM}
A^{(4),\partial,QM}&=& -({m_1}^2+{m_2}^2) g^{QM},\\
m A^{(4),\partial,QM}_{GS}&=&-\frac16({m_1}+{m_2})^2 g^{QM}\,.\nn
\eea
The quadratic dependence on the indices reflects the
six-dimensional origin of the derivative sector. Again, the naive perturbative
Green-Schwarz term, $A^{(4),\partial,QM}_{GS}$, does not cancel the anomaly.


\subsection{Effective Field Theory and Anomaly Cancellation from Flux Geometry}\label{subsec_effanomaly}

We now match the Green-Schwarz (GS) terms, as computed from the elliptic genus of F-theory compactifications, with
the anomaly cancelling terms that arise in the effective field theory from geometry,
The point is to understand how the different modular properties of the various contributions to the elliptic genus 
reflect the different geometrical origins of the anomaly cancelling terms. 

For four-dimensional perturbative heterotic string compactifications on smooth Calabi-Yau spaces  with general abelian gauge groups, 
the Green-Schwarz terms  have been analyzed  in \cite{Blumenhagen:2005ga}
and extended to non-perturbative models with heterotic 5-branes in \cite{Blumenhagen:2006ux}.
The perturbative GS mechanism does not only involve  the universal heterotic 2-form field $B^0$,
but in general also the 2-form fields dual to the axionic scalars that are obtained from $B^0$ by dimensional reduction.
In the presence of heterotic 5-branes, additional counterterms are induced by the self-dual tensor fields coupling to the 5-branes (see \cite{Lukas:1997fg,Lukas:1998ew,Carlevaro:2005bk} for their M-theory origin). This feature is present already in compactifications to six dimensions \cite{Honecker:2006dt}.

In this sub-section we will analyze the field theoretical realization of these Green-Schwarz terms in terms of the elliptic genera discussed in
 previous section.
Our main results can be summarised as follows.
Let us first recall how we have determined, in Section~\ref{subsec_Anomaliesmod}, the 1-loop anomaly coefficients~\eqref{Anom} in six and four dimensions, via the elliptic genus of the heterotic string. In general, the anomaly coefficients decompose according to the modular and the quasi-modular contributions to the elliptic genus: 
\be
A^{(d)} =  A^{(d) ,M}  + A^{(d) ,QM} \,,  
\ee
and we have established, for the modular contributions, that
\be
A^{(6), M} = m \, A_{GS}^{(6),M} \,,   \qquad  A^{(4) ,M} = m \, (A_{GS}^{ (4) ,M}  + 2 A_{GS}^{(4),\partial,M}  ) \,.
\ee 
Here the 1-loop Green-Schwarz term, $A_{GS}^{(d), M}$ is computed from the modular part of the elliptic genus, and in four dimensions we have indicated that there can be an additional derivative part (recall in particular eq.~(\ref{deridef})).
We will find, as expected,  that these ``modular'' Green-Schwarz  terms perfectly match the {\it perturbative} Green-Schwarz terms in the effective action that involve the universal heterotic B-field and, in four dimensions, also the dual axions which relate to the derivative part. Moreover, a possible deficit between this perturbative Green-Schwarz mechanism and the total anomaly will be
attributed to the quasi-modular part of the anomaly,~$A^{(d) ,QM}$. This deficit needs to be
cancelled via non-perturbative Green-Schwarz terms involving the 5-brane tensor fields as discussed in \cite{Honecker:2006dt,Blumenhagen:2006ux}.

Now let us go into the details. In order to
avoid the complication  of
finding heterotic duals \cite{Anderson:2014gla,Cvetic:2015uwu,Cvetic:2016ner,Anderson:2019axt} for F-theory models with abelian gauge groups,
we will first perform the computation in the Type IIB/F-theory duality frame. 
In this way we will arrive at an intersection theoretic interpretation of the Green-Schwarz terms that we computed from the elliptic genus in the previous sections.
The form of the Green-Schwarz mechanism in this duality frame has been derived in detail in the literature, beginning with \cite{Sadov:1996zm} and substantially extended in \cite{Grassi:2011hq,Park:2011wv,Park:2011ji,Taylor:2011wt,Cvetic:2012xn,Grimm:2012yq,Weigand:2017gwb}. We can hence be brief and focus on comparing the results of the elliptic genus to
the anomaly of the effective action.

In F-theory compactified to $d$ dimensions, the Green-Schwarz counterterms are encoded in the Chern-Simons-type couplings of the Ramond-Ramond 4-form ${\mathbf C}_4$ to the 7-branes,
\be
S_{\rm CS-7} =  - \frac{2\pi}{2} \int_{\mathbb R^{1,d-1} \times D_A}   {\mathbf C}_4 \wedge {\rm tr} \, e^{F_A} \,  \sqrt{\hat A(R)} \,.
\ee
Here we are assuming that a stack of 7-branes wraps a divisor $D_A$ on the internal space and carries a gauge group $G_A$.
Our conventions for the normalization of the action follows the discussion in \cite{Weigand:2017gwb}.
To avoid clutter of notation, we will right away specialise to the situation where the gauge group is given by
$G=U(1)$, with field strength $F$.
In this case, the divisor $D_A$ is to be identified with the height-pairing $b$ defined in (\ref{bU1def}) on the internal space.
We will furthermore focus purely on gauge anomalies, neglecting gravitational ones.
The relevant part of the Cherns Simons couplings then reads
\be \label{SCS-72}
S_{\rm CS-7} =  - \frac{2 \pi}{2} \int_{\mathbb R^{1,d-1} \times b} {\mathbf C}_4 \wedge \frac{1}{2}  \left(F \wedge F\right)  + \ldots
\ee
We will first discuss the implications of this Chern-Simons coupling in 
comparison with the elliptic genus of the dual heterotic string for compactifications to six dimensions. Most of this material
is well known.

\subsubsection{GS Mechanism in Six Dimensions in Relation to (Quasi-)Modularity}

The compactification space of a Type IIB/F-theory in six dimensions is given by the
 base, $B_2$, of an elliptic Calabi-Yau threefold, $Y_3$.
Let us fix a basis for its (co-)homology,
\be
\tW_\alpha \in H^{2}(B_2) \,, \qquad \quad \tW^\alpha \in H_{2}(B_2)\,,
\ee
with intersection form
\be
\Omega_{\alpha \beta} = \int_{B_2} \tW_\alpha \wedge  \tW_\beta  =: \tW_\alpha \cdot \tW_\beta \,.
\ee
The intersection form on $H_{2}(B_2)$ is determined by the inverse matrix, $\Omega^{\alpha \beta}$. 
In terms of the basis of $H^{2}(B_2)$ we expand the Ramond-Ramond  4-form field as 
\be
{\mathbf C}_4 = B^\alpha \wedge \tW_\alpha \,.  \qquad 
\ee
Note that a D3-brane wrapping some curve $C^I = C^I_{\alpha} \tW^\alpha$ gives rise to a string that couples locally to the
combination
\be
B^I = C^I_\alpha \, B^\alpha \,.
\ee
By dimensional reduction of the 7-brane Chern-Simons couplings (\ref{SCS-72}), we obtain the following Green-Schwarz interactions:
\bea \label{SGS6ddef}
S_{\rm GS}  &=& -  \frac{2\pi}{2} \theta_\alpha \int_{\mathbb R^{1,5}} B^\alpha \wedge F \wedge F \,, \qquad \theta_\alpha =    \frac{1}{2} \int_{B_2} b \wedge \tW_\alpha  \,,
\eea
where $b$ is the height-pairing associated to the $U(1)$ gauge group.
By standard arguments these couplings lead to a gauge variation of the effective action which is encoded in an anomaly eight-form, $I_{\rm eff} = {\cal A}^{(6)}_{\reff}  F^4$.  This then cancels the 1-loop anomaly via:
\be
{\cal A}^{(6)}_{\reff} = - A^{(6)}   \,.
\ee
This tree-level contribution from the exchange of the various $B$-fields from the effective action is given by
\bea\label{Acounter6d1}
{\cal A}^{(6)}_{\reff} &=& -{ \frac{1}{2}} \theta_\alpha\, \Omega^{\alpha \beta} \theta_\beta \,.
\eea

We will now separate the different contributions to ${\cal A}^{(6)}_{\reff}$ as
based on their respective geometric origin (and later, modularity properties). For this we consider a simplified but prototypical situation that captures all relevant contributions.
Following the general philosophy of Section~\ref{subsec_P1fibered}, we specialise to an F-theory base $B_2$ which is by itself a $\mathbb P^1$ fibration. 
While every F-theory base with a standard heterotic dual is a blow-up of a rational fibration, the simplification we are going to make is to consider only a single blowup over a point on the base of this fibration; this of course is easily generalised to an arbitrary sequence of blowups.
This means that we consider the blowup of a Hirzebruch surface, which we denote by $B_2 = {\rm Bl}^1 \mathbb F_n$. For this the  homology group of curves is spanned by the classes of the generic fiber $f$, the exceptional section $S_- = h$ and an exceptional curve $C^2_E$.  A consequence of the blowup is that the generic fiber splits over a point into two effective curves:
\be
f = C^1_E + C^2_E =(f - C^2_E) + C^2_E \,.
\ee
The non-zero intersection numbers between $f$, $h,$ and $C^2_E$ are:
\be
f \cdot h = 1 \,,   \qquad h \cdot h = -n \,, \qquad C^2_E \cdot C^2_E = -1  \,.
\ee

As will become clear momentarily, a convenient choice of basis for $H_2(B_2)$, which is adapted to the duality with the heterotic string, is
given by
\be \label{TWcurveB21}
\{\tW^\alpha\} = \{C^0, \tilde C^0, C^E \} \,,
\ee
where
\be \label{basischoice6d}
C^0 = f \,, \qquad \tilde C^0 = n \, f + 2h - C_E^2 \,, \qquad C^E = C_E^1 - C_E^2 = f - 2 C_E^2\,.
\ee
The intersection form for this basis is 
\be\label{Omega6}
\Omega^{\alpha \beta} = \left( \begin{matrix}  0 & 2 & 0 \\ 2 & -1 & 0 \\ 0 & 0 & -4    \end{matrix} \right) \,.
\ee

To understand the significance of this basis, note first that the dual basis of $H^2(B_2)$ is given by $\{ \tW_\alpha\} = \{\Omega_{\alpha \beta} \tW^\beta\}$ with $\Omega_{\alpha \beta} \Omega^{\beta \gamma} = \delta_\alpha^\gamma$.
It reads explicitly:
\be \label{dualbasis6d}
\tW_0 = \frac{1}{4} (C^0 + 2 \tilde C^0) \,, \qquad \tilde \tW_0 = \frac{1}{2} C^0 \,, \qquad \tW_E = - \frac{1}{4} C^E \,.
\ee
The point is now that 
a D3-brane wrapping the curve class $C^0$ with $C^0 \cdot C^0=0$ gives rise to a heterotic string, which can become asymptotically
tensionless if the volume of  $C^0$ shrinks to zero size. Moreover this curve class can split further into effective classes as $C^0 = C^1_E + C_E^2$, and a D3-brane wrapping the two classes gives rise to two non-critical E-strings, respectively.
If we correspondingly expand the RR 4-form in the dual divisor classes given in (\ref{dualbasis6d}),
\be\label{C4decomp}
{\mathbf C}_4 = B^0 \wedge \tW_0 + \tilde B^0 \wedge  \tilde \tW_0 + B^E  \wedge \tW_E \,,
\ee 
it follows from the intersection form  (\ref{Omega6}) that the 2-form field $B^0$ couples only to the heterotic string, and is hence identified with the universal perturbative heterotic B-field. 

On the other hand,
the 2-form field $B^E$ maps to the anti-self-dual tensor field associated with a heterotic 5-brane that is located at the point on the base of the Hirzebruch surface over which the blowup has been performed.
Indeed, we will see that this tensor field couples to the linear combination of E-strings from $C^E=C^1_E - C^2_E$ in the right way to be identified with the tensor field on the heterotic 5-brane.
While $B^E$ is anti-self-dual, the perturbative universal B-field  $B^0$ is neither self-dual nor anti-self-dual but can rather be written as a sum of a self-dual and anti-self-dual tensor field. The field $\tilde B^0$ is related to the field dual to $B^0$.
 More precisely, if we were  to consider just the Hirzebruch surface $B_2 = \mathbb F_n$ without a blowup, the dual heterotic string compactification would be purely perturbative.
The analogue of $\tilde C^0$ on $\mathbb F_n$ would then be
$\tilde C^0_{\rm pert} = n f + 2h$ with $\tilde C_{\rm pert}^0 \cdot \tilde C_{\rm pert}^0=0$,
 and the associated field $\tilde B_{\rm pert}^0$ is the dual of $B^0$. 
It is this property that has motivated the choice of basis (\ref{basischoice6d}).

In terms of the decomposition (\ref{C4decomp}),
we  can now read off the Green-Schwarz couplings defined in (\ref{SGS6ddef}) as follows:
\bea
\theta_0 &=& \frac{1}{2}   \omega_0 \cdot b =   \frac{1}{4} m + \frac{1}{4} \tilde C^0 \cdot b \,, \\
 \tilde \theta_0 &=& \frac{1}{2} \tilde \omega_0 \cdot b = \frac{1}{4} C^0 \cdot b = \frac{1}{2} m \,, \\
 \theta_E &=& \frac{1}{2} \omega_E \cdot b = - \frac{1}{4} (m_1 - m_2) \,.
\eea
Here we have defined
\be
m = \frac{1}{2} C^0 \cdot b \,, \qquad m_1 = \frac{1}{2} C^1_E \cdot b \,,\qquad m_2 = \frac{1}{2} C^2_E \cdot b,
\ee
with $m = m_1 + m_2$.

\begin{figure}[t!]
\centering
\includegraphics[width=17cm]{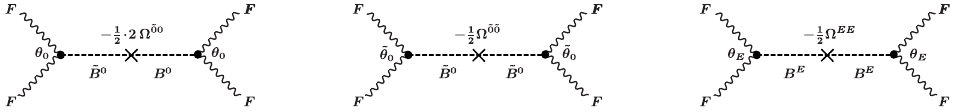}
\caption{ 
Contributions to the Green-Schwarz mechanism of a six-dimensional F-theory compactification on $B_2 = {\rm Bl}^1 \mathbb F_n$. Only the leftmost contribution is perturbative from the perspective of the dual heterotic string,
and matches the fully modular contribution from the elliptic genus.
}
\label{fig_GS6d}
\end{figure}

As shown in Figure \ref{fig_GS6d}, the couplings $\theta_\alpha$ give rise  to the following tree-level contributions to the anomaly coeffcient (\ref{Acounter6d1}):
\be \label{Acounter2}
{\cal A}^{(6)}_{\rm eff} =  -{ \frac{1}{2}} \theta_\alpha \Omega^{\alpha \beta}  \theta_\beta = {\cal A}_{\tilde B^0-B^0} +  {\cal A}_{\tilde B^0- \tilde B^0} +  {\cal A}_{B^E - B^E} \,,
\ee
where
\bea
{\cal A}_{\tilde B^0-B^0}  &=& - 2 \tilde \theta_0 \, \theta_0  = -m \, (\frac{1}{4} m + \frac{1}{4} \tilde C^0 \cdot b) \,, \\
{\cal A}_{\tilde B^0-\tilde B^0}  &=& \frac{1}{2} \tilde \theta_0 \, \tilde \theta_0  = {\frac{1}{2}} (\frac{1}{2} m)^2 \,, \\
{\cal A}_{B^E-B^E} &=&  2 \theta_E \, \theta_E = {\frac{1}{2}} (\frac{1}{2} (m_1 - m_2))^2   \,.
\eea

On the other hand, recall from the previous section that the anomaly (\ref{Anom}) as encoded in the elliptic genus
 can be written as a sum of two pieces
\be
A^{(6)} = A^{{(6)}, M} + A^{(6){,QM}} \,,
\ee
which are associated with the modular and the quasi-modular parts of the  elliptic genus, respectively.
The quasi-modular part of the six-dimensional anomaly has been computed in full generality in (\ref{A6QMellgen}), and is cancelled by the sum ${\cal A}_{\tilde B^0- \tilde B^0} +  {\cal A}_{B^E - B^E}$.  
This then provides 
the following match between couplings of the effective action and the modular and quasi-modular parts of the elliptic genus:
\bea
{\cal A}_{\tilde B^0-B^0}  &=&  -  A^{(6) ,M}  \,, \label{AmodvsF6d}\\
 \qquad \quad {\cal A}_{\tilde B^0- \tilde B^0} +  {\cal A}_{B^E - B^E} &=&  -  A^{(6) ,QM}  \,.\label{AquasimodvsF6d}
\eea
Moreover, from (\ref{anomalycancell}) 
 we recall that
\be \label{AGSmodvsF6d}
A^{{(6)}, M}   =   { m}   \,  A_{GS}^{{(6)},M} \,.
\ee
By comparison with the above expression for ${\cal A}_{B^0-\tilde B^0}$, this then
identifies the field theoretic coupling $\theta_0$ with the modular contribution to the Green-Schwarz term,
\be \label{AGStheta0id}
 A^{(6),M}_{GS} =   \theta_0 =  \frac{1}{4} m + \frac{1}{4} \tilde C^0 \cdot b \, .
\ee

In turn, these identifications also allow to link
the  ``quasi-modular'' pieces of the anomaly, $A^{(6) ,QM}$,  to the presence of non-perturbative
heterotic 5-branes \cite{Honecker:2006dt} as follows:
The Green-Schwarz term $\theta_0$ is, up to overall normalization, the Green-Schwarz term of the perturbative heterotic B-field, as obtained by standard dimensional reduction of the perturbative 10-dimensional Green-Schwarz terms. The analogue of this term is present also in heterotic backgrounds without heterotic 5-branes, and consequently is determined by the purely modular piece of the elliptic genus.

By contrast, the anomaly cancelling contributions  ${\cal A}_{\tilde B^0- \tilde B^0} +  {\cal A}_{B^E - B^E}$ are non-perturbative from the heterotic point of view:
The first term, ${\cal A}_{\tilde B^0- \tilde B^0} $, arises from the exchange of the dual of the heterotic B-field,  $\tilde B^0$. 
This coupling is absent in the 
perturbative cousin of this geometry where there is no blowup and the base $B_2$ is given by the Hirzebruch surface, $\mathbb F_n$.
As noted above, in this case we have $\tilde C^0_{\rm pert} \cdot  \tilde C^0_{\rm pert} = 0$
and hence the coupling of $\tilde B^0_{\rm pert}$ to itself in (\ref{Acounter2}) vanishes. Thus the origin of such a term must be attributed to non-perturbative heterotic 5-branes in the dual heterotic picture, as explained in \cite{Honecker:2006dt}.

Similarly, the second term,  ${\cal A}_{B^E - B^E}$, must originate from couplings that involve the extra anti-symmetric tensor field $B^E$ living in the worldvolume of the 5-brane.
The form of these anomalies likewise matches the results in \cite{Honecker:2006dt}.

In summary, eq.~(\ref{AquasimodvsF6d}) shows that the two non-perturbative contributions to the anomaly are beautifully matched by the quasi-modular part of the anomaly, precisely as encoded in the elliptic genus.
On the other hand, the corresponding Green-Schwarz terms, $\tilde\theta_0$ and $\theta_E$, are not reproduced by
 $A^{(6),QM}_{GS}$ from the elliptic genus, as the latter is a one-loop quantity that is agnostic about the non-perturbative sector.
 Only the perturbative, modular contribution  $A^{(6),M}_{GS}$ of the elliptic genus matches,
as per (\ref{AGStheta0id}),  the Green-Schwarz term $\theta_0$ in the effective action.

\subsubsection{GS Mechanism in Four Dimensions versus Modularity and Derivative Terms}  \label{GS4d}

After this preparation,  we now turn to the Green-Schwarz mechanism in four dimensions.
 The new ingredient, in line with the main theme of this paper, 
will be the derivative sector of
the elliptic genus. As we will see, it encodes the Green-Schwarz mechanism associated with extra, non-universal 2-form fields $B^i$ which arise from the perturbative B-field in ten dimensions by dimensional reduction.

In four dimensions, we reduce the Ramond-Ramond four-form field, ${\bf C}_4$, into 2-form fields $B^{\alpha}$ and their dual, zero-form  axions $c_\alpha$ by expanding
\be \label{expansionC4}
{\bf C}_4 = B^{\alpha} \wedge \tW_{\alpha} + c_\alpha  \, \tW^\alpha \,.
\ee
Here $\{ \tW_{\alpha}\}$ is a basis of $H^{1,1}(B)$ and $\{\tW^\alpha\}$ the dual basis of  $H^{2,2}(B_3)$, which are defined such that 
\be
\int_{B_3} \tW_\alpha \wedge \tW^\beta =: \tW_\alpha \cdot \tW^\beta  = \delta_\alpha^\beta \,.
\ee
Plugging this expansion into the 7-brane action (\ref{SCS-72}),  we can read off the Green-Schwarz terms involving the 2-form fields $B^\alpha$ and the Chern-Simons terms involving their dual axions $c_\alpha$ as follows:
\bea
S_{\rm CS-7} &=& S_{\rm GS} + S_{CS}  \\
S_{\rm GS}  &=&  -  \frac{2\pi}{2} \theta_{\alpha}  \int_{\mathbb R^{1,3}} B^\alpha \wedge F \,, \qquad \theta_{\alpha} = b \cdot F \cdot \tW_\alpha \,,\\
S_{CS} &=&  -  \frac{2\pi}{2} m^\alpha  \int_{\mathbb R^{1,3}}c_\alpha \,  F \wedge F \,, \qquad m^{\alpha} = \frac{1}{2}  b \cdot \tW^\alpha \,.   
\eea
The tree-level exchange of the $B$- and $c$-fields then gives rise to a gauge variation of the effective action,
 with anomaly six-form given by $I_{\rm eff} = {\cal A}_{\reff}^{(4)} F^3$, where
\be \label{Acounter4d}
{\cal A}_{\reff}^{(4)}=  - {  m^\alpha \, \theta_{\alpha}}  \,.
\ee
This precisely cancels the field theoretic 1-loop anomaly $A^{(4)}$ as normalised in (\ref{A1loopnorm}):
\be \label{Acounter1}
{\cal A}_{\reff}^{(4)} = - A^{(4)} \,.
\ee

As in the six-dimensional setting, we now specialize to a prototypical example that captures all variants of anomaly cancellation.
For this we consider a fourfold base, $B_3\subset Y_4$,  which is by itself a $\mathbb P^1$-fibration over some base space
 $B_2$, and for simplicity of presentation we assume here again that there is only one blow-up divisor, $E$. 
We refer to Section~\ref{subsec_P1fibered} for our notation for this type of geometries.
A basis of $H_2(B_3)$ that is convenient for comparison with the dual heterotic geometry is given by
\be
\{ \tW^\alpha \}   =   \{ \tW^0: = C^0 \,, \quad   \tW^E :=   C^1_E  - C^2_E \,, \qquad \tW^i :=   \quad S_+ \cdot p^\ast({C}^i) \} \,,
\ee
where $\{C^i\}$ is a basis of divisor  classes on $B_2$.
A D3-brane along the rational fiber $C^0$ of $B_3$, or along the exceptional curves $C^{1,2}_E$, gives rise to a heterotic string
 or two copies of E-strings, respectively.
The 2-form fields coupling to these strings are obtained by expanding $\mathbf C_4$ with respect to the basis of dual divisors:
\be \label{divisorbasis}
\{ \tW_\alpha\} = \{  \tW_0:= S_- - \frac{1}{2} E \,, \quad  \tW_E :=    \frac{1}{2} E \,,  \quad \tW_i :=p^*(C_i) \} \,.
\ee
Here, $S_-$ is the exceptional section of the $\mathbb P^1$-fibration, $E$ is the blowup divisor, and $\{C_i\}$ is a basis of $H^{1,1}(B_2)$ related to $\{C^i\}$ via 
\beq
{C}^i \cdot_{B_2} C_j = \delta^{i}_j\,. 
\eeq
The corresponding expansion
\be
{\mathbf C}_4 = B^\alpha \wedge \omega_\alpha =  B^0 \wedge (S_- - \frac{1}{2} E) + B^E \wedge (\frac{1}{2} E) + B^i \wedge p^*(C_i)   + \ldots
\ee
then defines the various 2-form fields that are relevant for us. First,
the field $B^0$  maps to the perturbative heterotic B-field in four dimensions. 
Furthermore $B^E$ represents the 2-form 
field in four dimensions which lives on the worldvolume of the spacetime-filling heterotic 5-brane,
which is the geometry that is dual to the blowup on the F-theory side.
Finally, the 2-form fields $B^i$ correspond to the 2-forms obtained on the heterotic side by expanding the ten-dimensional magnetic dual six-form, $B_6$,
on 4-forms $p^\ast(C_i)$, which are the pullbacks of the 2-form divisor classes on $B_2$.

In the basis (\ref{divisorbasis}), the coefficient of the anomaly (\ref{Acounter4d}) induced by the Green-Schwarz terms then 
reads:
\bea \label{Acounter2}
{\cal A}_{\reff}^{(4)}=  -  (F \cdot b \cdot (S_- - \frac{1}{2} E))   \, m  -  (F \cdot b \cdot \frac{1}{2} E)  \,  (m_1 - m_2)  - (F \cdot b \cdot p^\ast(C_i)) \,  m^{(i)} \,,
\eea
where
\bea\label{mdefs}
 m&=&  \frac{1}{2}  (b \cdot C^0) \,,\nn\\
  m_1 - m_2   &=&  \frac{1}{2} (b \cdot (C^1_E - C^2_E)) \,,
  \\
   m^{(i)} &=& \frac{1}{2}  (b \cdot S_+ \cdot p^\ast({C}^i) ) \,.\nn
 \eea
  Now our task is to compare ${\cal A}_{\reff}^{(4)}$  with the various modular/quasi-modular/non-modular components of the anomaly $A^{(4)}$, as encoded in the decomposition~(\ref{Zhetstr1}) of the elliptic genus.
For this we rewrite the latter in terms of a new basis  (signified by a tilde) that
is adapted to (\ref{Acounter2}). This geometrically motivated  basis
is however not well aligned with modularity, and thus will generically mix the
various components of the elliptic genus. Concretely,
in our specific prototypical example, and also in all other
examples studied in this work, we find that the elliptic genus can be equivalently written as
\begin{equation}
\begin{aligned} 
 \label{ZHformhet}
Z_{{\bf G},C^0} &=
{ \tilde g}^0  \, \tilde Z_{-1,m}^{0} +   {  \tilde g}^E \, \tilde Z_{-1,m}^{E} + \sum_i    { \tilde g}^i    \left(  \tilde Z_{-1,m}^{i} +   \frac{1}{2m} \xi \partial_\xi \tilde Z^i_{-2,m} \right)
\,,\cr
\end{aligned} 
\end{equation}
with the following flux-dependent coefficients:
\begin{equation}
\begin{aligned} \label{gcoeffs}
{ \tilde g}^0  = (F \cdot b \cdot (S_-  - \frac{1}{2} E)) \,, \qquad 
{ \tilde g}^E =  (F \cdot b \cdot \frac{1}{2} E) \,, \qquad 
 {  \tilde g}^i  =  (F \cdot b \cdot p^\ast(C_i)) \,.
\end{aligned} 
\end{equation}
Here, $\tilde Z_{-1,m}^{0}$ is a modular Jacobi form of weight $-1$ and index $m$ which is necessarily of the form 
(\ref{UniqueZ}), 
while the remaining contributions mix both modular and quasi-modular pieces. Following our previous notation, we can thus write:
\bea
{ \tilde {Z}^0_{\ast,m}} &=& {  \tilde {Z}^{0,M}_{\ast,m} \,,  \nn  }  \\ 
 \tilde {Z}^E_{-1,m} &=&  \tilde {Z}^{E,M}_{-1,m}  + \tilde {Z}^{E,QM}_{-1,m} \,, \label{ZiQM4d} \\
 \tilde {Z}^i_{\ast,m} &=&   \tilde  Z^{i,M}_{\ast,m}  + \tilde  {Z}^{i,QM}_{\ast,m} \label{ZEQM4d} \,. \nn
\eea
After these preparations we can now easily compare the various terms
 of ${\cal A}_{\reff}^{(4)}$ in (\ref{Acounter2}) with the coefficients (\ref{ZHformhet}) of the elliptic genus, $Z_{{\bf G},C^0}$,
 by filtering the individual $\tilde Z^\ast_{*,m}$ through the formula (\ref{Anom}).
This leads to the following decomposition of the total anomaly:
\bea \label{4danomaly1}
A^{(4)} 
=:   {\tilde g}^0 A_{\tilde { Z}_{-1,m}^0} + { \tilde g}^E A_{ \tilde {Z}_{-1,m}^E} + \sum_i { \tilde g}^i (A_{ \tilde{ Z}^i_{-1,m}}  +  A_{\tilde{ Z}_{-2,m}^i} ) \,.
\eea
Imposing ${\cal A}_{\reff}^{(4)} = - A^{(4)}$ we can thus identify\footnote{Note that one can write the flux-dependent triple-intersections on $B_3$ directly on $Y_4$ by exploiting the identity $ - \int_{Y_4} \bG \wedge \sigma \wedge \pi^\ast D = F \cdot b \cdot D$ for $\bG = \sigma \wedge \pi^\ast F$.}  
\bea
{ \tilde g}^0 A_{\tilde{Z}_{-1,m}^0}  
&=&
 (F \cdot b \cdot (S_- - \frac{1}{2} E))  \, m\,, \label{finalmatch1}
\\ \
{ \tilde g}^E A_{\tilde{Z}_{-1,m}^E}     
 & =&
  (F \cdot b \cdot \frac{1}{2} E) \, (m_1 - m_2)     \,, \label{finalmatch2}
 \\
 { \tilde g}^i A_{ \tilde Z^i_{-1,m}}  + { \tilde g}^i A_{\tilde {Z}^i_{-2,m}}        
  &=&   
 (F \cdot b \cdot p^\ast(C_i))    \, m^{(i)} \,  \label{finalmatch3}
\,.    
\eea
This generic match between geometric, flux dependent
quantities pertaining to the effective action on the one hand,
and the various contributions to the elliptic genus on the other, is a central point of the present paper.  While we have worked it out here  for a prototypical  situation which is dual to a heterotic string, it applies also to much more
general geometries. However it is difficult to refine these statements without
specifying more data explicitly, and this is why we will present in 
Section~\ref{sec_examples} some detailed computations for explicit examples.
Most importantly, we will see directly how the derivative part of the elliptic genus, encoded in the $\tilde {Z}^i_{-2,m}$, relates to the Green-Schwarz mechanism involving the additional four-dimensional 2-form fields $B^i$.

Before concluding this section, let us present some further remarks about the
Green-Schwarz terms, which are determined by the 1-loop computation
shown in eq.~(\ref{ACT}).  Like the anomaly  (\ref{4danomaly1}),
also the Green-Schwarz terms receive contributions 
from both the modular and the quasi-modular parts of the elliptic genus:
\bea\label{full4dano}
A^{(4)}_{GS}  &=& A^{(4), M}_{GS}  + A^{(4), QM} _{GS}\,. \nn
\eea
By (\ref{anomalycancell}), the Green-Schwarz terms computed from the purely modular contributions are guaranteed to match the field theoretic Green-Schwarz terms, which are perturbative in the sense that they are independent of the heterotic 5-branes. Indeed, in the present context,
where $Z_{{\bf G},C^0}$ is
 written in the special basis (\ref{ZHformhet}), anomaly cancellation in the
 perturbative sector takes the form:
  \be\label{AGSA}
A^{(4), M} =  { {m}} \left( {  \tilde g}^0 A^{\rm GS}_{{\tilde Z}_{-1,m}^0} + \sum_i  {\tilde g}^i  (A^{\rm GS}_{\tilde Z_{-1,m}^{i,M}} + 2 A^{\rm GS}_{{\tilde  Z}_{-2,m}^{i,M}})  \right) \,.
\ee
From the perspective of the dual heterotic string, this equation comprises the complete
perturbative part of the anomaly cancellation mechanism, involving the universal B-field $B^0$ in the first term and the additional four-dimensional 2-form fields $B^i$, which are obtained by dimensional reduction of the ten dimensional six-form field, $B_6$.

Note the factor of 2 in the last expression, which reflects the relation (\ref{AGS 54d6d}) for the derivative part of the elliptic genus. 
Note also that in (\ref{AGSA}) we did not write any modular contributions from     
 $\tilde {Z}^{E,M}_{-1,m}$, because they must vanish anyway:
 \be \label{ZEMgen1}
A_{{\tilde  Z}_{-1,m}^{E,M}} = 0 \,, \qquad    A^{\rm GS}_{{\tilde  Z}_{-1,m}^{E,M}} = 0   \,.
\ee
This is because a fully modular contribution will always produce both the anomaly and its acompanying
Green-Schwarz term at the same time, but the latter cannot arise in perturbation theory from the elliptic genus, since
it is localized on the brane and thus cannot be captured by the bulk theory. 
We will find this general expectation confirmed in the
computation of explicit examples, in particular by equation  (\ref{ZEMod}) below.

Similar to the anomalies discussed before, we can also match the
Green-Schwarz terms as computed in (\ref{ACT}) explicitly
in terms of the flux geometry, as long as we stay in the fully modular sector.
Concretely we find
\be\label{AGS4deffactM}
\tilde {g}^0 A^{\rm GS}_{{\tilde Z}_{-1,m}^0} =  \, (F \cdot b \cdot (S_- - \frac{1}{2} E))  \,.
\ee
The right-hand side is the coefficient of the standard Green-Schwarz term
that involves the universal $B^0$-field of the heterotic string.

More interesting are the Green-Schwarz terms associated with the derivative,
secretly six-dimensional subsector of the theory whose properties are encoded in the embedded threefolds,~$\mathbb Y_3^i$.
These terms, as computed from the elliptic genus,
are expected to agree with the actual counterterms in the effective theory,
as long as  $\tilde { Z}^{i}_{-1,m}$  and $\tilde { Z}^{i}_{-2,m}$ 
are both fully modular. Following the same line of arguments as before,
we find the following identifications between the Green-Schwarz terms as computed from the modular expressions in (\ref{ACT}) with the intersection numbers of the F-theory flux geometry:
\bea \label{AGS4deffact}
 {\tilde g}^i (A^{\rm GS}_{{\tilde Z}_{-1,m}^{i,M}} + 2 A^{\rm GS}_{{\tilde Z}_{-2,m}^{i,M}}) &=&  \, \frac{m^{(i)}}{m} \,  (F \cdot b \cdot p^\ast(C_i))
\,.
\label{AGS4deffact2} 
\eea
Note that this identification follows from the requirement of anomaly cancellation, i.e., from~(\ref{finalmatch3}) and {\it imposing}
${\cal A}_{\reff}^{(4)} = - A^{(4)}$. It does not {\it prove} that the
anomalies are actually cancelled. To complete a proof, one would need to
show that in general the value for $A^{\rm GS}$  in (\ref{AGS4deffact}) as computed from the elliptic genus does actually come out right
such as to match the $m^{(i)}$ as determined 
 in (\ref{mdefs}) from the flux geometry in the effective action.
 While we do not have a general proof for this, we have
checked this to be true for the examples presented in the next section. Turning tables round, {\it assuming} anomaly cancellation gives a prediction for the modular part of $A^{\rm GS}$ in terms of the geometric intersection numbers $m^{(i)}$.

Note that these considerations, and in particular
the relation (\ref{AGS4deffact2}), do not hold if 
the elliptic genus has quasi-modular contributions,
${\tilde Z}_{\ast,m}^{i,QM}  \neq 0$: In this case the left-hand side
 of (\ref{AGS4deffact2}),  which depends only on the modular parts, is still expected to yield the perturbative Green-Schwarz terms for the fields $B^i$, but there will be in general additional non-perturbative contributions from the heterotic 5-branes. 

In this case, the quasi-modular part of the anomaly, $A^{(4), QM} $,  will be non-zero and generically receive  two types of contributions:
\be\label{A4QM}
A^{(4), QM}  = { \tilde g}^E A_{\tilde{Z}_{-1,m}^{E,QM}}  + \sum_i  {\tilde g}^i  (A^{\rm GS}_{\tilde Z_{-1,m}^{i,QM}} + 2 A^{\rm GS}_{{\tilde  Z}_{-2,m}^{i,QM}})   \,.
\ee
 
 The first term encodes the part of the anomaly which is cancelled by the Green-Schwarz mechanism involving the B-field $B^E$ from the heterotic 5-branes.
 From  (\ref{finalmatch2}) we see that the anomaly associated with ${\tilde Z}^{E,QM}_{-1,m}$ is proportional to $m_1 - m_2$.
This agrees beautifully with the form of the anomaly computed from the elliptic genus in (\ref{A4QMm1-m2}), provided ${\tilde Z}^{E,QM}_{-1,m}$  is of the form (\ref{antiDZQM}).
In a sense this predicts that it \emph{must} be possible to write ${\tilde Z}^{E,QM}_{-1,m}$ {\it must be} as in (\ref{antiDZQM}), as far as its expansion up to order $\zh^3$ is concerned. 

The second type of contributions in (\ref{A4QM}) must be cancelled by non-perturbative contributions to the Green-Schwarz terms for the additional $B^i$-fields in four dimensions.
 Note that $\tilde Z_{-1,m}^{i,QM}$  and $\tilde Z_{-2,m}^{i,QM}$ are non-zero only if the threefold $\mathbb Y^i_3$, 
 whose BPS invariants are encoded in $\tilde Z_{-2,m}^{i}$, 
contains the exceptional fibral curves $C^1_E$ and $C^2_E$. 
This is the case if the class of the blowup curve $\Gamma$, when expanded with respect to a chosen basis $\{C_j\}$, has a non-trivial contribution from $C_i$. In fact, we may assume without loss of generality that $\Gamma$ is proportional to $C_i$. 
 By construction this curve maps to the curve wrapped by the NS5-brane on the dual heterotic 3-fold.
We can thus conclude a bit more sharply that:
\be \label{A4QMhetcomp}
A^{(4), QM}   =    (F \cdot b \cdot \frac{1}{2} E) \, (m_1 - m_2) +     (F \cdot b \cdot p^\ast(\Gamma))  \, m_\Gamma \,,
\ee 
where
$m_\Gamma$ is a parameter which we cannot determine from the above considerations alone. However it is reassuring that
the general structure of these terms indeed matches the architecture of 
 the abelian heterotic Green-Schwarz mechanism in the presence of 5-branes, as was discussed in \cite{Blumenhagen:2006ux}.

\section{Elliptic Genera of 4d Heterotic Strings}\label{sec_examples}  

So far we have presented generic and prototypical results concerning the modularity of elliptic genera in relation
to the background fourfold geometries, $Y_4$, and flux configurations, $\bf G$.  This general structure will now be illustrated by detailed computations for
a few examples, for which sharper statements can be made, in particular concerning the structure and r\^{o}le
of the embedded six-dimensional, derivative sector.

We begin in this section with the elliptic genus of four-dimensional heterotic strings.
In the example of Section \ref{sec_Ex1het}, the embedded threefolds that encode the derivative sector are by themselves Calabi-Yau spaces, while in Section \ref{sec_Ex2het}  we discuss an example where the derivative sector encodes the relative BPS numbers of a non-Calabi-Yau threefold.

We will label the examples by the geometry of the base, $B_3,$ of a given elliptic fourfold fibration, $Y_4$.
In order to curb mathematical overload of this section,
we have relegated further details about the geometry to Appendix \ref{app_modelS}.

 \subsection{Example 1: $B_3 = dP_2 \times \mathbb P^1_{l'}$}\label{sec_Ex1het}

As our first example we consider F-theory on an elliptic fibration $Y_4$ whose base $B_3$
 is given by the blow-up 
 \be
 B_3 = dP_2 \times \mathbb P^1_{l'}   \,,
 \ee
 of the rational fibration $\mathbb F_1 \times \mathbb P^1_{l'}$, 
where the del Pezzo surface $dP_2$ is viewed as the blow-up of the Hirzebruch surface $\mathbb F_1$ in one point: 
 Since the Hirzebruch surface is a rational fibration with generic fiber $C^0=\mathbb P^1_f$ and  base $\mathbb P^1_h$, we can view $B_3$ as the blowup of the rational fibration over the base $B_2 = \mathbb P^1_h  \times  \mathbb P^1_{l'}$\,,
\bea \label{pproj-Ex1}
p :\quad C^0 \ \ \longrightarrow & B_{3}& \cr 
&  \downarrow &\cr 
& \mathbb P^1_h  \times \mathbb P^1_{l'} &(=: B_2) \,,
\eea
where the blowup locus is homologous to $\IP^1_{l'}$ in $B_2$. 
This type of base spaces fits into the class of geometries described in Section \ref{subsec_P1fibered}.
In particular, the rational fibration has an exceptional section $S_-$, which embeds $B_2$ into $B_3$ and which satisfies the relation
\be
S_- \cdot S_+ = 0  \,,\qquad S_+ = S_- +  p^\ast( \mathbb P^1_{l'}  ) \, .
\ee
As a result of the blow-up inherited from $dP_2$, the generic rational fiber $C^0$ splits into a sum of two exceptional curves $C^1_E + C^2_E$,
where $C^1_E$ and $C^2_E $ are distinguished  by their intersection numbers with $S_-$ as in (\ref{E1E2int}).
The splitting  occurs over a point on $\mathbb P^1_h$ within $B_2$.   
The exceptional divisor $E$ associated with this blowup is thus a fibration of $C^2_E$ over $\mathbb P^1_{l'}$, and since the rational fibration of $B_3$ is trivial over $\mathbb P^1_{l'}$, it is in fact a direct product: 
\be
E =C^2_E \times  \mathbb P^1_{l'} \,.
\ee
As a convenient basis for $H_2(B_3)$  we pick the curve classes 
\be
C^0 = \mathbb P^1_f \,, \qquad C^E = C^1_E - C^2_E \,, \qquad C_1 = \mathbb P^1_{l'}     \,, \qquad C_2 = \mathbb P^1_{h}   \,.
\ee
Since we will also need the curve classes dual to $C_1$ and $C_2$ on $B_2$, note that the only non-zero entries of the intersection form $\eta_{ij} = C_i \cdot_{B_2} C_j$ are $\eta_{12} =\eta_{21}=1$, so that the dual curves $C^i = \eta^{ij} C_j$ on $B_2$ are 
\be \label{C^ijS}
C^1 = C_2 \,, \qquad C^2 = C_1 \,. 
\ee
The choice of basis $\{C_1, C_2\}$ for $H_2(B_2)$ made here corresponds to the generators of the simplicial Mori cone of $B_2$, which was advertised as a basis with special properties in Section \ref{subsec_Y3}; the dual curves, $C^1$ and $C^2$, hence generate the K\"ahler cone of $B_2$. 
As basis for the divisor group on $B_3$ we introduce  $\{D_\alpha\}_{\alpha=1}^4$, which are defined as
\be
 D_1 = p^\ast(C_1) \, , \qquad  D_2 = S_-   \,,  \qquad  D_3 = p^\ast(C_2) \,, \qquad D_4 =  p^\ast(C_1) + S_- - E \,.
\ee
In terms of these basis elements $D_\alpha$, the intersection polynomial and the anti-canonical class are 
\bea \label{interB3-Model S} 
I(B_3) &=& D_1 D_2 D_3 - D_2^2 D_3 + D_1 D_3 D_4\,,\\
\bar K_{B_3} &=& 2D_1 + D_2 + 2D_3 + D_4 \,. 
\eea

Having specified some properties of the base, $B_3 = dP_2 \times \mathbb P^1_{l'}$, 
we now turn to the structure of the elliptic fibration $Y_4$. It has an additional rational section in addition to the zero-section. 
As reviewed in Section~\ref{subsec_relBPS}, this leads to gauge group $G=U(1)$. 
Moreover we choose a specific fibration \cite{Grimm:2010ez,Morrison:2012ei} for which the
height-pairing is given by
\be
b = 2 \bar K_{B_3} \,.
\ee
Further details of the fibration are provided in Appendix \ref{app_modelS}.
The massless matter spectrum comprises states of charge $r=\pm 1$ ,
which are localised in the fiber along the curve $\Sigma_{r=1}$ of class
\be \label{SigmarmodS}
\Sigma_{r=1} = 12 \bar K_{B_3} \cdot \bar K_{B_3} = - 84 C^0 + 21 (b \cdot p^{\ast}(C_1))  +  24 (b  \cdot  p^{\ast}(C_2)) \,.
\ee
Furthermore we note that the $U(1)$ flux in $H^{2,2}_{\rm vert}(Y_4,\mathbb R)$ can be expanded into the divisors $D_\alpha$ as
follows:
\be\label{vU(1)flux}
{\bf G}  \equiv {\bf G}_{U(1)} = \sigma \wedge \pi^\ast F   \,, \qquad {\rm where \ \ } F =: \sum_{\alpha=1}^4 c_\alpha D_\alpha \,.
\ee
With respect to this flux,  given the matter curve (\ref{SigmarmodS})  the chiral index of massless chiral spectrum evaluates to
\be\label{ciIndex}
\chi_{\bG, r=1}  = F \cdot \Sigma_{r=1} = 96 c_1  + 48 c_2  + 84 c_3  +96 c_4 \,.
\ee

\subsubsection{Relative BPS Invariants and Elliptic Genus}

As outlined  in Section~\ref{subsec_relBPS}, we can use
mirror symmetry, suitably adapted to fourfolds \cite{Greene:1993vm,Hosono:1993qy,Hosono:1994ax}, to compute a finite number of relative BPS invariants $N_{\bG;C^0}(n,r)$. Recall from (\ref{pproj-Ex1}) that $C^0$ is the fiber of $B_3$ which is wrapped by a D3-brane
such as to produce the solitonic heterotic string. These invariants can then be packaged into the elliptic genus and extrapolated
to all orders by modular completion.

The results of this computation are sketched in  Appendix~\ref{app_modelS}.
We list here just the lowest order in $q$:
\bea\label{Het:S}
Z_{\bG;C^0} &=& -\frac{1}{q} \sum_{n,r} N_{\bG; C^0}(n,r) q^n \xi^r  \nn \\
&=& - [96 c_1 \xi^{\pm\bar1} + 48 c_2 \xi^{\pm \bar1} + 84 c_3 \xi^{\pm\bar1} +96 c_4 \xi^{\pm \bar 1} ] 
+\, \cO(q),
\eea
where $\xi^{\pm \bar n} := \xi^n - \xi^{-n}$.
Comparing to (\ref{ciIndex}), we confirm the
advertised relation (\ref{NGC0id}) between the relative BPS invariants at level one, $N_{\bG; C^0}(1,r)$, and the chiral index $\chi_{\bG, r=1}$.

Moreover, following the discussion in Section~\ref{sec_geommod},
we aim to identify the BPS invariants $N_{\bG; C^0}(n,r)$ as expansion coefficients of Jacobi or quasi-modular Jacobi forms (or their derivatives) of $U(1)$ fugacity index 
\be
m= \frac{1}{2} C^0 \cdot b = 2 \,.
\ee
Given a sufficient number of known $N_{\bG; C^0}(n,r)$, we can uniquely determine the 
elliptic genus as follows:
\be \label{ZfullmodelS}
Z_{\bG;C^0} =( F \cdot C^0) Z^0_{-1,2} + (F \cdot b \cdot p^\ast(C_1))  \frac{1}{4} \xi \partial_\xi Z^1_{-2,2} + (F \cdot b \cdot p^\ast(C_2))  \frac{1}{4} \xi \partial_\xi Z^2_{-2,2} \,,
\ee
where
\bea
Z^0_{-1,2} &\equiv& Z^{0,M}_{-1,2} =  84 \phi_{-1,2} \\
Z^2_{-2,2}  &\equiv& Z^{2,M}_{-2,2} 
= \frac{1}{12}\frac1{\eta^{24}} \left(14 E_4 E_{6,2}+10 E_{4,2}E_6\right)   \label{ModelSZs2}\\
Z^1_{-2,2}  &\equiv& Z^{1,M}_{-2,2}  + Z^{1,QM}_{-2,2}\,,\ \ \ \ {\rm where} \label{ModelSZs1} \\
Z^{1,M}_{-2,2} &=&  Z^2_{-2,2}   - \frac{1}{12}\frac 1{\eta^{24}}E_{4,1}E_{6,1} 
\,, \qquad Z^{1,QM}_{-2,2}  = \frac{1}{12}\frac 1{\eta^{24}}E_2{E_{4,1}}^2\,. \nn\ 
\eea
For the definition of the Eisenstein and the Eisenstein-Jacobi forms, recall Appendix~\ref{app_modular}.
Moreover the flux dependent coefficients in (\ref{ZfullmodelS}) evaluate to
\bea\nn 
(F \cdot C^0) &=&    c_2 + c_4  \,, \\ 
(F \cdot b \cdot p^\ast(C_1)) &=& 4 (c_2 + c_3 + c_4) \,, \\ \nn   
(F \cdot b \cdot p^\ast(C_2)) &=& 2 (2c_1 + c_2 + 2c_4) \,.
\eea
Thus the elliptic genus $Z_{\bG;C^0}$ in (\ref{ZfullmodelS}) does have the general form as advertised in~(\ref{Zconf2}), 
or more specifically in (\ref{Zhetstr1}) and (\ref{fluxcoeffs}),
with the special feature that for the example at hand there happens to be no four-dimensional quasi-modular term, $Z^E_{-1,m}$. This, in fact, mirrors the absence of a term proportional to $C^1_E$ in (\ref{SigmarmodS}).

Let us now take a closer look at the derivative sector, ie., at the $Z^i_{-2,2}$ for $i=1,2$.
Acccording to the arguments given in Section~\ref{sec_geommod},
the $Z^i_{-2,2}$ are generating functions for  the relative  BPS invariants of certain elliptic threefolds $\mathbb Y^i_3$ within $Y_4$.
To verify this in the present example, we follow
the general prescription of Section~\ref{sec_geommod} and consider the threefolds
\be \label{Y13defmodS}
\mathbb Y^1_3   =  Y_4|_{p^\ast(C^1)} \,, \qquad \mathbb Y^2_3   =  Y_4|_{p^\ast(C^2)} \,,
\ee
where, according to (\ref{C^ijS}), $C^1=C_2$ and $C^2=C_1$. 
Both threefolds are elliptically fibered with respective base spaces 
\be
\mathbb B_2^1 = p^\ast(C^1) \simeq dP_2 \, ,\qquad \mathbb B_2^2 = p^\ast(C^2) \simeq C^0 \times  \mathbb P^1_{l'} \,.
\ee
Note that only $\mathbb B_2^1$ contains the blowup locus of $B_3$, whereas the rational fiber of $\mathbb B^2_2$ never splits into two exceptional curves. 

Furthermore, the normal bundles of both ${\mathbb Y^i_3}$ happen to be trivial:
\be
N_{{\mathbb Y^i_3}/Y_4} = \cO_{\mathbb Y_3^i} \,.
\ee
This is a consequence of the fact that the $\mathbb Y^i_3$ are simply the restriction of the elliptic fibration to $p^\ast(C^i)$, which in turn have vanishing self-intersection on $B_3$ and hence trivial normal bundle:
\be
N_{{\mathbb B_2^i}/B_3}= {\cal O}_{\mathbb B_2^i} \,.
\ee
By the adjunction formula 
this implies that 
\be
c_1(\bar K_{\mathbb Y^i_3}) = c_1(\bar K_{Y_4}|_{\mathbb Y^i_3}) - c_1(N_{{\mathbb Y^i_3}/Y_4}) = 0 \,,
\ee
and therefore both threefolds are themselves Calabi-Yau spaces;
we emphasize that this is a special property of the example at hand. Following the discussion in Section \ref{subsec_Y3},
we can therefore view $Y_4$ as a fibration in two ways, namely with generic fiber given by either
$\mathbb Y^i_3$ over its respective base $C_i$, for $i=1,2$.
Furthermore we can conclude from the adjunction formula that the anti-canonical bundle of the base spaces of ${\mathbb Y^i_3}$ is simply the restriction of $\bar K_{B_3}$:
\be
\bar K_{\mathbb B_2^i} = \bar K_{B_3} |_{\mathbb B_2^i} \otimes N^{-1}_{\mathbb B_2^i/B_3} = \bar K_{B_3} |_{\mathbb B_2^i} \,.
\ee
The structure of the elliptic fibration of ${\mathbb Y^i_3}$ is inherited by restriction from $Y_4$.
 This means that both threefolds are fibrations with an extra section whose height-pairing is simply given by the restriction of $b = 2 \bar K_{B_3}$ to the respective bases. See Appendix \ref{app_modelS} for details.
 
  With this preparation we can now compare the expressions $Z^i_{-2,2}$ given in eqs.~(\ref{ModelSZs2}) and (\ref{ModelSZs1}) with the elliptic genera of the heterotic strings that are dual to six-dimensional F-theory compactifications
on the Calabi-Yau threefolds ${\mathbb Y^i_3}$.
In more detail, the first terms in the expansion of $Z^i_{-2,2}$ are
\bea
\label{Het:Sb}
Z^1_{-2,2} \!\!\! &=&\!\!\!  \frac{2}{q} - \left( 252 + 84 \xi^{\pm 1}\right)  
-q\left( 116580  + {65164} \xi^{\pm 1} + 9448 \xi^{ \pm 2} + 84 \xi^{\pm 3} -2 \xi^{\pm 4}\right) 
+...
\\
\label{Het:Sa}
Z^2_{-2,2} \! \!\!  &=&\!\! \!  \frac{2}{q} - \left(288 + 96 \xi^{\pm 1}\right) 
-q \left(123756  + {69280} \xi^{\pm 1} + 10192 \xi^{\pm 2} + 96 \xi^{\pm 3} - 2 \xi^{\pm 4}\right) 
+..
\eea
We have checked by direct computation in Appendix \ref{app_modelS} that these expansions perfectly match the low-lying relative BPS invariants for $C^0$, for the elliptic  Calabi-Yau threefold  fibrations ${\mathbb Y^i_3}$ over $dP_2$ and $\mathbb F_0$, respectively. This explicitly demonstrates how the six-dimensional structure encoded in the flux-dependent, four-dimensional elliptic genus (\ref{ZfullmodelS}) manifests itself, here in terms 
of two embedded Calabi-Yau geometries that can be independently controlled by dialing the flux background.

Alternatively, in terms of the dual heterotic
language, one can recognize from  (\ref{ModelSZs2})
that  $Z^2_{-2,2}$ is nothing but the modular, 
$U(1)$ refined elliptic genus of a perturbative heterotic string compactification on $K3$ 
(with some specific bundle background turned on such that only the $U(1)$ gauge symmetry is unbroken). 
This is most visible when switching off the background gauge field:
\be
Z^2_{-2,2}(q,\xi=1)\ =\ \frac2{\eta^{24}}E_4 E_6\ \equiv\ Z_{K3}(q)\,.
\ee
On the other hand, note that  $Z^1_{-2,2}$ is only quasi-modular, in agreement with the above observation that ${\mathbb Y^1_3}$ contains the blowup locus.  In the dual heterotic language, this corresponds to a non-perturbative background. More precisely,
the difference,
\be
Z^1_{-2,2} -Z^2_{-2,2}\ =\ \frac{1}{12}\frac{E_2E_{4,1}^2}{\eta^{24}}-\frac{1}{12}\frac{E_{4,1}E_{6,1}}{\eta^{24}}\,,
\ee
{is suggestive of a transition where a small instanton has been traded against a heterotic NS5-brane.}

Finally, we note that the decomposition of the matter curve $\Sigma_{r=1}$ defined in  (\ref{SigmarmodS}) perfectly reproduces the general pattern advertised in eq.~(\ref{Sigmar-relation}), i.e.,
\be
\Sigma_{r=1} = N^0_{C^0}(1,1) \, C^0 + \frac{1}{4} N^1_{C^0}(1,1) \, (b \cdot  p^{\ast}(C_1))   + \frac{1}{4} N^2_{C^0}(1,1) \,(  b \cdot  p^{\ast}(C_2))\,,
\ee
with $N^1_{C^0}(1,1) = 84$ from (\ref{Het:Sb}), $N^2_{C^0}(1,1) = 96$ from (\ref{Het:Sa}) and $N^0_{C^0}(1,1) = - 84$.
The latter follows from $Z^0_{-1,m=2} =  84 \phi_{-1,2} = - \frac{1}{q} \sum N^0_{C^0}(n,r) q^n \xi^r  = 84 (\xi - \xi^{-1}) + \ldots$.

Let us close this part of the discussion by demonstrating the claim of Section~\ref{sec_m2flux} that the embedded threefold invariants relative to $C^0$ have another interpretation in terms of non-transversal
 $(-2)$-fluxes. This allows us to explicitly verify the special form of the elliptic holomorphic anomaly equation as given in (\ref{holellanom-2}). Recall that the  non-transversal
$(-2)$-fluxes lie in the  space $H^{(2,2)}_{(-2)}(Y_4,\mathbb R)$ defined in (\ref{H22vertdecomp}), and do not lift to fluxes in F-theory. We first need to fix a concrete basis (\ref{basism2gen})  for it. Out of the set of all products of elements $\{S_-, E, p^\ast(C^1), p^\ast(C^2) \}$,  the following four non-vanishing fluxes can be taken as a maximal linearly independent set:
\be
\begin{aligned}   \label{m2basemodelS}
\bG^{1} &= \pi^\ast(S_-) \wedge \pi^\ast(p^\ast(C^1))    \, , \qquad 
\bG^{2} &=& \pi^\ast(S_-) \wedge \pi^\ast(p^\ast(C^2)) \,,  \\
\bG^{E} &= \pi^\ast(E) \wedge \pi^\ast(p^\ast(C^1)) \, , \qquad 
\bG^{0} &=& \pi^\ast(p^\ast(C^1)) \wedge \pi^\ast(p^\ast(C^2)) \,.
\end{aligned}
\ee
By performing
analogous computations in mirror symmetry as sketched in Appendix~\ref{app_modelS}, we can compute the relative BPS invariants with respect to $C^0$ in each of these backgrounds.
The result is 
\be
\begin{aligned} \label{modelSinvm2}
& {\cal F}_{\bG^{1}; C^0} &=& - q Z^{1}_{-2,2} \,, \qquad     &{\cal F}_{\bG^{2}; C^0} &= - q Z^{2}_{-2,2}  \,,\\
& {\cal F}_{\bG^{E}; C^0} &=& \,\,  0\,, \qquad     &{\cal F}_{\bG^{0}; C^0} &=  0 \,,\\
\end{aligned}
\ee
where the $Z^{i}_{-2,2}$  encode the relative BPS invariants for $C^0$ as a curve within the Calabi-Yau threefolds $\mathbb Y_3^i$,
 as displayed in (\ref{Het:Sb}) and (\ref{Het:Sa}).
This confirms the claims stated in eq.~(\ref{Zifromm21}) and (\ref{Z0ZEvanishing}).
In particular we confirm the relation (\ref{NiGicomp1}) for the lowest degeneracies in $Z^{i}_{-2,2}$, because $C^i\cdot_{B_2} K_{B_2} = -2$ for both base curves.

\subsubsection{Anomalies and Counterterms}

We now show explicitly how the general structure of the $U(1)$ gauge anomaly and its counterterms, as laid out
in generality in Section~\ref{sec_anomalies}, works in this example.
First we rewrite the elliptic genus in (\ref{ZfullmodelS}) in the form (\ref{ZHformhet}) as
\bea \label{ZHformhet-S}
Z_{\bG;C^0} 
&=:& { \tilde g}^0  \, \tilde Z_{-1,2}^{0}   
+ \sum_{i=1}^2    { \tilde g}^i    \left(  \tilde Z_{-1,2}^{i} +   \frac{1}{4} \xi \partial_\xi \tilde Z^i_{-2,2} \right) \,,
\eea
where the flux dependent parameters 
\bea
{ \tilde g}^0 &=& (F \cdot b \cdot (S-\frac{1}{2} E)) = 4 c_1 - 4 c_2 + c_3 - 2 c_4 \,, \\
 {  \tilde g}^1  &=& (F \cdot b \cdot p^\ast(C_1)) = 4 (c_2 + c_3 + c_4) \,, \\
{\tilde g}^2& =& (F \cdot b \cdot p^\ast(C_2)) = 2 (2 c_1 + c_2 + 2 c_4) 
\eea
multiply the modular Jacobi forms 
\be
\begin{aligned}
\tilde Z_{-1,2}^{^\ast} &\equiv \tilde Z_{-1,2}^{^\ast, M} \, \ \ \ {\rm with}\cr
\tilde Z_{-1,2}^{0, M}  &=- 12  \phi_{-1,2}\,,\qquad  
 \tilde Z_{-1,2}^{1,M} &=  3  \phi_{-1,2}   \,,  \qquad   
\tilde Z_{-1,2}^{2,M}  &=12  \phi_{-1,2} \,.       
 \end{aligned}
\ee
Moreover we write
\be
\begin{aligned}
\tilde Z^{2}_{-2,2} = Z^{2,M}_{-2,2} \,,  \qquad \qquad 
\tilde Z^{1}_{-2,2} = Z^{1,M}_{-2,2}  + Z^{1,QM}_{-2,2} \,,
 \end{aligned}
\ee
as defined in (\ref{ModelSZs1}).

In line of what we pointed out before, the
 inherently four-dimensional contributions to the elliptic genus are all proportional to the modular Jacobi form $ \phi_{-1,2}$. 
The anomaly and Green-Schwarz term associated with  $\phi_{-1,2}$   follow from (\ref{4dcancell}) and are given by:
\be \label{AGSphi-1}
 \phi_{-1,2}: \qquad A^{(4)} =  - \frac{1}{6} \,, \qquad     A_{GS}^{(4)} = - \frac{1}{12}   \,.
\ee
Anomaly cancellation, $A=m  A_{GS}$, is of course automatic as a consequence of modularity.
 
Consider next the more interesting derivative contributions to the elliptic genus.
We have seen that we can formally associate six-dimensional anomalies and Green-Schwarz terms with them.
The contributions to the four-dimensional anomaly and GS terms then follow  from these via (\ref{4d6d1}) and (\ref{4d6d2}).

Specifically, recall  that  $Z^2_{-2,2}$,  associated with the embedded threefold $\mathbb Y^2_3$, is purely modular. Consistent
with this, we obtain, by direct evaluation of (\ref{Anom}) and (\ref{ACT}), for
\bea
\tilde Z^{2}_{-2,2} = Z^{2,M}_{-2,2}: \qquad   A^{(6),M} =  4 \,, \qquad  \quad A_{GS}^{(6),M} = 2 \label{AGS62S}
\eea
in agreement with the general formula (\ref{A6gen}).
By contrast, $Z^1_{-2,2}$, which is associated with $\mathbb Y^1_3$,  receives both a modular and a quasi-modular contribution. 
The modular part works out along the same lines  as for  $Z^2_{-2,2}$.  The quasi-modular anomaly and Green-Schwarz
 terms follow from (\ref{A6QMellgen}) and (\ref{AGS6QMellgen}), and altogether we find:
\be
\begin{aligned}
\tilde Z^{1,M}_{-2,2}:    \quad  A^{(6),M} &=4 \,,               
      \qquad         &A^{(6),M}_{GS}  &=2    \,,   \cr
\tilde Z^{1,QM}_{-2,2}:  \quad    A^{(6),QM} &= -\frac{1}{4}(m_1^2 + m^2_2) = - \frac{1}{2}   \,,    \qquad         &A^{(6),QM}_{GS} &= -\frac{1}{12 m} (m_1 + m_2)^2 = - \frac{1}{6}    \,,
\end{aligned}
\ee
where we used $m_1=m_2=1$ and $m=m_1+m_2=2$.

With everything combined and recalling in particular (\ref{4d6d1}), the four-dimensional anomaly thus evaluates to
\bea
A^{(4)} &=&  A^{{(4),M}} +  A^{{(4),QM}} \\
A^{{(4),M}} &=& (F \cdot b \cdot (S-\frac{1}{2} E)) \, m +( F \cdot b \cdot p^\ast(C_2)) ( m^2-m) +( F \cdot b \cdot p^\ast(C_1) )(m^2-\frac{m}{4}  ) \nn  \\
A^{{(4),QM}} &=&(F \cdot b \cdot p^\ast(C_1)) (- \frac{1}{2})  \,.\nn
\eea
In total, when all variables are substituted by their definite expressions,  this indeed correctly reproduces the 1-loop anomaly:
\be
A^{(4)} = 16 c_1 + 8 c_2 + 14 c_3 + 16 c_4  = \frac{1}{3!} \chi_{\bG, r=1} \, ,
\ee
where $\chi_{\bG, r=1}$ is given in (\ref{ciIndex}).

Let us now have a closer look at the Green-Schwarz terms and check anomaly cancellation in particular for the derivative terms
of the elliptic genus. We mentioned already that
 in the fully modular, non-derivative sector, the Green-Schwarz terms are 
guaranteed to cancel the
perturbative anomaly from the perspective of the heterotic string. 
Indeed the relation (\ref{AGS4deffactM}) is obviously satisfied by means of (\ref{AGSphi-1}):
\be
{\tilde g}^0 A^{\rm GS}_{{\tilde Z}_{-1,m}^0}=  \frac{12}{12}\,( F \cdot b \cdot (S_- - \frac{1}{2} E))  = (F \cdot b \cdot (S_- - \frac{1}{2} E)  )\,.
\ee
A bit less trivial is the anomaly cancellation in the derivative, but modular subsector. 
This sector is associated just with $\mathbb Y_3^2$,
since $\mathbb Y_3^1$ involves a quasi-modular piece. Let us thus check the 
relation (\ref{AGS4deffact2}) which arises from $\tilde Z_{-1,2}^{2}$ and  $\tilde Z_{-2,2}^{2}$ in the elliptic genus.
Explicitly, (\ref{AGSphi-1}), (\ref{AGS62S}) and (\ref{4d6d2}) when taken together yield
\bea
{ \tilde g}^2 (A^{\rm GS}_{{\tilde Z}_{-1,m}^{2,M}} + 2 A^{\rm GS}_{{\tilde Z}_{-2,m}^{2,M}}) = (-\frac{12}{12} + 2 \times  \frac{1}{4} \times 2 \times 2) \, (F \cdot b \cdot p^\ast(C_2) )= (F \cdot b \cdot p^\ast(C_2)) \,.
\eea
To compare this to (\ref{AGS4deffact2}), which posits that
\bea
{\tilde g}^2   (A^{\rm GS}_{{\tilde Z}_{-1,m}^{2,M}} + 2 A^{\rm GS}_{{\tilde Z}_{-2,m}^{2,M}})  = \frac{m^{(2)} }{m}  \, (F \cdot b \cdot p^\ast(C_2)) \,,
\eea
we need to know the parameter $m^{(2)}$. While we could not offer in 
Section~\ref{GS4d}
an a priori argument why it always takes the correct value, we find here in the current example by direct computation that  
\bea
m^{(2)} = \frac{1}{2} (b \cdot S_+ \cdot p^\ast(C^2)) = 2 \,.
\eea
Hence indeed the anomaly is cancelled as expected.

By contrast, for the Green-Schwarz terms derived from $\tilde Z_{-1,m}^{1}$ and $\tilde Z_{-2,m}^{1}$, which exhibit a quasi-modular contribution,
we find, using among others,  (\ref{4d6d2}), 
\bea\label{nineteen}
{\tilde g}^1(A^{\rm GS}_{{\tilde Z}_{-1,m}^{1}} \!\!+ 2 A^{\rm GS}_{{\tilde Z}_{-2,m}^{1}}) =  (-\frac{3}{12} + 2 \!\times\!  \frac{1}{4}\! \times\! 2 \!\times \!(2-\frac{1}{6})) \, (F \cdot b \cdot p^\ast(C_1) )= \frac{19}{12} \,(F \cdot b \cdot p^\ast(C_1))
\eea
while
\be
 m^{(1)} = \frac{1}{2} (b \cdot S_+ \cdot p^\ast(C^1))\ =\ 3\,.
 \ee
 Therefore we get a mismatch as compared to (\ref{nineteen}):
\be  
\frac{m^{(1)}}{m}  (F \cdot b \cdot p^\ast(C_1) )   = \frac{3}{2}\,(F \cdot b \cdot p^\ast(C_1)) \,,
\ee
which reflects that further Green-Schwarz terms, localized on the heterotic NS5-branes, are needed to fully cancel the anomaly.
 Nevertheless, the quasi-modular contribution to the anomaly as such,  $A^{{QM}}$, is of the general form given in (\ref{A4QMhetcomp}). In our specific example, where $m_1 = m_2 =1$, the first term in 
 (\ref{A4QMhetcomp}) actually vanishes.

\subsection{Example 2: $B_3 = \rm{Bl}^1 \mathbb H_1$}   \label{sec_Ex2het}  

As our second example we take the base of the F-theory elliptic fourfold $Y_4$ to be a blowup of 
the space $\mathbb H_1$, which is defined as a $\mathbb P^1$-fibration over $B_2 = \mathbb P^2$ with twist bundle ${\cal L} = {\cal O}(p^\ast H)$. Here $H$ is the hyperplane class on $B_2$.
Since this geometry was discussed already in detail in \cite{Lee:2019tst}, we can be brief in explaining it.

The blowup is performed over a curve in the class $C_1 = H$ on $B_2$ and leads to a splitting of the rational fiber $C^0$ into two exceptional curves, $C_E^1$ and $C_E^2$.
As in the case study of Section~\ref{subsec_P1fibered}, the blowup divisor $E$ is given by a fibration of $C_E^2$ over $C_1$.
For our basis of divisors on $B_3$ we pick
\be
D_1 = S_- - E \,, \qquad D_2 = p^\ast(C_1) - E \,, \qquad  D_3 = E \,.
\ee
In this basis, the anti-canonical bundle reads
\be
\bar K_{B_3}  = 2 D_1 + 4 D_2 + 5 D_3 \,,
\ee
and the intersection form becomes
\be
I_{B_3}= 4 D_1^3 - 3 D_2^3 - 2 D_1^2 D_3 + 2 D_2^2 D_3 +  D_1 D_3^2 -  D_2 D_3^2      \,. 
\ee
The elliptic fibration over the base $B_3$ is chosen
as in the example of Section~5 of \cite{Lee:2019tst}, to which we refer for further details. 
It leads to a gauge group $G=U(1)$, and the associated height-pairing takes the form
\be
b = 6 \bar K_{B_3}  - 2 \beta = 8 D_1  + 20 D_2 +22 D_3 \,,
\ee
where  $\beta =  2 D_1 + 2 D_2 + 4 D_3$. Moreover
its intersection numbers with the rational fiber and the exceptional curves are as follows:
\bea \label{mm1m2modelV}
m &=& \frac{1}{2} b \cdot C^0 = 4 \,, \nn
\\
 m_1 &=& \frac{1}{2} b \cdot C^1_E = 1 \,, 
 \\
  m_2 &=& \frac{1}{2} b \cdot C^2_E = 3 \,. \nn
\eea

This fibration gives rise to two types of charged massless matter fields, namely of charges 
$r=1$ and $r=2$.  The matter excitations are localised on two curves on $B_3$ in the respective classes
\be
\begin{aligned}
\Sigma_{r=1} &=  3 (6 \bar K_{B_3} - 2 \beta) \cdot (6 \bar K_{B_3} - 2 \beta) - 16 (2 \bar K_{B_3} - \beta) \cdot  (3 \bar K_{B_3} - \beta)  \cr
&= - 123 \, C^0 - 87 \, C^1_E + \frac{45}{2} (b \cdot p^\ast(C_1)) \,,  \cr
 \Sigma_{r=2} &= (2 \bar K_{B_3} - \beta) \cdot  (3 \bar K_{B_3} - \beta) \cr
&=   - \frac{33}{2} C^0 +\frac{3}{2} C^1_E + \frac{15}{4}( b \cdot p^\ast(C_1))  \,. \label{Sigma12dec}
\end{aligned}
\ee
Geometrically, the elliptic fiber over these two curves degenerates such as to contain the rational curves $C^{\rm f}_{r=1}$ and $C^{\rm f}_{r=2}$, respectively.
Switching on a general transversal $U(1)$ flux in $H^{2,2}_{\rm vert}(Y_4)$,
\be\label{ex2flux}
\bG = \bG_{U(1)} = \sigma \wedge \pi^\ast F \,, \qquad  F =: \sum_{\alpha=1}^3 c_\alpha D_\alpha  \,,
\ee
produces a chiral spectrum with indices in the respective charge sectors given by
\bea\label{chirdefs}
\chi_{r=1} &=&12 c_1 + 132 c_2 + 48 c_3, 
\\
\chi_{r=2} &=&12 c_1 + 12 c_2 + 48 c_3\,.\nn
\eea

\subsubsection{Relative BPS Invariants and Elliptic Genus}

After this preparation we turn to discussing
 the elliptic genus for the heterotic string that arises from wrapping a D3-brane on $C^0$ in presence of the $U(1)$ flux, $\bG$.
The relative BPS invariants $N_{\bG;C^0}(n,r)$ for small values of $n$  have already been computed by mirror symmetry in \cite{Lee:2019tst}.  It was noted there that for the most general choice of $U(1)$ flux,
the elliptic genus is neither modular nor a quasi-modular meromorphic form. This observation was in fact
the motivation for the present work.

The new ingredient, according to the general discussion in Section~\ref{sec_geommod}, is that the ``non-modular'' component
of the elliptic genus is actually a derivative of a quasi-modular form.
More specifically, the relative invariants $N_{\bG;C^0}(n,r)$, as computed in  \cite{Lee:2019tst}, allow to 
uniquely determine, by modular completion, the elliptic genus as follows:
\be \label{ZhetstrmodV}
Z_{\bG;C^0} = {g}^{0}  \, Z_{-1,m}^{0} +  { g}^{E}  \, Z_{-1,m}^{E} +    { g}^1  \frac{1}{2m} \xi \partial_\xi \, Z_{-2,m}^1 \,,
\ee
where the index is, as per (\ref{mm1m2modelV}), given by $m=4$.
Here the flux-dependent coefficients are
\be
{ g}^{0} =  F \cdot C^0 = c_1\,, \qquad 
{g}^{E}  =    F \cdot C^1_E = - c_2 + c_3  \,, \qquad 
{g}^1  =   F \cdot b \cdot p^\ast(C_1) = 6 c_1 + 2 c_2 + 6 c_3\,
\ee
and the quasi-modular or modular forms are given by
\begin{align}
Z_{-1,4}^{0}(q,\xi) &=  \frac{1}{16}\phi_{-1,2}({21\phi_{0,1}^2-23 E_4 \phi_{-2,1}^2}+2E_2\phi_{0,1}\phi_{-2,1})\nn
\\
&= 123 \xi^{\pm \bar 1} + 33\xi^{\pm \bar 2} + q (981  \xi^{\pm \bar 1} + 144 \xi^{\pm \bar 2} - 423 \xi^{\pm \bar 3})   + \mathcal{O}(q^2) \label{eq:Z0MV}   
\\
Z_{-1,4}^{E}(q,\xi )&= \frac{1}{16}   \phi_{-1,2} (9\phi_{0,1}^2+13 E_4\phi_{-2,1}^2-22E_2\phi_{0,1}\phi_{-2,1})\nn
\\
&=  87 \xi^{\pm \bar 1}  - 3  \xi^{\pm \bar 2}   + q(2169 \xi^{\pm \bar 1}- 1584 \xi^{\pm \bar 2} + 333 \xi^{\pm \bar 3})  + \mathcal{O}(q^2)    \label{eq:ZEMV} 
\\
Z_{-2,4}^1 (q,\xi) &=\frac{1}{12\,  \eta^{24}}
(10 E_{4,3}E_{6,1}+6 E_{4,1}F_{6,3}+19E_{4,1}G_{6,3}+E_2 E_{4,1}E_{4,3})\nn
\\
&=\frac{3}{q}-30\,(8 +6 \xi^{\pm1}+\xi^{\pm 2})+\mathcal{O}(q) \,.
\label{eq:zv6d}
\end{align}
As always,
$\xi^{\pm \bar n} := \xi^n - \xi^{-n}$, $\xi^{\pm n} := \xi^n + \xi^{-n}$
and the Jacobi forms are defined in Appendix~\ref{app_modular}.

The form (\ref{ZhetstrmodV}) of parametrizing the elliptic genus is tuned to mirror 
the decomposition (\ref{Sigma12dec}) of the matter curves $\Sigma_{r=1}$ and $\Sigma_{r=2}$ 
in terms of $C^0$, $C^1_E$ and $(b\cdot p^\ast(C_1))$, and
follows the general pattern advertised in (\ref{Sigmar-relation}).
The BPS invariants entering this latter equation are 
of course defined in terms of the partition functions (\ref{eq:Z0MV}),  (\ref{eq:ZEMV}) and 
(\ref{eq:zv6d}) via
\be \label{Nstardef}
Z_{w,m}^{\ast} = - \frac{1}{q} N^\ast_{C^0}(n,r) q^n \xi^r \,.
\ee

Of particular interest for us is the quasi-Jacobi form $Z^1_{-2,4}$ which appears with a derivative in the elliptic genus.
In line with our general arguments, we expect that
it encodes the relative BPS invariants with respect to $C^0$ viewed as a curve within an elliptic threefold,
 $\mathbb Y^1_3$, in $Y_4$. Recall from Section~\ref{subsec_Y3}  that
this threefold is constructed as the restriction of the elliptic fibration of $Y_4$ 
to the divisor $p^\ast(C^1) = p^\ast(C_1)$ on $B_3$, which in the present example
 is topologically a del Pezzo surface, $dP_2$.
The threefold is therefore elliptically fibered with base  $\mathbb B_2^1=p^\ast(C_1)$, and we denote the projection by
 \be \label{Y31modelV}
 \pi_1: \mathbb Y^1_3 \rightarrow \mathbb B_2^1  \,.
 \ee
By the adjunction formula we find that $\mathbb Y^1_3$ has a non-zero first Chern class:
\be
c_1(\bar K_{\mathbb Y^1_3}) = c_1(\bar K_{Y_4}|_{\mathbb Y^1_3}) - c_1(N_{{\mathbb Y^1_3}/Y_4}) =  - \pi_1^\ast (p^\ast(C^1)|_{p^\ast(C^1)}) \,,
\ee
because the divisor $p^\ast(C^1)$ has non-vanishing self-intersection on $B_3$.

From the point of view of the elliptic fibration, this can also be seen as follows.
The anti-canonical class is related to the pullback of the discriminant of the fibration from the base, $\Delta_{\mathbb Y^1_3}$, via the  relation
\be
c_1(\bar K_{\mathbb Y^1_3}) = \pi_1^\ast (c_1(\bar K_{\mathbb B_2^1})) - \frac{1}{12} \Delta_{\mathbb Y^1_3} \,.
\ee
The discriminant $\Delta_{\mathbb Y^1_3}$ in turn  is inherited from the discriminant of the fibration of the ambient space $Y_4$,
\be
\Delta_{\mathbb Y^1_3} = \Delta_{Y_4}|_{\mathbb Y^1_3} = 12 \pi_1^\ast(c_1(\bar K_{B_3})|_{\mathbb B_2^1}) \,.
\ee
By the adjunction formula we have
\be
c_1(\bar K_{\mathbb B_2^1}) = c_1(\bar K_{B_3}|_{\mathbb B_2^1}) - c_1(N_{{\mathbb B_2^1}/B_3}) \,,
\ee
where $c_1(N_{{\mathbb B_2^1}/B_3}) = p^\ast(C^1)|_{ p^\ast(C^1)}$. 
Thus altogether we have
\be \label{DeltaY31}
\Delta_{\mathbb Y^1_3} = 12  \pi_1^\ast(c_1(\bar K_{\mathbb B_2^1})) + 12 \pi_1(  p^\ast(C^1)|_{ (p^\ast(C^1))}) \,,
\ee
rather than just $\Delta_{\mathbb Y^1_3} = 12  \pi_1^\ast(c_1(\bar K_{\mathbb B_2^1}))$ which would be required for
a Calabi-Yau threefold.

Even though ${\mathbb Y^i_3}$ has a negative anti-canonical bundle, 
the concept of {relative} Gromov-Witten invariants still makes sense.
Unlike for Calabi-Yau spaces, however, we cannot as easily compute BPS invariants via mirror symmetry and compare them to the invariants $N_{C^0}^1(n,r)$ that are encoded in the expansion of $Z_{-2,4}^1(q,\xi)$.

Despite this technical complication, we can provide some evidence for our conjecture that $Z^1_{-2,4}$ encodes the relative invariants for ${\mathbb Y^i_3}$.
The following discussion is an illustration of the general arguments in Section \ref{subsec_Y3}.
Recall first that 
since the elliptic fibration of $\mathbb Y^1_3$ is inherited from $Y_4$, we know that it exhibits an extra rational section and that the height-pairing associated with this section is the restriction of the height-pairing $b$ to~$p^\ast(C^1)$. 
The elliptic fiber of  ${\mathbb Y^1_3}$ degenerates over a number of points on $p^\ast(C^1)$ in such a way as to contain rational curves $\Cr$ of $U(1)$ charges $r=1$ and $r=2$. Recall Figure~\ref{fig_fibr-degen} for a visualization.
The number of points where this happens equals the number of holomorphic fibral curves in class $\Cr$ on ${\mathbb Y^1_3}$, which in turn is computed by the Gromov-Witten invariants $N(\Cr)$.
For Calabi-Yau threefolds  whose base is a rational fibration with fiber $C^0$, these invariants $N(\Cr)$ agree with the relative BPS invariants at level $n=1$  , i.e.,
\be \label{3foldGWrelBPS}
N^1_{C^0}(n=1,r) = N(\Cr) \,.
\ee
This is the threefold analogue of the relation (\ref{NGC0id}) and follows from
F-theory--heterotic duality.
While strictly speaking we cannot invoke this duality for non-Calabi-Yau threefolds, it is reasonable to assert that (\ref{3foldGWrelBPS}) holds more generally whenever the base of the elliptic threefold is a rational fibration.
If we assume this, we have a simple means to test the hypothesis that the invariants defined by $Z^1_{-2,m}$ compute the relative BPS invariants of the non-Calabi-Yau threefold $\mathbb Y^1_3$, at least at level $n=1$.
This is because the number of degeneration points and hence the BPS invariants $N(\Cr)$ are easy to determine.

Concretely, as explained in Section \ref{subsec_Y3},
the number of degeneration points on $\mathbb Y^1_3$ leading to fibral curves $\Cr$  is simply given by the number of intersection points between the corresponding degeneration locus of $Y_4$ on $B_3$, i.e., the curves $\Sigma_r$, and the base of $\mathbb Y^1_3$, i.e., the divisor $p^\ast(C^1)$. This yields the BPS invariants
\bea
N(C^{\rm f}_{r=1})& =& \Sigma_{r=1} \cdot p^\ast(C^1) = 180 \,, \\
N(C^{\rm f}_{r=2}) &=& \Sigma_{r=2} \cdot p^\ast(C^1) = 30 \,.
\eea
These values are in perfect agreement with the relative invariants $N_{C^0}^1(n=1,1) = 180$ and $N_{C^0}^1(n=1,2) = 30$ 
which appear in the expansion (\ref{eq:zv6d}) of $Z_{-2,4}^1$. We can also reproduce the remaining invariant, $N_{C^0}^1(n=1,r=0)$,
which according to our conjecture must coincide with the Gromov-Witten invariant for the fibral class $\mathbb E_\tau$.
For an elliptic Calabi-Yau threefold,
this in turn would be given by the negative of the Euler characteristic of the threefold, i.e. by $-\chi({\rm CY3}) = - \int_{{\rm CY3}} c_3({\rm CY3}) =- 2 \, {\rm ch}_3({\rm CY3})$, where the Chern character is given by ${\rm ch}_3 = \frac{1}{2} c_3 - \frac{1}{2} c_1 c_2 + \frac{1}{6} c_1^3$.
For our {\it non}-Calabi-Yau threefold $\mathbb Y^1_3$ at hand, we find explicity that 
\be
 N^1_{C^0}(n=1,r=0)  = -  2 \,  {\rm ch}_3(\mathbb Y^1_3) =  - \int_{Y_4} c_3(Y_4) \wedge \pi^\ast (p^\ast(C^1)) = 240   \,,
\ee
which indeed matches the index of the uncharged states in (\ref{eq:zv6d}).

Thus, all in all we have verified that the identity (\ref{3foldGWrelBPS})  holds for our example, and this lends further support
to the conjecture that {\it all} invariants in $Z^1_{-2,4}$ match the relative BPS invariants of the non-Calabi-Yau threefold, $\mathbb Y^1_3$.

Let us present yet another, more speculative perspective on the significance of the partition function $Z^1_{-2,4}$. 
Recall  that  the non-Calabi-Yau space $\mathbb Y^1_3$ is elliptically fibered over the base $\mathbb B^1_2 = p^\ast(C^1)$, which in turn is the blowup of a  Hirzebruch surface $\mathbb F_1$ at one point, with its own base $C^1$ and generic fiber $C^0$.
This implies that 
$\mathbb Y^1_3$ also
admits a K3-fibration over $C^1$, 
\bea \label{K3FfibrF3}
\rho :\quad {\rm K3}_F \ \rightarrow & \   \mathbb Y^1_3 \cr 
& \ \ \downarrow \cr 
& \ \  C^1  
\eea
whose fiber ${\rm K3}_F$ is elliptically fibered over $C^0$.

Let us  formulate a Weierstrass model for this fibration, and
introduce the notation $[a'_1 : a'_2]$ for the homogenous coordinates on the base $C^1$,
 and $[a_1 : a_2]$ for the coordinates on $C^0$. Then a
Weierstrass model for the elliptic fibration of $\mathbb Y^1_3$ can be written, away from the exceptional curve of the blown up Hirzebruch surface, as
\be
y^2= x^3 + f(a_i, a_j') \, x \, z^4  +  g(a_i, a_j') \, z^6   \,,
\ee
where $f(a_i, a_j')$ is a section of $\Delta^{1/3}_{\mathbb Y^1_3}$ and $g(a_i, a_j')$ is a section of $\Delta^{1/2}_{\mathbb Y^1_3}$.
The degrees of the discriminant $\Delta_{\mathbb Y^1_3}$ on the base and fiber of  $\mathbb B^1_2$ follow
 from   (\ref{DeltaY31}) as
\be
\Delta_{\mathbb Y^1_3} |_{\mathbb B_2 ^1} = 48 \, C^0   +  24 \, C^1  \,.    
\ee
Taking into account that $C^0$ is fibered over $C^1$ with twist $1$ (which is the case for the Hirzebruch $\mathbb F_1$), 
this means that we can expand $f$ and $g$ as
\bea
f(a_i, a_j') &=& \sum_k f_{12 - k}(a_1',a_2')  \,  a_1^{4-k}   \,  a_2^{4+k}   \,, \\
g(a_i, a_j') &=& \sum_k g_{18 - k}(a_1',a_2')  \,  a_1^{6-k}  \,  a_2^{6+k}   \,, 
\eea
where the subscripts denote the degrees of the polynomials on the fiber coordinates.

Let us pause for a moment and consider instead an  elliptic fibration over $\mathbb B^1_2$ which is Calabi-Yau. For this we would have polynomials given by $f_{8 - k}(a_1',a_2')$ and $g_{12 - k}(a_1',a_2')$, respectively.
  This Calabi-Yau threefold could serve as a standard F-theory 
 compactifiation space to six dimensions.  This theory would in turn have a
six-dimensional heterotic dual defined in terms of an elliptic surface, $K3_{\rm het}$, with base $C^1$. A Weierstrass model for
$K3_{\rm het}$ would then be
 obtained by keeping the polynomials $f_{8 - k}(a_1',a_2')$ and $g_{12 - k}(a_1',a_2')$ for $k=0$ \cite{Morrison:1996na,Morrison:1996pp}, i.e.,
\be
{\rm K3}_{\rm het}:  y^2 = x^3 + f_{8}(a_1',a_2') x z^4 + g_{12}(a_1',a_2') z^6 \,.
\ee

Now we return to our presently considered geometry, where we deal with the non-Calabi Yau threefold,  $\mathbb Y^1_3$. 
A priori, F-theory on this space is not well-defined. Let us nonetheless
formally define a dual heterotic background,  
by keeping the middle polynomials in analogy to the well-established Calabi-Yau case.  
This leads to the following Weierstrass model
\be
\cE:\,  y^2 = x^3 + f_{12}(a_1',a_2') x z^4 + g_{18}(a_1',a_2') z^6 \,,
\ee
which is an elliptic fibration with $36$, rather than $24$ singular fibers. This defines an elliptic surface $\cE$ with 
\be
\bar K_{\cE} = - \mathbb E_\tau \,,
\ee
where $\mathbb E_\tau$ is the fiber.

It is tempting to interpret $Z_{-2,4}^1(q,\xi)$ 
as a generalized, refined elliptic genus associated with this elliptic surface.
More specifically, if we switch off the $U(1)$ background field, we find
\be
Z_{-2,4}^1(q,\xi=1)= \frac1{12\eta^{24}} E_4(E_2 E_4+35E_6)  = \frac3{\eta^{24}}E_4E_6 +  \frac1{12\eta^{24}} (E_2E_4^2-E_4E_6)\,,
\ee
which is suggestive of a non-perturbative instanton/NS5 brane transition of such a geometry.

Let us come back to a more concrete, definite property of  $Z_{-2,4}^1(q,\xi)$ as given in (\ref{eq:zv6d}). Recall the  general
relationship between the derivative part of the elliptic
genus and the partition functions associated to non-transversal  $(-2)$-fluxes, which was proposed in Section~\ref{sec_m2flux}.
If we take as basis for  the fluxes in $H^{(2,2)}_{(-2)}(Y_4,\mathbb R)$
\be
\begin{aligned} \label{m2basismodelV}
\bG^{1} &= \pi^\ast(S_-) \wedge \pi^\ast(p^\ast(C^1))  \,, \\ 
\bG^{E} &= \pi^\ast(E) \wedge \pi^\ast(p^\ast(C^1)) \, , \\ 
\bG^{0} &= \pi^\ast(p^\ast(C^1)) \wedge \pi^\ast(p^\ast(C^1)) \,,
\end{aligned}
\ee
then the associated BPS invariants assemble into the generating functions
\be
\begin{aligned} \label{modelVinvm2} 
& {\cal F}_{\bG^{1}; C^0} &=& - q Z^{1}_{-2,4} \,,\qquad 
 {\cal F}_{\bG^{E}; C^0} &=& \,\,  0\,, \qquad     &  {\cal F}_{\bG^{0}; C^0} &=  0     \,.\\
\end{aligned}
\ee
Thus we can indeed 
confirm the purported relationship between the two partition functions, which are associated with transveral $U(1)$ flux and non-transversal~$(-2)$ flux, respectively. 
Moreover,
while the vanishing of the last two generating functions is clear on general grounds as for the previous model, it is noteworthy 
to mention that the lowest BPS number in $Z^{(1)}_{-2,m}$, as read off from  (\ref{eq:zv6d}), perfectly matches the intersection theoretic expression (\ref{NiGicomp1}) because  $C^1 \cdot_{B_2} K_{B_2} = -3$.

\subsubsection{Anomalies and Green-Schwarz Terms}

In order to discuss the anomalies in a $U(1)$ flux background, we rewrite the elliptic genus (\ref{ZhetstrmodV}) into the form (\ref{ZHformhet}),
which is more suitable for comparison with the Green-Schwarz mechanism in the heterotic duality frame.
Concretely, 
\bea \label{ZHformhet-V}
Z_{\bG;C^0} 
&=:& { \tilde g}^0  \, \tilde Z_{-1,m}^{0} +   {  \tilde g}^E \, \tilde Z_{-1,m}^{E} +    { \tilde g}^1    \left(  \tilde Z_{-1,m}^{1} +   \frac{1}{2m} \xi \partial_\xi \tilde Z^1_{-2,m} \right) \,,
\eea
for $m=4$, with the flux dependent coefficients:
\bea
 { \tilde g}^0 &=& F \cdot b \cdot (S-\frac{1}{2} E) = -9 c_1  + 9 c_2   \,, 
\qquad  {  \tilde g}^E = F \cdot b \cdot (\frac{1}{2} E) = 3 c_1  + 9 c_2  - 6 c_3 \,,   \\
  { \tilde g}^1  &=& F \cdot b \cdot p^\ast(C_1) =  6 c_1 + 2 c_2  + 
 6 c_3 \,. \quad 
\eea
Moreover each term,  $\tilde Z_{w,m}^{^\ast}\equiv \tilde Z_{w,m}^{^\ast, M} + \tilde Z_{w,m}^{^\ast,QM}$, generically 
splits into a modular and quasi-modular piece.  Explicitly:
\be\label{explictZV}
\begin{aligned}
\tilde Z_{-1,4}^{0, M} &= -\frac{1}{12}\phi_{-1,2}  ( \phi_{0,1}^2 - E_4 \phi_{-2,1}^2) \,, \qquad  &\tilde Z_{-1,4}^{0, QM}    &=0    \,,  \cr
\tilde Z_{-1,4}^{E, M}  &= - \frac{1}{6} \phi_{-1,2}  \phi_{-2,1}^2 E_4     \,,                                        \, \qquad   &\tilde Z_{-1,4}^{E, QM}   &=     \frac{1}{6}  \phi_{-1,2}  \phi_{-2,1}  E_2 \phi_{0,1}  \,,\cr
\tilde Z_{-1,4}^{1,M} &= \frac{1}{32} \phi_{-1,2} (3 \phi^2_{0,1}   - E_4 \phi_{-2,1}^2    )    \,,  \, \qquad   &\tilde Z_{-1,4}^{1,QM}  &= -  \frac{1}{16}  \phi_{-1,2}\phi_{-2,1} E_2 \phi_{0,1}   \,, \cr
\tilde Z^{1,M}_{-2,4} &= \frac{1}{12\,  \eta^{24}}(10E_{4,3}E_{6,1}+6 E_{4,1}F_{6,3}+19E_{4,1}G_{6,3})   \,,  \qquad   &\tilde Z_{-2,4}^{1,QM}  & = \frac{1}{12\,  \eta^{24}}    E_2 E_{4,1}E_{4,3} \,.
\end{aligned}
\ee
Note that all these partition functions have, depending on their   modularity properties, the  respective general form as
advertised in (\ref{generalZstructure}).

Let us first discuss the anomalies and Green-Schwarz terms derived from the non-derivative (quasi) modular forms in $Z_{\bG;C^0}$.
These receive contributions only from three independent
building blocks, which are readily computed following the general prescription outlined in Sections~\ref{subsec_Anomaliesmod} and~\ref{subsec_AnomaliesQuasimod}:
\be
\begin{aligned}
& \phi_{-1,2} \phi_{-2,1}^2 E_4 :    \qquad &A^{(4),M} &= 0 \, \qquad  \qquad & A^{(4),M}_{GS}  &= 0  \cr
&\phi_{-1,2} \phi_{0,1}^2:            \qquad  &A^{(4),M}  &=   -48 \,, \qquad &A^{(4),M}_{GS} &=   \frac{1}{m} A^{(4),M}   = -12   \cr
& \phi_{-1,2} \phi_{-2,1} E_2 \phi_{0,1} : \qquad  &A^{(4),QM}  &=  6 (m_1 - m_2) = -12  \,, \qquad &A^{(4),QM}_{GS} & =0 \,.   \cr
\end{aligned}
\ee
Recall that the values $m=4$ and $m_1=1$, $m_2 =3$ were determined from the geometry in (\ref{mm1m2modelV}), but we
leave the values sometimes unassigned in order to exhibit the structure of the various terms.

Note also in passing that the modular component related to the exceptional curve,  $\tilde Z_{-1,m}^{E}$,  is proportional to 
\be\label{ZEMod}
\phi_{-1,2}  \phi_{-2,1}^2 E_4 \ =\ \cO(\zh^5)\,,
\ee
and therefore contributes neither to the anomaly nor to the Green-Schwarz term, in line with an argument made after
equation (\ref{ZEMgen1}). 

Now we turn to the derivative sector.
From $\tilde Z^{1}_{-2,4}$ we first compute the contributions to the six-dimensional anomaly as follows:
\be
\begin{aligned}
&\tilde Z^{1,M}_{-2,4} :   \qquad &  A^{(6),M}  &= 30  \,, \qquad                                                                      &A^{(6),M}_{GS}  &= \frac{1}{4}  \, A^{(6), M}  \,, \cr
& \tilde Z^{1,QM}_{-2,4} :   \qquad &  A^{(6),QM}  &= -\frac{1}{4}(m_1^2 + m^2_2) = - \frac{5}{2}     \,, \qquad        &A^{(6),QM}_{GS}  &= -\frac{1}{12 m} (m_1 + m_2)^2 = - \frac{1}{3} \,.
\end{aligned}
\ee
This then determines, via (\ref{4d6d1}) and (\ref{4d6d2}), the contributions  $\xi \partial_\xi \tilde Z_{-1,4}^{1}$ to the four-dimensional anomaly and Green-Schwarz terms.

Altogether, the complete anomaly becomes 
\bea
A^{(4)} &=&  A^{{(4),M}} +  A^{{(4),QM}}   \nn \\
A^{{(4),M}} &=& m \, (F \cdot b \cdot (S-\frac{1}{2} E))  + { (\frac{30}{2} - 48 \frac{3}{32})} \,( F \cdot b \cdot p^\ast(C_1))    \\
A^{{(4),QM}} &=&        {(m_1 - m_2)} \, (F \cdot b \cdot (\frac{1}{2} E))  +    {  (-\frac{1}{8}(m_1^2 + m^2_2) - \frac{6}{16} (m_1 - m_2))}  \, (F \cdot b \cdot p^\ast(C_1) ) \,\nn
\eea
and this correctly reproduces the 1-loop anomaly as required:
\be
A^{(4)} =  18 c_1 + 38 c_2 + 72 c_3 = \frac{1}{3!} \sum_{r=1} ^2 r^3 \chi_{r}  \, .
\ee
Here $\chi_{r}$ refers to the chiral index in the charge sector $r$ as given in (\ref{chirdefs}).

The structure of the anomaly reflects the Green-Schwarz mechanism in the dual heterotic frame explained in Section~\ref{GS4d}. In particular,  the purely modular contribution from $\tilde Z_{-1,m}^{0}$, 
\be
{\tilde g}^0  A_{\tilde{Z}_{-1,m}^0}   = m  \,  (F \cdot b \cdot (S_- - \frac{1}{2} E)) \,, 
\ee
maps to the part of the anomaly that is cancelled by the perturbative universal Green-Schwarz term.
Moreover,  the quasi-modular contribution to the anomaly,  $A^{{QM}}$, turns out to be of the expected form (\ref{A4QMhetcomp}).
For the derivative terms, the analogue of this would be the relation (\ref{AGS4deffact2}), but this can be checked not to hold due to the quasi-modular contributions,  in line with general expectations.

 \section{Elliptic Genera of 4d Non-Critical Strings} \label{sec_non-criticalstrings}

 Our conjecture of Section \ref{sec_geommod} on the form of four-dimensional elliptic genera
 is supposed to hold not only for heterotic strings, but more generally also for other, in particular
 non-critical solitonic strings.
In this section we illustrate this by presenting the elliptic genus of two different types of non-critical strings in four dimensions.
 We begin by discussing the four-dimensional E-string in the next section, followed by a non-critical string obtained in F-theory compactified on $B_3 = \mathbb P^3$ in Section~\ref{sec_P3string}.

 \subsection{Four-Dimensional E-Strings} \label{sec_Estring}

 The six-dimensional non-critical E-string \cite{Seiberg:1996vs,Ganor:1996mu,Witten:1996qb,Klemm:1996hh,Lerche:1996ni,Minahan:1997ch,Minahan:1997ct,Minahan:1998vr} arises in F-theory 
 by wrapping a D3-brane on a rational curve $C_E$ of self-intersection $-1$ which lies in the base $B_2$ of some
 elliptic Calabi-Yau threefold, $Y_3$.
 Such curves have normal bundle $N_{C_E/B_2} = {\cal O}_{C_E}(-1)$ and can arise in two different settings: Either $C_E$ is the exceptional section of a Hirzebruch surface $B_2 = \mathbb F_1$,
 or it appears after blowing up a general Hirzebruch surface at a point. After the blowup the rational fiber over the point in the base splits into two exceptional curves $C_E^1$ and $C^2_E$ with normal bundle 
 $N_{C^i_E/B_2} = {\cal O}_{C^i_E}(-1)$.
 
 We can generalise the notion of an E-string to F-theory compactifications to four dimensions by wrapping a D3-brane on a curve with normal bundle 
 \be \label{NbundleE}
 N_{C_E/B_3} = {\cal O}_{C_E}(-1) \oplus {\cal O}_{C_E} \,.
 \ee
{ In this paper we will call the strings obtained from D3-branes that wrap such curves {\it four-dimensional E-strings} and study their elliptic genera. Aspects of four-dimensional analogues of E-string have previously been considered in \cite{Mayr:1996sh} (see also \cite{Morrison:2016nrt} and \cite{Apruzzi:2018oge}). Clearly such strings are special cases of a multitude of non-critical strings that can arise from much more general types of curves. More recently, the compactification of six-dimensional $N=(1,0)$ SCFTs on Riemann surfaces with fluxes has been a subject of intense study \cite{Gaiotto:2015usa,Franco:2015jna,Hanany:2015pfa,Coman:2015bqq,Razamat:2016dpl,Bah:2017gph,Mitev:2017jqj,Kim:2017toz,Bourton:2017pee,Kim:2018bpg,Razamat:2018gro,Razamat:2018zus,Razamat:2019vfd,Chen:2019njf,Pasquetti:2019hxf,Razamat:2019ukg,Sela:2019nqa}, and it would be worthwhile to relate our setup to the field theoretic approach, though this will not be the focus of this paper.}

 The trivial summand ${\cal O}_{C_E}$ in (\ref{NbundleE}) implies that the curve $C_E$ is fibered over a distinguished normal direction within $B_3$, thereby tracing out a rationally fibered divisor which we call $D_E$.
The four-dimensional geometry probed  by the E-string is hence a fibration at least locally, where the fiber $C_E$ is either the base of a Hirzebruch surface $\mathbb F_1$ or one of the blowup curves in a rational fibration.
 We will exemplify both types of geometries and their associated E-string elliptic genera. 
 
 From the perspective of geometry alone, one might think that the properties of the resulting four-dimensional E-string are entirely inherited from 
 the six-dimensional E-string which is locally fibered. According to the logic of this paper, this would mean that the 
 four-dimensional E-string elliptic genus would be a derivative of the six-dimensional E-string elliptic genus.
 This, however, is in general not the case. The definition of the model requires specifying the background flux, and we will see that the latter can introduce genuinely four-dimensional, non-derivative contributions to the elliptic genus, precisely as expressed in the most general form of Conjecture~\ref{Conj2} in Section~\ref{sec_geommod}.
This phenomenon is independent of the geometric realisation of the E-string curve, either as the base of a locally fibered Hirzebruch surface or due to a blowup in the fiber of the rational fibration.
 
 \subsubsection{E-Strings of Six-Dimensional Origin}

As our first example we consider an F-theory compactification for which the base space, $B_3$, is of the type introduced in Section~\ref{subsec_P1fibered}, namely it is a rational fibration by itself: $p: B_3 \to B_2$. 
After we perform a blowup over a curve $\Gamma$ on the base $B_2$, 
the rational fiber $C^0$ over $\Gamma$ splits into two exceptional curves $C^1_E$ and $C^2_E$.
Each of the two $C^i_E$ is therefore fibered over the curve $\Gamma$ and the fibration defines two divisors, $D_E^i$,
 of the following form:
\bea \label{pprojDEi}
p^i_E :\quad C^i_E \ \rightarrow & \  \ D_E^i \cr 
& \ \ \downarrow \cr 
& \ \  \Gamma  
\eea
Note that in Section \ref{subsec_P1fibered} we had called the divisor $D_E^2=:E$.
Furthermore, as long as we  perform only one blowup, which is what we assume from now on, 
we have $D_E^1 + D_E^2 = p^\ast(\Gamma)$.

For simplicity we now take the gauge group in four dimensions to be $U(1)$ and consider a background with 
transversal flux, ${\mathbf G} = {\mathbf G}_{U(1)}\in H^{2,2}_{\rm vert}(Y_4)$.
We will discuss an example with non-abelian gauge group later in Section \ref{sec_4dEstring}.
We claim that for such a setup, the elliptic genus of the E-string associated with either $C^1_E$ or $C^2_E$ can be brought into the following universal form:
\be \label{ZEstringgen1}
Z_{\bG_{U(1)};C^i_E}(q,\xi) = \frac{1}{2m_i} (F \cdot b \cdot D_E^i)  \,  \xi \partial_\xi  Z_{-2,m_i}(q,\xi)  \,, \qquad m_i = \frac{1}{2} b\cdot C^i_E \,,
\ee
 where $Z_{-2,m_i}$ is the elliptic genus  \cite{Klemm:1996hh,Minahan:1998vr} of a six-dimensional E-string with $U(1)$ fugacity index $m_i$, i.e.: 
 \be
  Z_{-2,m_i}(q,\xi)  = \frac{1}{\eta^{12}(q)} E_{4,m_i}(q,\xi)  \,.
 \ee
Furthermore,  the degeneracies contained in $Z_{-2,m_i}(q,\xi)$ are the relative BPS invariants for the curve $C_E^i$ within an elliptic threefold embedded into $Y_4$.

 Let us illustrate this general formula for the E-strings that arise (besides the heterotic string) 
in the two examples that were discussed in Sections~\ref{sec_Ex1het} and~\ref{sec_Ex2het}.
 First, consider the base $B_3 = dP_2 \times \mathbb P^1_{l'}$.
In the notation of Section~\ref{sec_Ex1het}, the two divisors $D_E^i$ defined via (\ref{pprojDEi}) are immediately identified as
 \be
 D_E^1 = p^\ast(C_1) - E = D_4 - D_2\,, \qquad     D_E^2 = E = D_1 + D_2 - D_4 \,,
 \ee
 with
 \be
 F \cdot b \cdot D_E^1 = 4 c_2 + 2 c_3 \,, \qquad F \cdot b \cdot D_E^2 = 2 c_3 + 4 c_4   \,.
 \ee
 We recall furthermore that in this geometry the geometric intersection numbers $m_i = \frac{1}{2} b\cdot C^i_E=1$ for both $i=1,2$.

 As detailed in Appendix \ref{app_modelS},  we can compute by mirror symmetry the lowest-lying relative BPS invariants $N_{C_E^i; \bG}(n,r)$ and the associated elliptic genera 
 \be
 Z_{\bG_{U(1)};C^i_E} = - q^{-1/2} \sum_{n,r} N_{\bG_{U(1)};C^i_E}(n,r) q^n \xi^r
 \ee
  as follows:
 \bea
 Z_{\bG_{U(1)};C^1_E}(q,\xi) &=& -(2 c_2 + c_3) q^{-1/2}\Big(
 (56 \xi^{\pm \bar 1} + 2 \xi^{\pm \bar 2}) + q (1248 \xi^{\pm \bar 1} +276 \xi^{\pm \bar 2}) + \cO(q^2) \Big) \\ \nn
 Z_{\bG_{U(1)};C^2_E}(q,\xi) &=& -(c_3 + 2 c_4) q^{-1/2}\Big(
 (56 \xi^{\pm \bar 1} + 2 \xi^{\pm \bar 2}) + q (1248 \xi^{\pm \bar 1} +276 \xi^{\pm \bar 2}) + \cO(q^2 )  \Big) \,.\\ \nn
 \eea
 These are uniquely completed into the exact expressions:
  \be
   Z_{\bG_{U(1)};C^i_E}(q,\xi) = \frac{1}{2} (F \cdot b \cdot D_E^i)   \, \frac{1}{\eta^{12}(q)}  \, \xi \partial_\xi   E_{4,1}(q,\xi)   \,,
  \ee
which are  in perfect agreement with general pattern (\ref{ZEstringgen1}).
 The threefold whose relative BPS invariants are contained in $Z_{C^i_E; \mathbf G}(q,\xi)$ is the Calabi-Yau $\mathbb Y_3^1 = Y_4 |_{p^\ast(C^1)}$ introduced already in (\ref{Y13defmodS}).
 Note that $C^1$ is the curve dual on $B_2$ to $C_1$, whose class in turn corresponds to the class of the curve $\Gamma$ over which the E-string curves are fibered as in (\ref{pprojDEi}). 
 Formula (\ref{ZEstringgen1}) is hence very much analogous to the derivative  contributions to the heterotic string elliptic genus (\ref{ZfullmodelS}) in the same geometry. The difference is that the E-string curves $C^i_E$ are fibered only over $\Gamma$, and hence there appears only a single derivative contribution. For the heterotic string, on the other hand, the curve $C^0$ is fibered over all of $B_2$ so that we must sum over several contributions, each corresponding to one basis element $C_i$ of $H_2(B_2)$.

 As for the  example of Section \ref{sec_Ex2het}, we consider the two E-strings in the geometry $B_3= {\rm Bl}^1 \mathbb H_1$
  with
  \be
 D_E^1 = p^\ast(C_1) - E = D_2\,, \qquad     D_E^2 = E = D_3 \,,
 \ee
  and 
  \be
( F \cdot b \cdot D_E^1) = -16 c_2 + 18 c_3 \,, \qquad (F \cdot b \cdot D_E^2) = 6 c_1 + 18 c_2 - 12 c_3   \,.
 \ee
 The geometric intersection numbers are $m_1 = 1$ and $m_2=3$.
 By application of mirror symmetry, we have computed the lowest relative BPS numbers, which assemble into the following expansions:
\allowdisplaybreaks
\begin{align} \nn
 Z_{\bG_{U(1)};C^1_E}(q,\xi) &= (- 8c_2 + 9 c_3) q^{-1/2} \Big(
(56 \xi^{\pm \bar 1} + 2 \xi^{\pm \bar 2} ) \,\\  
&\qquad +\, q(1248 \xi^{\pm \bar 1} + 276 \xi^{\pm \bar 2}  )\,\\ \nn
&\qquad +\, q^2(13464 \xi^{\pm \bar 1} + 4716 \xi^{\pm \bar 2} + 168 \xi^{\pm \bar 3} ) \,\\ \nn
&\qquad +\, q^3(103136 \xi^{\pm \bar 1} + 46008 \xi^{\pm \bar 2} + 3744 \xi^{\pm \bar 3} + 4 \xi^{\pm \bar 4} ) \,\\ \nn
&\qquad +\, \cO(q^4) 
 \Big) \,, \\ \nn 
 Z_{\bG_{U(1)};C^2_E}(q,\xi) &= (c_1 + 3 c_2 - 2 c_3 ) q^{-1/2}  \Big(
(54 \xi^{\pm \bar 1} + 54 \xi^{\pm \bar 2} + 6 \xi^{\pm \bar 3}) \,\\ 
&\qquad +\, q(1080 \xi^{\pm \bar 1} + 1188 \xi^{\pm \bar 2} + 504 \xi^{\pm \bar 3} + 108 \xi^{\pm \bar 4})\,\\ \nn
&\qquad +\, q^2(11016 \xi^{\pm \bar 1} + 13122 \xi^{\pm \bar 2} + 7350 \xi^{\pm \bar 3} + 2376 \xi^{\pm \bar 4} + 270 \xi^{\pm \bar 5} + 6 \xi^{\pm \bar 6} ) \,\\ \nn
&\qquad +\, q^3(81216 \xi^{\pm \bar 1} + 102060 \xi^{\pm \bar 2} + 66264 \xi^{\pm \bar 3} + 26244 \xi^{\pm \bar 4} + 5400 \xi^{\pm \bar 5} + 516 \xi^{\pm \bar 6} ) \,\\ \nn
&\qquad +\, \cO(q^4) 
 \Big)  \,.
\end{align}
 These uniquely determine the elliptic genera as follows:
 \bea
Z_{\bG_{U(1)};C^1_E}(q,\xi) &=& \frac{1}{2} (F \cdot b \cdot D_E^1) \, \frac{1}{\eta^{12}(q)}  \, \xi \partial_\xi   E_{4,1}(q,\xi) \,, \\
Z_{\bG_{U(1)};C^2_E}(q,\xi) &=& \frac{1}{6} (F \cdot b \cdot D_E^2) \, \frac{1}{\eta^{12}(q)}  \, \xi \partial_\xi   E_{4,3}(q,\xi) \,,
 \eea
which again illustrates the general claim (\ref{ZEstringgen1}). We expect that the invariants encoded in $Z_{\mathbf G;C^i_E}(q,\xi)$ are the relative BPS invariants with respect to $C^i_E$ within the non-Calabi-Yau threefold $\mathbb Y^1_3$, which was discussed around
eq.~(\ref{Y31modelV}) in the context of the heterotic string.
 
 The two examples discussed so far are special to the extent that the four-dimensional E-string is completely determined by its six-dimensional cousin, as reflected in the purely derivative structure of the elliptic genus.
As we will see in the next section,  this is no longer the case if the E-string curve lies inside a 7-brane and is threaded by gauge flux.

 Before coming to this point, however, we take a brief digression to understand the invariants encoded in $Z_{-2,m_i}(q,\xi)$ via  (\ref{ZEstringgen1}) also in terms of non-transversal $(-2)$ fluxes, analogous to what we found for the heterotic string.
 Irrespective of the details of $B_2$, we can give the general pattern by 
 first considering the relative BPS invariants  for the E-string curves $C_E^1$ and $C_E^2$ at $n=0$ and $r=0$.
 The curves  $C_E^1$ and $C_E^2$ are each fibered over the curve $\Gamma$ on $B_2$. The class of the moduli space of both curves with one point fixed is hence the surface $D_E^i$ on $B_3$ traced out by this fibration, i.e. 
  \be
  \mu(C_E^i) =  S_0  \wedge   D_E^i    \,, \qquad i=1,2 \,.
  \ee
A non-vanishing BPS number for $C_E^i$ is only possible in a flux background which intersects this surface.  
It is easy to see which types of 4-form fluxes on $B_3$ have this property,
because the fibration structure of $B_3$ implies that the only non-vanishing intersections with $D_E^i$ are 
\bea \label{Estringint}
D_E^1  \cdot S_- \cdot p^\ast(C) = C \cdot_{B_2} \Gamma  \,, \quad 
\qquad D_E^i  \cdot  D_E^j \cdot p^\ast(C) = (1 - 2 \delta^{ij})  \, C \cdot_{B_2} \Gamma  \,.
\eea
 Here $C$ represents some curve class on $B_2$, and  we used the fact that only $C^1_E$ is intersected by $S_-$.
 Whenever $C$ intersects $\Gamma$ transversally, it is natural to infer from the local nature of the fibration 
 that the 4-forms $S_- \cdot p^\ast(C)$ and  $D_E^i \cdot p^\ast(C) $   restrict the moduli space of the more general curve $C^i_E + n \mathbb E_\tau  +  \Cr$ in $Y_4$ to $Y_4|_{p^\ast(C)}$. 
 Hence up to normalisation, which is given by the intersection numbers (\ref{Estringint}), we expect to exactly reproduce the BPS numbers as encoded in $Z_{-2,m_i}(q,\xi)$.
 
 This general expectation is perfectly confirmed by computation.
 For  the two E-strings of the model with base $B_3 = dP_2 \times \mathbb P^1_{l'}$
of Section~\ref{sec_Ex1het} we find, for the basis (\ref{m2basemodelS}) of non-transversal $(-2)$ fluxes
\be
\begin{aligned}
& {\cal F}_{\bG^{1}; C^1_E} &=& - q^{1/2} \frac{1}{\eta^{12}(q)}  \,   E_{4,1}(q,\xi)  \,, \qquad     &{\cal F}_{\bG^{2}; C^1_E} &= 0 \,,\\
& {\cal F}_{\bG^{E}; C^1_E} &=&- q^{1/2} \frac{1}{\eta^{12}(q)}  \,   E_{4,1}(q,\xi)  \,\,  \,, \qquad     &{\cal F}_{\bG^{0}; C^1_E} &=  0 \\
\end{aligned}
\ee
 and 
 \be
\begin{aligned}
& {\cal F}_{\bG^{1}; C^2_E} &=& \, 0 \,, \qquad     &{\cal F}_{\bG^{2}; C^2_E} &= 0 \,,\\
& {\cal F}_{\bG^{E}; C^2_E} &=& + q^{1/2} \frac{1}{\eta^{12}(q)}  \,   E_{4,1}(q,\xi)  \,\,  \,, \qquad     &{\cal F}_{\bG^{0}; C^2_E} &=  0 \,.\\
\end{aligned}
\ee
 The extra sign in in the expression for ${\cal F}_{\bG^{E}; C^2_E}$ reflects the fact that $E \cdot C_E^2 = D_E^2 \cdot C_E^2 = -1$. 
 Similarly, for $B_3= {\rm Bl}^1 \mathbb H_1$ of Section  \ref{sec_Ex2het} 
 for the basis (\ref{m2basismodelV}) of non-transversal $(-2)$ fluxes we have
 \be
\begin{aligned}
& {\cal F}_{\bG^{1}; C^1_E} &=& - q^{1/2} \frac{1}{\eta^{12}(q)}  \,   E_{4,1}(q,\xi)  \,, \qquad     \\
& {\cal F}_{\bG^{E}; C^1_E} &=& - q^{1/2} \frac{1}{\eta^{12}(q)}  \,   E_{4,1}(q,\xi)  \,\,  \,, \qquad     &{\cal F}_{\bG^{0}; C^1_E} &=  0 \\
\end{aligned}
\ee
and 
 \be
\begin{aligned}
& {\cal F}_{\bG^{1}; C^2_E} &=& \, 0 \,, \qquad     \\
& {\cal F}_{\bG^{E}; C^2_E} &=& + q^{1/2} \frac{1}{\eta^{12}(q)}  \,   E_{4,3}(q,\xi)  \,\,  \,, \qquad     &{\cal F}_{\bG^{0}; C^2_E} &=  0 \,.\\
\end{aligned}
\ee

Hence again certain threefold BPS invariants - here of the E-string curves $C_E^i$ within $\mathbb Y_3^1$  - are obtained as fourfold invariants in suitable, non-transversal $(-2)$ flux backgrounds.
This does not mean, however, that {\it all} relative BPS invariants within a given threefold are generated by one and the same $(-2)$ flux.
For instance, the flux $\bG^{1}$ reproduces the relative BPS invariants of $C^0$ and $C_E^1$ within  $\mathbb Y_3^1$, but not of  $C_E^2$, and $\bG^{E}$ gives the relative BPS invariants for 
$C_E^1$ and $C_E^2$ in $\mathbb Y_3^1$, but not of $C^0$, as comparison with (\ref{modelVinvm2}) and (\ref{modelSinvm2}) shows.

 \subsubsection{Genuinely Four-Dimensional E-String}\label{sec_4dEstring}

 While in the previous class of constructions the E-string genera are entirely of six-dimensional origin,  we now present an example where  one finds
 in addition  a genuinely four-dimensional contribution.
 The E-string lives inside a stack of 7-branes carrying a gauge sector with $G=SU(2)$.
 The  background geometry, $B_3 = \mathbb F_1 \times \mathbb P^1$, is detailed in Appendix~\ref{sec:Het-A1}. 
 The base is rationally fibered with projection $p: B_3 \to B_2$ over $B_2 = C_E \times \mathbb P^1$, where $C_E$ is the base of $\mathbb F_1$.
 A D3-brane wrapping $C_E$ gives rise to a four-dimensional E-string. The E-string curve is trivially fibered over the extra  $\mathbb P^1$, and this fibration traces out the divisor $D_E = C_E \times \mathbb P^1$ on $B_3$.
 
  The elliptic fibration $Y_4$ over $B_3$ models a stack of 7-branes with gauge group $G=SU(2)$ wrapped on the divisor
  \be
  b= C_E \times \mathbb P^1 \,.
  \ee
 Since the E-string curve $C_E$ lies inside the 7-brane divisor $b$,
its intersection number with $b$ is negative,
  \be \label{mA1model}
b \cdot C_E = - 1 \,.
  \ee
In the elliptic fibration over $B_3$, the elliptic fiber over $b$ splits into two rational curves which intersect like the nodes of the affine Dynkin diagram of $SU(2)$.  Fibering both curves over $b$ defines two exceptional divisors $e_0$ and $e_1$ with $e_0 + e_1 = \pi^\ast(b)$, where the fiber of $e_0$ is associated with the affine node. 
The generator of the Cartan $U(1)$ within $SU(2)$
can be identified with the divisor $-e_1$. To understand the normalisation, note that matter excitations
in the fundamental representation arise from M2-branes wrapping a  curve $C^{\rm f}$ in the fiber over a curve on $b$. The holomorphic curve $C^{\rm f}$ satisfies
$-e_1 \circ C^{\rm f} = 1$, so that matter excitations from an M2-brane along $C^{\rm f}$ have charge $r=1$
 under the $U(1)$ generated by $-e_1$.
 
For this geometry we consider a transversal flux background of the form
  \be
  \bG = (-e_1) \wedge \pi^\ast(F) \,, \qquad F = c_1 D_1 + c_2 D_2 + c_3 D_3 \,,
  \ee
  which corresponds to the Cartan $U(1)$ of $SU(2)$. 
  Here
   \be
   D_1 = C^0 \times \mathbb P^1 \,, \qquad  D_2 = C^0 \times \mathbb P^1 + C_E \times \mathbb P^1    \,, \qquad D_3 = p^\ast(C_E) \,.\ee
As before, we determine via mirror symmetry a number of
relative BPS invariants  $N_{C_E; \bG}(n,r)$ and so obtain 
the first terms in the expansion of the elliptic genus. While we refer to
eq.~(\ref{ellgenEexp1}) in the Appendix for more data, we present here just the lowest order in $q$:\footnote{Note that the flux labelled by $c_2$ does not contribute.}
\bea
q^{1/2}  \, Z_{C_E;\bG}(q,\xi)  &=& \nn
q [ c_1 (16 \xi^{-2}  -160 \xi^{-1} + 768 -2400 \xi + 5616 \xi -10752\xi^3 + 17920 \xi^4 \\ 
&&\qquad  -27136\xi^5 + 38400 \xi^6 -51712 \xi^7 + 67072 \xi^8 + \cdots ) \\ \nn
&&\quad + c_3 ( 6\xi^{-2} -40 \xi^{-1} -472\xi + 2042 \xi^2 -4608 \xi^3+ 8192 \xi^4 -12800 \xi^5 \\ \nn
&&\qquad + 18432 \xi^6 -25088\xi^7  + 32768 \xi^8 + \cdots) ] +\cO(q^2)\,.  \nn
\eea
The noteworthy feature is that at any given order of $q$ there is an infinite series in $\xi$. 
This points to the expansion of  a $\xi$-dependent denominator. Such behaviour is expected from experience with non-critical strings of six-dimensional $N=(1,0)$ SCFT's with nontrivial gauge symmetry \cite{Haghighat:2013tka,Hohenegger:2013ala,Haghighat:2014vxa,Kim:2015fxa,Kim:2016foj,Gadde:2015tra,DelZotto:2016pvm,DelZotto:2017mee,Kim:2018gak,Kim:2018gjo,DelZotto:2018tcj,Gu:2019dan}, which are obtained when D3-branes wrap a 
curve that is also wrapped by 7-branes. In six dimensions, such strings are interpreted as instanton strings with respect to the non-abelian gauge group localized on the stack of 7-branes. The point is that
for a coincident 7-brane/D3-brane system, the open strings in the 3-7 sector lead to massless charged  chiral fields the $N=(0,2)$ supersymmetric worldsheet theory, and in turn to charged bosonic excitations in spacetime.
These are responsible for the appearance of an $\xi$-dependent denominator in the elliptic genus.

Our goal is to understand the four-dimensional elliptic genus as the derivative of a six-dimensional elliptic genus, possibly augmented by a genuinely four-dimensional piece. To find the six-dimensional contribution, recall that 
the curve $C_E$ is trivially fibered over the $\mathbb P^1$ factor in $B_3$, thereby tracing out the divisor $D_E$ within $B_3$.
This suggests that we must consider the E-string within a threefold, $\mathbb Y_3$,  obtained by restricting $Y_4$ to the fiber of this fibration. This threefold is simply
\be
\mathbb Y_3 = Y_4 |_{D_3} \,,
\ee
where the divisor $D_3$ cuts out the Hirzebruch surface $\mathbb F_1$ on $B_3$.
This threefold happens to be a Calabi-Yau space, and thus we can
easily compute the relevant BPS invariants via mirror symmetry. The first terms in the expansion of the elliptic genus for the six-dimensional E-string in $\mathbb Y_3$ leads to the BPS numbers presented in (\ref{E:A1b}).
These BPS numbers can be identified as the expansion coefficients of 
\be \label{Z6dinststring}
Z_{-2,-1}(q,\xi) =   -  {  \frac{1}{\eta(\tau)^{12}}\sum_{i=2}^4\frac{\theta_i(0,\tau)^{10}}{\theta_i(2z,\tau)^2}}   \,.
\ee
In accord with our expectations, this coincides with the elliptic genus of an instanton string of a six-dimensional $N=(1,0)$ SCFT with $SU(2)$ gauge symmetry and $SO(20)$ flavor symmetry \cite{Kim:2015fxa}, in the limit where the chemical potentials of its $SO(20)$ flavor symmetry have been switched off and the only nonvanishing chemical potential is the one for the Cartan generator of $SU(2)$.

While we expect (\ref{Z6dinststring}) to contribute to the elliptic genus of the four-dimensional E-string with a derivative, it turns out that for general Cartan flux $\bG$, the four-dimensional elliptic genus receives in addition a fully modular, non-derivative contribution without a six-dimensional origin.
In fact, the elliptic genus of our four-dimensional E-string can be brought into the 
 following form, in full agreement with Conjecture~\ref{Conj2} in Section~\ref{sec_geommod}:
  \be \label{Zestring4d6d}
  Z_{C_E; \mathbf G}(q,\xi)  = (F \cdot C_E)   Z_{-1,m}(q,\xi) + \frac{1}{m} (F \cdot b \cdot D_E)   \xi \partial_\xi  Z_{-2,m}(q,\xi)  \,,
  \ee
  where\footnote{Note that in the definition of $m$ we have not included a factor of $\frac{1}{2}$. This reflects the different normalisation
of the $U(1)$ as Cartan subgroup of $SU(2)$, as compared to the non-Cartan $U(1)$'s studied in the other examples in this work.
The value of $m$ defined in this way gives the $U(1)$ fugacity index of the modular forms.
Correspondingly, the prefactor of the derivative term in (\ref{Zestring4d6d})  is $\frac{1}{m}$, rather than $\frac{1}{2m}$. 
} 
  \be
  m = b \cdot C_E = - 1
  \ee
and the flux-dependent parameters are
  \be
  F \cdot C_E = c_1 \,, \qquad F \cdot b \cdot D_E = - c_3   \,.
  \ee
Note again that the flux labelled by $c_2$ does not contribute.  The first term, 
\be
Z_{-1,-1}(q,\xi) = {   16 \, i \frac{\theta_1(z,\tau)^{10}}{\theta_1(2z,\tau)^3\eta(\tau)^9}} \,,
\ee
represents a genuinely four-dimensional contribution to the elliptic genus. 
Evidently, fluxes for which $F \cdot C_E \neq 0$ affect the spectrum of the solitonic string in a more drastic way if the string lies inside a 7-brane stack, and as a result the four-dimensional E-string can no longer
simply be viewed as a fibered version of a six-dimensional E-string. The final expression (\ref{Zestring4d6d}) then nicely disentangles via the fluxes the universal six-dimensional contribution and the genuinely four-dimensional contribution to the elliptic genus.

\goodbreak
\subsection{Four-Dimensional String on $B_3 = \mathbb P^3$} \label{sec_P3string}
  
As our final example, we present the elliptic genus of a non-critical string that probes a base geometry without any fibration structure. 
Nonetheless, we will be able to write the elliptic genus in the general form advertised in Section~\ref{sec_geommod}. 
  
  We will consider the simplest possible elliptic fourfold fibration, namely where the base is
given by $B_3=\mathbb P^3$. We will focus on the elliptic genus of the non-critical string obtained by wrapping a D3-brane along some curve $C_{\rm b}$, whose class is given in terms of the hyperplane class by  
\be
C_{\rm b}  = H \cdot H \,.
\ee
To keep things simple, we engineer a gauge group $G=U(1)$ with associated height-pairing
 $b = 2 \bar K_{B_3} = 8 H$. The details of the geometry can be found in Appendix~\ref{app_P3string}.
 The transversal $U(1)$ flux is of the form 
\be\label{cdef}
 \bG_{U(1)} = \sigma \wedge \pi^\ast(F)\,, \qquad {\rm with\ } F = c \, H\,.
\ee
  The elliptic genus of the non-critical string is expected to encode the relative BPS invariants $N_{C_{\rm b};\bG}(n,r)$ via
 \be
  Z_{\bG_{U(1)};C_{\rm b}}(q,\xi)  =  -\frac{1}{q^2} \sum_{n,r} N_{\bG_{U(1)};C_{\rm b}}(n,r)   q^n \xi^r   \,,
 \ee
 where the prefactor $q^{-2} = q^{-E_0/2}$ reflects the vacuum energy $E_0 = C_{\rm b} \cdot \bar K_{B_3} = 4$ on the string.
 The lowest invariants can be computed by mirror symmetry as discussed in Appendix \ref{app_P3string}.
Exploiting modularity, we can completely determine the elliptic genus as
 \be\label{P3ellgen}
 Z_{\bG_{U(1)};C_{\rm b}}(q,\xi)  =(F \cdot C_{\rm b})   \, Z_{-1,m}(q,\xi)  + \frac{1}{2m} (F \cdot b \cdot H)  \xi \partial_\xi   Z_{-2,m}(q,\xi) \,,
 \ee
where  $Z_{-1,m}(q,\xi)$ and  $Z_{-2,m}(q,\xi)$ are meromorphic Jacobi forms for which explicit expressions  are presented in  eqs.~(\ref{P2Z2}) and (\ref{P2Z1}).
Moreover,  the $U(1)$ fugacity index is given by
 \be
 m = \frac{1}{2}  b \cdot C_{\rm b} = 4 \,,
 \ee
while
 \be
F \cdot C_{\rm b} = c \,, \qquad F \cdot b \cdot H = 8 \, c\,.
 \ee
Recall that the single flux parameter $c$ is defined in (\ref{cdef});  since $h^{1,1}(B_3) =1$ there is not enough room for different types of $U(1)$ fluxes to 
separate out the modular and derivative components, so that just a single linear combination appears.

We conclude from this example that the four-dimensional elliptic genus can receive a derivative contribution even if the base,  $B_3$,
 is not a fibration. 
 The invariants contained in the derivative piece $Z_{-2,4}(q,\xi)$
 once more appear in the background of a 
non-transversal $(-2)$ flux.
In the present case, the only non-trivial $(-2)$ flux is of the form
\be  \label{GHbackground}
\bG_{H} = \pi^\ast(H) \wedge  \pi^\ast(H) \,.
\ee 
 For an elliptic fibration over $\mathbb P^3$ without $U(1)$ gauge group, such fluxes were investigated
 in~\cite{Haghighat:2015qdq,Cota:2017aal}. 
 As we show in Appendix~\ref{app_P3string}, the generating function for the BPS invariants in this flux background precisely encodes the invariants contained in  $Z_{-2,4}(q,\xi)$, i.e.
 \be
{ \cal F}_{\bG_{H};  C_{\rm b}} = - q^2  Z_{-2,4}(q,\xi)   \,.
 \ee

It is interesting to wonder if the invariants  $Z_{-2,4}(q,\xi)$ are again the invariants of a certain embedded threefold but 
we have no evidence that this is the case in this example. In fact, in \cite{toappear} we will give very direct arguments for the relation between certain $(-2)$ flux invariants and the invariants of embedded threefolds,
but these do not apply to the flux background  (\ref{GHbackground}).

\section{Conclusions and Outlook}

In this work we have investigated the rich interplay between the modular properties of elliptic genera, the enumerative geometry of genus zero relative BPS invariants on elliptic fourfolds with background fluxes, and the structure of $U(1)$
anomalies for effective field theories in four dimensions.
Our analysis has been distilled to Conjecture \ref{Conj2} in Section \ref{sec_geommod}, which disentangles the flux-induced
elliptic genus in four dimensions into a sum of modular Jacobi forms, quasi-modular Jacobi forms, and derivatives thereof.
Our conjecture applies in particular to critical heterotic strings and to non-critical solitonic strings in four dimensions with $N=(0,2)$ worldsheet supersymmetry. Such theories can be engineered in F-theory as worldvolume theories of 
D3-branes compactified on curves which lie in the base, $B_3$, of an elliptic fourfold $Y_4$.

That the elliptic genus in four dimensions is not necessarily
a (quasi-)modular Jacobi form {\it per se} had been explicitly observed already in \cite{Lee:2019tst}, {and is in agreement with the general results of \cite{libgober2009elliptic} and with the conjecture \cite{Oberdieck:2017pqm} that the generating functions of BPS invariants of elliptic fibrations are captured in terms of quasi-Jacobi forms.}
According to Conjecture \ref{Conj2} of the present paper, the non-(quasi-)modular components of the elliptic genus are, in fact, of a simple derivative form. Apart from breaking modular invariance, they also break the elliptic shift symmetriy which corresponds to spectral flow in
the $U(1)$ current algebra. These anomalies can be remedied at the cost of introducing a  non-holomorphic derivative as in (\ref{pznonhol}).
This results in eq.~(\ref{eq:ell_anomaly}), which can be seen as a concrete realization
of the elliptic holomorphic anomaly equation that was proposed in \cite{Oberdieck:2017pqm}, thereby giving further support of it.

 Our  arguments are based on a central observation of this paper, namely that the derivative contributions to the elliptic genus in the presence of transversal flux can be obtained in two a priori independent ways:
 First, they arise from the partition function of BPS invariants in the background of non-transversal fluxes. This is in accordance
 with the elliptic holomorphic anomaly equation that was introduced in \cite{Oberdieck:2017pqm}.
 A main result in our work concerns a second, novel interpretation of these invariants, namely as BPS invariants of certain   threefolds  $\mathbb Y_3^i$ that are embedded in $Y_4$. 
 Taking the second point of view in fact allowed us to independently derive the elliptic holomorphic anomaly equation.

In special cases, the threefolds $\mathbb Y_3^i$ are themselves Calabi-Yau manifolds and their BPS invariants can easily be computed by means of mirror symmetry. By analysing a variety of examples, we find that these BPS invariants indeed match the derivative sector of elliptic genera. This lends strong support to our claims.
More generally, the concept of relative BPS invariants is well-defined also for non-Calabi-Yau geometries, and we gather 
circumstantial evidence that the observed pattern persists for such $\mathbb Y_3^i$.  It would be intriguing to develop techniques to
 compute these invariants more directly,  from first principles, in order to compare them with our predictions.

In the special case of a heterotic string, we have been able to prove the geometric interpretation of the BPS invariants in the derivative sector, at least at the first energy level $n=1$ of the elliptic genus.
This rests on the geometric interpretation of the fourfold BPS invariants in terms of the moduli space of curves in the fourfold.
It would be desirable to extend this proof to all levels $n$ \cite{toappear}. This would open up the fascinating possibility of using mirror symmetry on {\it Calabi-Yau fourfolds} to determine relative BPS invariants of {\it non-Calabi-Yau threefolds}.

From a physics perspective, the embedded threefolds $\mathbb Y_3^i$ can be thought of as formally 
defining six-dimensional sectors in the following sense.
If the $\mathbb Y_3^i$ are Calabi-Yau manifolds,
 the derivative part of the elliptic genus can be literally interpreted as a collection of six-dimensional elliptic genera. 
 For instance, for the case of a heterotic solitonic string, we can go to the dual weakly coupled eigenframe
which describes a compactification on a $K3$ surface with gauge bundle. In this case
the six-dimensional elliptic genus we talk about is precisely the elliptic genus of such a compactification. 
 We have also discussed the more
generic situation where $\mathbb Y_3^i$ is not Calabi-Yau: while the arguments are not as sharp, 
we can still make a formal analogy and consider a dual heterotic geometry associated with some elliptic surface, which however is not a $K3$
surface any longer.

We have illustrated Conjecture \ref{Conj2} by a number of examples covering a variety of four-dimensional, critical as well as non-critical strings, and their elliptic genera.
For the special case of the critical heterotic string, the modular properties of the elliptic genus reflect the intricacies of the Green-Schwarz anomaly cancelling mechanism  in four dimensions. {As described, contrary to what happens in six dimensions, the elliptic genus in four dimensions} is not necessarily a
(quasi-)modular Jacobi form, and we show in detail how this ties in with anomaly cancellation involving extra $B$-fields.
Specifically, the modular part of the elliptic genus is linked to the universal $B$-field, while the other sectors
reflect the presence of further $B$-fields, in general of both perturbative and non-perturbative nature.

We have furthermore analysed two types of examples of non-critical strings in four dimensions. The first is a generalisation of the concept of an E-string to four dimensions: 
Its elliptic genus is generally the sum of a derivative piece (which is related to the familiar six-dimensional E-string), plus a genuinely four-dimensional contribution, whose details depend on the flux background.
While the E-strings are special in that they can be decoupled from gravity in four dimensions, we have  tested Conjecture \ref{Conj2}  also for a non-critical string which is associated with a non-shrinkable curve in $\mathbb P^3$, so that gravity
cannot be decoupled.
Clearly the handful of prototypical examples provided in this work only form the beginning of a much more systematic study of elliptic genera of strings in four dimensions.

\subsection*{Acknowledgements} 
We thank Matteo Costantini and Thorsten Schimannek for discussions on related topics. The work of SJL is supported by IBS under the project code, IBS-R018-D1.  
WL and TW thank the KITP at UC Santa Barbara for hospitality during important stages of this work. This part of the 
research was supported by the National Science Foundation under Grant No.~PHY-1748958.
  
\appendix

\appendix

 \section{Modular Forms} \label{app_modular}
\subsection{Rings of Jacobi ($\Jac_{*,*}$), Quasi-Modular Jacobi ($\Jac_{*,*}^{QM}$), and Quasi-Jacobi  ($\Jac_{*,*}^{QJ}$)  Forms}  \label{app_jacobi}

Jacobi forms  \cite{EichlerZagier}, as holomorphic functions of two variables, $\Phi(\tau,z):  \IH\times\IC\rightarrow \IC$, 
are primarily characterized by their simple transformation properties under the modular group:
\bea\label{jacobitrApp}
\Phi_{w,m}  \left(\frac{a \tau + b}{c \tau +d}, \frac{z}{c \tau +d} \right) &=& (c \tau+d)^w e^{2\pi  i  \frac{m c}{c\tau +d}  z^2}    \Phi_{w,m}(\tau,z)\ \ \,{\rm for}    
 \left(\begin{matrix} 
      a & b \\
      c & d \\
   \end{matrix}\right) \in SL(2,\IZ),
\\
\Phi_{w,m}\left( \tau , z + \lambda \tau + \mu \right) &=& e^{-2 \pi i   m (\lambda^2 \tau  + 2  \lambda  z )  }   \Phi_{w, m} (\tau,  z)\,,
\quad \lambda, \mu \in \mathbb Z \,.\label{periodicity}
\eea
They possess a Fourier expansion 
\be
\Phi_{w,m}\ =\ \sum_{n\geq0}\sum_{r^2\leq4mn}c(n,r)\,e^{2 \pi i(n\tau+ rz)}\,,
\ee
and are the natural building blocks \cite{Kawai:1993jk,Gritsenko:1999fk,Gritsenko:1999nm}   of elliptic genera (\ref{Zellgendef}) that are refined by an extra $U(1)$ current. There exists an extensive literature about Jacobi forms (for example, besides the original work \cite{EichlerZagier}, also  \cite{Kawai:1998md,Dabholkar:2012nd,gritsenko2018graded}), so we can be brief. We just mention here some aspects that are important in this work.
A Jacobi form  $\Phi_{w, m}(\tau, { {z}})$ is called 
\begin{itemize}
\item a holomorphic Jacobi form if $c(n,r) = 0$ unless $4 {m} n \geq r^2$,
\item a Jacobi cusp form  if $c(n,r) = 0$ unless $4 {m} n > r^2$,
\item a weak Jacobi form if $c(n,r) = 0$ unless $ n \geq 0$ \,.
\end{itemize}
One furthermore defines a weakly (or nearly) holomorphic Jacobi form by requiring $c(n,r) = 0$ unless $ n \geq n_0$ for a negative integer $n_0$.

Jacobi forms form a bi-graded ring which we denote by
\be
\Jac_{\ast,\ast}\ = \oplus_{w,m}\Jac_{w,m}\,.
\ee
For even weight and integer index,  $\Jac_{2k,m}$  is freely generated by
$\phi_{0,1}$ and $\phi_{-2,1}$ with coefficients given by polynomials in the Eisenstein series
$E_4$ and $E_6$.  The Eisenstein series, as well as the generators $\phi_{0,1}$ and $\phi_{-2,1}$, can be expressed in terms of the 
Dedekind function $\eta(\tau) = q^{\frac{1}{24}} \prod_{n=1}^\infty (1 - q^n)$ and the 
familiar Jacobi theta functions as follows:\footnote{Note that we adopt conventions different to those of \cite{Lee:2019tst}.}
\bea\label{defJacobieven}
&& E_4(\tau) = \frac{1}{2}\sum_{\ell=1}^4\vartheta_\ell(\tau,0)^8,
\\
&& E_6(\tau) = \frac{1}{2}\left(\vartheta_2(\tau,0)^8(\vartheta_3(\tau,0)^4+\vartheta_4(\tau,0)^4)+\vartheta_3(\tau,0)^8(\vartheta_2(\tau,0)^4-\vartheta_4(\tau,0)^4)\right.\\
&&\left.\hspace{.7in}-\vartheta_4(\tau,0)^8(\vartheta_2(\tau,0)^4+\vartheta_3(\tau,0)^4)\right),\nn
\\
&& \phi_{-2,1}(\tau, z) = -\frac{\vartheta_1(\tau,z)^2}{\eta^6(\tau)},  \label{phim21}
\\
&& \phi_{0,1}(\tau, z) = 4 \left(  \frac{\vartheta_2(\tau,z)^2}{\vartheta_2(\tau,0)^2}  + \frac{\vartheta_3(\tau,z)^2}{\vartheta_3(\tau,0)^2}  + \frac{\vartheta_4(\tau,z)^2}{\vartheta_4(\tau,0)^2}  \right).
\eea
For odd weight and integer index, $\Jac_{2k+1,m}$ has a single extra generator 
\bea\label{defJacobiodd}
&& \phi_{-1,2}(\tau, z) = \frac{i\vartheta_1(\tau,2z)}{\eta^3(\tau)}\,,
\eea
which implies that any odd-weight Jacobi form of integer index must be proportional to $\phi_{-1,2}$. This is of great significance in the present work, since the relevant elliptic genus of
four-dimensional theories has weight $w=-1$ and integer $m$.  Note that the odd generator obeys the relation
\be\label{phirelation}
432\phi_{-1,2}^2 = {\phi_{-2,1}}\left( \phi_{0,1}^3-3 E_4  \phi_{-2,1}^2\phi_{0,1}+2 E_6  \phi_{-2,1}^3\right)
\ee
so that effectively it appears at most linearly. 
So altogether the ring of weak Jacobi forms is generated by
\be
\Jac_{*,*}^{}: \ \left\{E_4,E_6,\phi_{0,1},\phi_{-2,1},\phi_{-1,2}\right\},
\ee
modulo the relation (\ref{phirelation}).

For reference,  we define $q=e^{2\pi i\tau}$,  $\xi=e^{2\pi i z}$, $\xi^{\pm n}=\xi^{ n}+\xi^{-n}$, $\xi^{\pm \bar n}=\xi^{ n}-\xi^{- n}$,
and  $\zh=2\pi i z$, and list the following expansions:
\bea\label{jacexpansions}
\phi_{0,1}(\tau, z)
 &=&    10+ \xi^{\pm1} +  (108-64   \xi^{\pm1}+10 \xi^{\pm2}    )q+ ...    
\\
&&= 12+E_2\zh^2+\frac1{24} ({E_2}^2+E_4)\zh^4+...,
\nn\\
\phi_{-2,1}(\tau, z) 
 &=&    -2+\xi^{\pm1}-(12-8   \xi^{\pm1}+2 \xi^{\pm2}    )q+ ...  
\\  
&&=\zh^2+\frac1{12}E_2\zh^4+\frac1{1440}(5{E_2}^2-E_4)\zh^6 +... ,
\nn\\
\phi_{-1,2}(\tau, z) 
&=& \xi^{\pm\bar1}+(3\xi^{\pm\bar1}-\xi^{\pm\bar3})q+...
\\
&&=  2\zh+\frac13E_2\zh^3+\frac1{180}(5{E_2}^2-2E_4)\zh^5+...\,.
\nn
\eea

Moreover, an important r\^ole is played by holomorphic quasi-modular forms. The ring of such forms is generated by the Eisenstein series
$E_4$ and $E_6$ plus
\be
E_2(\tau) = \frac1{2\pi i}\frac{\Delta'(\tau)}{\Delta(\tau)} = 1-24q-72q^2+\cO(q^3)\,,
\ee
where $\Delta=\eta^{24}=\frac1{1728}(E_4^3-E_6^2)$.
As is well-known, it is not fully modular but transforms with an extra piece
\be\label{E2quasi}
E_2\left(\frac{a\tau+b}{c\tau+d}\right)\ =\  (c\tau+d)^2E_2(\tau)-\frac{6i}\pi c (c\tau+d)\,.
\ee
This modular anomaly can be remedied at the expense of holomorphicity by defining
\be\label{e2hat}
\hat E_2(\tau)\ =\ E_2(\tau)-\frac3{\pi {\rm Im}\tau}\,.
\ee

In this paper we consider two extensions: a simpler one and a more complicated one which contains the first.\footnote{An even more drastic
extension would be in terms of more general mock-modular Jacobi forms, see for example ref.~\cite{Dabholkar:2012nd}. However this appears not to be relevant for the compact geometries we consider.}
The simple one is what we call
the ring $\Jac_{*,*}^{QM}$ of {\it quasi-modular} Jacobi forms. It is defined similar to
$\Jac_{*,*}$ except that the coefficients are polyomials in $E_2$ as well as in $E_4$ and $E_6$. That is, its generators are
\be
\Jac_{*,*}^{QM}: \ \left\{E_2,E_4,E_6,\phi_{0,1},\phi_{-2,1},\phi_{-1,2}\right\}.
\ee
The more complicated extension is obtained by
first defining an ``almost holomorphic'' function 
on $\IH\times\IC$ of the form
\be\label{almostholo}
\Phi(\tau,z)\ =\ \sum_{i,j\geq0} \Phi_{i,j}(\tau,z)\left(\frac1{{\rm Im}\tau}\right)^i\left(\frac{{\rm Im}z}{{\rm Im}\tau}\right)^j\,,
\ee
where the sum runs over finitely many terms and the $\Phi_{i,j}(\tau,z)$ are holomorphic (and appropriately convergent).
If the non-holomorphic function $\Phi(\tau,z)$ obeys the transformation laws of a Jacobi form as given in (\ref{jacobitrApp}) and 
(\ref{periodicity}), then the holomorphic first term in the sum is defined to be a {\it quasi-Jacobi} form:
\be
\Phi_{0,0}(\tau,z)\in \Jac_{*,*}^{QJ}   \,.
\ee
Note that $\Jac_{*,*}^{QM}\subset \Jac_{*,*}^{QJ}$ as it corresponds to the special case of
almost holomorphic Jacobi forms (\ref{almostholo})
for which $j\equiv0$.
For more thorough definitions, see  \cite{KanekoZagier,libgober2009elliptic,Oberdieck:2017pqm}.

Summarizing, {\it quasi-modular} Jacobi forms can be made modular by reparing the anomalous transformation 
behaviour of the generator $E_2$ in (\ref{E2quasi}) by adding a non-holomorphic piece as in (\ref{e2hat}).
This has many known manifestations in the physics literature.

On the other hand, {\it quasi-Jacobi} forms have a worse transformation behaviour that cannot be remedied 
in this simple way, and needs extra treatment in the form of $\frac{{\rm Im}z}{{\rm Im}\tau}$.
The quasi-Jacobi forms that appear in this paper all arise as $z$-derivatives of modular or quasi-modular Jacobi forms. More precisely,
if we start from a Jacobi form $\Phi_{w,m}(\tau,z)\in \Jac_{w,m}$ with given weight and index, then
\bea\label{quasijacdef}
\Phi_{w+1,m}(\tau,z) &=& \nabla_{z,m} \Phi_{w,m}(\tau,z)\,,
\\
\nabla_{z,m}&:=&\partial_z  + 4 \pi i m \frac{{\rm Im}z}{{\rm Im}\tau}\,,
\eea 
 is almost holomorphic while modular with weight $w+1$ under the transformations (\ref{jacobitrApp}) and  (\ref{periodicity}). 
 This follows from the transformation property of  $\alpha  \equiv \frac{{\rm Im}z}{{\rm Im}\tau}$:
 \bea
\alpha\left(\frac{a \tau + b}{c \tau +d}, \frac{z}{c \tau +d} \right) 
 &=&
 (c\tau+d) \alpha(\tau,z) -c\,z\,,
 \\
\alpha\left( \tau , z + \lambda \tau + \mu \right) 
 &=&
\alpha(\tau,z) +\lambda\,.
 \eea
 It then follows that by definition $\partial_z \Phi_{w,m}\in  \Jac_{w+1,m}^{QJ}$.  
 Analogous arguments apply if $\Phi_{w,m}\in\Jac_{w,m}^{QM}$.

\subsection{Eisenstein-Jacobi Forms}\label{app_Ekm}

In this section we identify a set of Jacobi forms which are closely related to the Eisenstein series $E_4(\tau)$ and $E_6(\tau)$.
This parametrization of the ring of Jacobi forms, $\Jac_{*,*}$, is naturally adapted to the geometries we consider,
and makes certain properties more manifest.
Specifically, we are interested in holomorphic Jacobi forms $\Phi_{w,m}(q,\xi)=\Phi_{w,m}(\tau,z)$, of weight $w=4 $ or $6$ and index $m$ 
characterized by the following properties:
\begin{itemize}
\item $\Phi_{w,m}(q,\xi=1) \equiv \Phi_{w,m}(\tau,z=0) = E_{w}(\tau)$,
\item $\Phi_{w,m}(q,\xi) = \sum_{n\geq 0}\sum_{r\in\mathbb{Z}} c(n,r) q^n \xi^r$,\ \  where $c(n,r)\in\mathbb{Z}$ and $c(0,r)=\delta_{0,r}$.
\end{itemize}
We further restrict our attention to Jacobi forms with index $m\leq 4$ and integral expansion coefficients which are relevant to the examples that appear in this paper. It is straightforward to construct explicitly all such forms within the ring $J_{*,*}$. Specifically, for any $w= 4, m \leq 3 $, as well as for $w=6, m\leq 2$, there is a unique such form, which coincides with the Jacobi-Eisenstein series  $E_{w,m}$ of \cite{EichlerZagier}. These are given by:
\begin{align}
E_{4,0} &= E_4=1+240q+\mathcal{O}(q^2),\label{eq:e4first}\\
E_{4,1}& =\frac1{12}\left({E_4\phi_{0,1}-E_6\phi_{-2,1}}\right)=1+(\xi^{\pm2}+56\xi^{\pm1}+126)q+\mathcal{O}(q^2),\\
E_{4,2}& =\frac1{12^2}\left({E_4\phi_{0,1}^2-2E_6\phi_{0,1}\phi_{-2,1}+E_4^2\phi_{-2,1}^2}\right)=1+(14\xi^{\pm2}+64\xi^{\pm1}+84)q+\mathcal{O}(q^2),\\
E_{4,3}& =\frac1{12^3}\left({E_4\phi_{0,1}^3-3E_6\phi_{0,1}^2\phi_{-2,1}+3E_4^2\phi_{0,1}\phi_{-2,1}^2-E_4 E_6\phi_{-2,1}^3}\right)\\
&=1+(2\xi^{\pm3}+27\xi^{\pm2}+54\xi^{\pm1}+74)q+\mathcal{O}(q^2)\,,\nn
\end{align}
and
\begin{align}
E_{6,0} &= E_6=1-504q+\mathcal{O}(q^2),\\
E_{6,1}& =\frac1{12}\left({E_6\phi_{0,1}-E_4^2\phi_{-2,1}}\right)=1+(\xi^{\pm2}-88\xi^{\pm 1}-330)q+\mathcal{O}(q^2),\\
E_{6,2}& =\frac1{12^2}\left({E_6\phi_{0,1}^2\!-\!2E_4^2\phi_{0,1}\phi_{-2,1}\!+\!E_4E_6\phi_{-2,1}^2}\right)=1-(10\xi^{\pm2}\!+\!128\xi^{\pm 1}\!+\!228)q+\mathcal{O}(q^2).\label{eq:e6last}
\end{align}
On the other hand, for $w=4, m=4$ and for $w=6,m=3$  and $4$, the sought-after Jacobi forms cannot be Jacobi-Eisenstein series, as the latter do not have integer coefficients. For each of these cases there is a one parameter family of Jacobi forms with the required properties, which we denote by $E_{w,m,t},\, t\in\mathbb{Z}$:
\begin{align}
\label{integralE4}
E_{4,4,t}&\!=\!\frac1{12^4}\Big({E_4\phi_{0,1}^4\!-\!4E_6\phi_{0,1}^3\phi_{-2,1}\!+\!6E_4^2\phi_{0,1}^2\phi_{-2,1}^2\!-\!4E_4 E_6\phi_{0,1}\phi_{-2,1}^3\!+\!(9E_4^3-8E_6^2)\phi_{-2,1}^4}\Big)\!-\!t \Delta\phi_{-2,1}^4\nn \\
&=1+(\xi^{\pm4}+56\xi^{\pm2}+126\xi^{\pm1}-t\frac{(1-\xi)^8}{\xi^4})q+\mathcal{O}(q^2),\\
E_{6,3,t}& =\frac1{12^3}\Big({E_6\phi_{0,1}^3-3E_4^2\phi_{0,1}^2\phi_{-2,1}+3E_4E_6\phi_{0,1}\phi_{-2,1}^2-E_6^2\phi_{-2,1}^3}\Big)-t \Delta\phi_{-2,1}^3\nn \\
&=1-(\xi^{\pm3}+27\xi^{\pm2}+135\xi^{\pm1}+178+t\frac{(1-\xi)^6}{\xi^3})q+\mathcal{O}(q^2),\\
E_{6,4,t}& =\frac1{12^4}\Big({E_6\phi_{0,1}^4\!-\!4E_4^2\phi_{0,1}^3\phi_{-2,1}\!+\!6E_4E_6\phi_{0,1}^2\phi_{-2,1}^2\!-\!4E_6^2\phi_{0,1}\phi_{-2,1}^3\!+\!E_4^2E_6\phi_{-2,1}^4}\Big)\!-\!t \Delta\phi_{0,1}\phi_{-2,1}^3\nn \\
&=1-(4\xi^{\pm3}+44\xi^{\pm2}+124\xi^{\pm1}+160+t\frac{(1-\xi)^6}{\xi^3}(\xi^{\pm1}+10))q+\mathcal{O}(q^2).
\label{integralE6}
\end{align}
In the paper, we employ these forms to express the $U(1)$ dependence numerator of the heterotic string elliptic genus on K3, which in the limit $z\to 0$ reduces to  $ 2 E_4(\tau)E_6(\tau).$ In fact, in the examples we make a further specialization and write the elliptic genera in terms of the following set of Jacobi forms:
\begin{align}
F_{6,3}:=E_{6,3,9},\quad\!\! G_{6,3}:=E_{6,3,-3},\quad\!\! F_{4,4}:=E_{4,4,1},\quad\! \!G_{4,4}:=E_{4,4,0},\quad\!\! F_{6,4}:=E_{6,4,0},\quad\!\! G_{6,4}:=E_{6,4,-1},
\end{align}
together with the Eisenstein-Jacobi forms (\ref{eq:e4first}-\ref{eq:e6last}).\\

We note that there exist a number of bilinear relations among this set of Jacobi forms:
\begin{align}
&E_{4,0} F_{4,4}-4E_{4,1}E_{4,3}+3E_{4,2}^2=0,\nn\\
&E_{4,0} E_{6,2}+E_{4,2}E_{6,0}-2E_{4,1}E_{6,1}=0,\nn\\
&E_{4,0} F_{6,3}+3E_{4,1}E_{6,2}-9E_{4,2}E_{6,1}+5E_{4,3}E_{6,0}=0,\nn\\
&E_{4,0} G_{6,3}-3E_{4,1}E_{6,2}+3E_{4,2}E_{6,1}-E_{4,3}E_{6,0}=0,\\
&E_{4,1} F_{6,3}-3E_{4,3}E_{6,1}+2F_{4,4}E_{6,0}=0,\nn\\
&E_{4,1} G_{6,3}-3E_{4,2}E_{6,2}+3E_{4,3}E_{6,1}-F_{4,4}E_{6,0}=0,\nn\\
&E_{4,0} F_{6,4}-3E_{4,2}E_{6,2}+2E_{4,3}E_{6,1}=0,\nn\\
&E_{4,0} G_{6,4}-2E_{4,1}G_{6,3}+2E_{4,3}E_{6,1}-G_{4,4}E_{6,0}=0.\nn
\end{align}
Moreover we find that the following relations hold:
\begin{align}
\label{nablarelations}
E_{4,1}\hat\nabla E_{4,2}&=12\Delta \phi_{-1,2}\phi_{-2,1},\nn\\
E_{4,1}\hat\nabla E_{4,3}&=2\Delta \phi_{-1,2}\phi_{0,1}\phi_{-2,1},\nn\\
E_{6,1}\hat{\nabla} E_{4,1}&=144\Delta \phi_{-1,2},\nn\\
E_{6,1}\hat{\nabla} E_{4,2}&=E_{6,2}\hat{\nabla} E_{4,1}=12\Delta \phi_{-1,2}\phi_{0,1},\\
E_{6,1}\hat\nabla E_{4,3}&=\Delta \phi_{-1,2}(\phi_{0,1}^2+E_4\phi_{-2,1}^2),\nn\\
E_{6,2}\hat\nabla E_{4,2}&=\Delta \phi_{-1,2}(\phi_{0,1}^2-E_4\phi_{-2,1}^2),\nn\\
F_{6,3}\hat\nabla E_{4,1}&=\Delta \phi_{-1,2}(\phi_{0,1}^2+9E_4\phi_{-2,1}^2),\nn\\
G_{6,3}\hat\nabla E_{4,1}&=\Delta \phi_{-1,2}(\phi_{0,1}^2-3E_4\phi_{-2,1}^2).\nn
\end{align}
The operator $\hat\nabla$ acts on a pair of Jacobi forms $\Phi_{w_1,m_1}$, $\Phi_{w_2,m_2}$ of nonzero index to produce the a Jacobi form of weight $w_1+w_2+1$ and index $m_1+m_2$, and has the following definition:
\be\label{nablaftright} 
\Phi_{m_1,k_1}\hat\nabla \Phi_{m_2,k_2} = \frac{1}{m_2}\Phi_{m_1,w_1}(\xi\partial_\xi)\Phi_{m_2,w_2}-\frac{1}{m_1}\phi_{m_2,w_2}(\xi\partial_\xi)\Phi_{m_1,w_1}.
\ee

\goodbreak

 \section{4d Heterotic and E-Strings for $B_3= dP_2 \times \IP^1_{l'}$ }   \label{app_modelS}

Here we will provide details of the geometry of the elliptic Calabi-Yau fourfold, $Y_4$, and of the embedded threefolds $\mathbb Y_3^i$, as discussed in Section~\ref{sec_Ex1het}.

 The toric coordinates of the threefold base
 \beq
  B_3= dP_2 \times \IP^1_{l'} 
 \eeq
 are listed in the upper-left part of Table~\ref{tb:Model S}, in terms of the $U(1)$ charges of a gauged linear sigma model (GLSM). 

As a basis of $H^{1,1}(B_3)$  we take
\beq
\begin{array} {lllllll}
D_1 &=& \nu_{z_0} &=& f &=& p^*(C_1) \,, \\ 
D_2 &=& \nu_{z_3} &=& h&=& S_- \,, \\ 
D_3 &=& \nu_{x_0} &=& l' &=& p^*(C_2)\,, \\ 
D_4 &=& \nu_{z_1} &=& f+h-E&=&p^*(C_1) + S_- -E\,. 
\end{array}
\eeq
The notation $f$ and $h$ refers to the pullback to $B_3$ of the respective
 classes of the fiber and base of the Hirzebruch surface $\mathbb F^1$ whose blowup constitutes $dP_2$; that is, $f \simeq C^0 \times \IP^1_{l'}$ and $h \simeq \IP^1_{h} \times \IP^1_{l'}$.
The rest of the notation has  been introduced in Section~\ref{sec_Ex1het}.  
\begin{table}
\begin{center}
\begin{tabular}{|c||ccccccc||cccc|}
\hline
& $\nu_{x_0}$ & $\nu_{x_1}$ & $\nu_{z_0}$ & $\nu_{z_1}$ & $\nu_{z_2}$ & $\nu_{z_3}$ & $\nu_{z_4}$ & $\nu_u$ & $\nu_v$ & $\nu_w$ & $\nu_s$ \\ \hline \hline
${f-E}$ &0&0&1&1&1& 0 & 0 & 0 & 3 &6&0 \\ \hline
${h}$ &0&0&0&1&0&1&0&0&2&4&0\\ \hline
${E}$ &0&0&1&0&0&0&1&0&2&4&0\\ \hline
${l'}$ &1&1&0&0&0&0&0&0&2&4&0\\ \hline \hline
$S_0$ &0&0&0&0&0&0&0&1&1&2&0\\ \hline
$S$ &0&0&0&0&0&0&0&0&1&1&1\\ \hline
\end{tabular}\end{center}
\caption{GLSM charges of the toric coordinates of the ${\rm Bl}_1 \mathbb P^2_{112}$ fibration over $B_3=dP_2 \times \IP^1_{l'}$. The upper-left part, as separated by the horizontal and the vertical double lines corresponds to the description of the base $B_3$ alone. } 
\label{tb:Model S}
\end{table}

We now turn to the geometry of the elliptic fibration $\pi: Y_4 \to B_3$. The elliptic fiber  is constructed as a general hypersurface of degree $4$ in ${\rm Bl}_1 \IP^2_{112}$, which is a convenient way of realising a $U(1)$ gauge symmetry via a rank-$1$ Mordell-Weil group of rational sections~\cite{Morrison:2012ei}. Specifically, the fourfold $Y_4$ is obtained by the vanishing locus of
\beq\label{P_MP}
P_{\rm MP} := sw^2 + b_0 s^2 u^2 w + b_1 s u v w + b_2 v^2 w + c_0 s^3 u^4 + c_1 s^2 u^3 v + c_2 s u^2 v^2 + c_3 u v^3 \,,
\eeq
where $u$, $v$, $w$ and $s$ are the four homogeneous coordinates of the ambient space ${\rm Bl}_1 \IP^2_{112}$ of the fiber. The coefficients $b_i$ and $c_i$ are sections of appropriate line bundles on $B_3$, whose degrees are parametrized by a cohomology class $\beta \in H^2(B_3, \IZ)$ as follows: 
\begin{center}
\begin{tabular}{|c||c|c|c|c|c|c|c|c|}
\hline
Sections & $b_0$ & $b_1$ & $b_2$ & $c_0$ & $c_1$ & $c_2$ & $c_3$ & $c_4$ \\
\hline 
& & & & & & & & \\[-1.1em] 
Classes & $\beta$ & $\bar K_{B_3}$ & $2\bar K_{B_3} - \beta$ & $2\beta$ & $\bar K_{B_3} + \beta$ & $2\bar K_{B_3}$ & $3\bar K_{B_3} - \beta$ & $4 \bar K_{B_3} - 2 \beta$ \\
\hline
\end{tabular}
\end{center}
If we denote by $\cL_u$ and $\cL_s$ the line bundles of which the coordinates $u$ and $v$ are sections, the remaining fibral ambient coordinates $w$ and $s$ are in turn sections of the bundles,
\bea
v &\in& H^0(Y_4, \cL_u \otimes \cL_s \otimes \cO(\beta -\bar K_{B_3}))\,, \\ \nn
w &\in& H^0(Y_4, \cL_u^2 \otimes \cL_s \otimes \cO(\beta))\,, 
\eea
which can be seen from the defining polynomial~\eqref{P_MP}. For the concrete fibration $Y_4$ we make the following choice:
\bea \label{beta-Model S}
\beta &=& 4D_1 + 2 D_2 + 4 D_3 + 2 D_4 \\ \nn
&=& 2 \bar K_{B_3} \,,
\eea
which leads to the toric description for $Y_4$ as described in Table~\eqref{tb:Model S}. 

Moreover, note that the height pairing divisor $b$ of the section $S$, defined as the image $\sigma(S)$ under the Shioda homomorphism $\sigma$~\cite{Park:2011ji}, 
\beq
b := -\pi_*(\sigma(S) \cdot \sigma(S)) \,,
\eeq
takes, taking into account~\eqref{beta-Model S}, the form:
\bea
b &=& 6 \bar K_{B_3}  - 2 \beta = 2 \bar K_{B_3} \,. 
\eea

Given this toric data for $Y_4$, upon performing appropriate combinatorial computations, e.g., by making use of PALP~\cite{Kreuzer:2002uu,Braun:2012vh} and SAGE~\cite{sagemath}, we can easily extract the generators $l^{(a)}$ of the Mori cone, ${\bf M}(Y_4)$: 
\beq\label{morigens-Model S}
\begin{array}{crrrrrrrrrrrrr}
l^{(1)}=&(&0,&     0,&     0,&     0,&     0,&     0,&     0,&     1,&     0,&    1,&   -1) \,, \\
l^{(2)}=&(&0,&     0,&     0,&     0,&     0,&     0,&     0,&    -1,&     1,&     0,&     2) \,, \\
l^{(3)}=&(&0,&     0,&     1,&     0,&     1,&    -1,&     0,&    -1,&     0,&     0,&     0) \,, \\
l^{(4)}=&(&0,&     0,&     0,&     0,&    -1,&     1,&     1,&    -1,&     0,&     0,&     0) \,, \\
l^{(5)}=&(&0,&     0,&     0,&     1,&     1,&     0,&    -1,&    -1,&     0,&     0,&     0) \,, \\
l^{(6)}=&(&1,&     1,&     0,&     0,&     0,&     0,&     0,&    -2,&     0,&     0,&     0) \,.
\end{array}
\eeq
These generators are described in terms of their intersection numbers with the toric divisors $d_\rho := \{ \nu_\rho = 0 \}$, for $\rho = x_0, x_1, \dots, w, s$, as ordered in Table~\ref{tb:Model S}. The quartic intersection numbers of $Y_4$ can be obtained via combinatorial computations as well, and we endode them in the following intersection polynomial:
\bea\label{interY4-Model S}\nn 
I(Y_4)&=&1302 J_1{}^4+672 J_1{}^3 J_2+336 J_1{}^2 J_2{}^2+168 J_1 J_2{}^3+84 J_2{}^4+120 J_1{}^3 J_3+64 J_1{}^2 J_2 J_3\\ \nn 
&&+\;32 J_1 J_2{}^2 J_3  +16 J_2{}^3 J_3+180 J_1{}^3 J_4+96 J_1{}^2 J_2 J_4+48 J_1 J_2{}^2 J_4+24 J_2{}^3 J_4\\ \nn
&&+\;14 J_1{}^2 J_3 J_4  +8 J_1 J_2 J_3 J_4 +4 J_2{}^2 J_3 J_4+14 J_1{}^2 J_4{}^2+8 J_1 J_2 J_4{}^2+4 J_2{}^2 J_4{}^2 \\ \nn
&&+\;120 J_1{}^3 J_5 +64 J_1{}^2 J_2 J_5  +32 J_1 J_2{}^2 J_5 +16 J_2{}^3 J_5+14 J_1{}^2 J_3 J_5+8 J_1 J_2 J_3 J_5 \\ 
&&+\;4 J_2{}^2 J_3 J_5 +14 J_1{}^2 J_4 J_5 +8 J_1 J_2 J_4 J_5+4 J_2{}^2 J_4 J_5 +105 J_1{}^3 J_6+56 J_1{}^2 J_2 J_6 \\ \nn
&&+\;28 J_1 J_2{}^2 J_6 +14 J_2{}^3 J_6 +14 J_1{}^2 J_3 J_6+8 J_1 J_2 J_3 J_6+4 J_2{}^2 J_3 J_6+21 J_1{}^2 J_4 J_6 \\ \nn
&&+\;12 J_1 J_2 J_4 J_6 +6 J_2{}^2 J_4 J_6 +3 J_1 J_3 J_4 J_6+2 J_2 J_3 J_4 J_6+3 J_1 J_4{}^2 J_6+2 J_2 J_4{}^2 J_6 \\ \nn
&&+\;14 J_1{}^2 J_5 J_6 +8 J_1 J_2 J_5 J_6 +4 J_2{}^2 J_5 J_6+3 J_1 J_3 J_5 J_6+2 J_2 J_3 J_5 J_6+3 J_1 J_4 J_5 J_6 \\ \nn
&&+\;2 J_2 J_4 J_5 J_6 \,, 
\eea
where $J_a$ are the generators of the K\"ahler cone that obey
\beq
\int_{l^{(a)}}J_b= \delta_b^a \,.
\eeq

Equipped with the topological data listed above, we can compute the BPS invariants, $N_{{\bf G}; C_{\rm b}}(n,r)$, to any given order for  curves of the form
\beq\label{bfC-AppB}
{\bf C} = C_{\rm b} + n \mathbb E_\tau + C^{\rm f}_r \,, 
\eeq
with respect to the transversal $U(1)$ flux 
\be\label{vU(1)flux-App}
{\bf G}  \equiv {\bf G}_{U(1)} = \sigma \wedge \pi^\ast F   \,, \qquad {\rm where \ \ } F =: \sum_{\alpha=1}^4 c_\alpha D_\alpha \,.
\ee
The base curves, $C_{\rm b}$, that are of interest in our examples are 
\beq\label{basecurves}
C_{\rm b}\ =\ \left\{C^0, \,C^1_E,\,C^2_E\right\} \,.
\eeq
These correspond respectively to the heterotic string featuring Section~\ref{sec_Ex1het}, and to the two types of E-strings discussed in Section \ref{sec_Estring}. 

The computation of the BPS invariants proceeds by mirror symmetry,
 in practice using A. Klemm's Mathematica package {\tt inst.m} and extensions thereof.
For this purpose we need to expand the various curves appearing on the RHS of~\eqref{bfC-AppB} over the Mori cone generators. This is achieved by making use of the explicit form for the curves~\eqref{morigens-Model S}, as well as of the intersection data~\eqref{interY4-Model S}. This leads to the identification
\bea \label{curves-Model S}
C^0 &=&l^{(4)}+l^{(5)} \,, \\
C^1_E &=& l^{(4)} \,, \\
C^2_E &=& l^{(5)} \,, \\
\mathbb E_\tau &=& 3l^{(1)}+2l^{(2)} \,, \\ 
C^{\rm f}_{r=1} &=& l^{(1)} + l^{(2)}  \,.
\eea  
We have calculated the relative BPS invariants up to certain finite degrees, and list them below in terms of the  generating functions
\be 
{\cal F}_{\bG; C_{\rm b}} = \sum_{n,r} N_{\bG; C_{\rm b}}(n, r) q^n \, \xi^r \,.
\ee
Concretely, for each of the three base curves~\eqref{basecurves} under consideration, we get
\bea
{\cal F}_{\bG; C^0} &=& \nn
q [96 c_1 \xi^{\pm\bar1} + 48 c_2 \xi^{\pm \bar1} + 84 c_3 \xi^{\pm\bar1} +96 c_4 \xi^{\pm \bar 1} ] \\ \nn
 &&+ \,q^2 [
 c_1 (69280  \xi^{\pm \bar 1} + 20384\xi^{\pm \bar 2} + 288 \xi^{\pm \bar 3} -8\xi^{\pm \bar 4})   \\ 
 &&\qquad+c_2 (99552  \xi^{\pm \bar 1} +29088\xi^{\pm \bar 2}+ 480 \xi^{\pm \bar 3}  -12\xi^{\pm \bar 4}  )    \\  \nn
 &&\qquad+c_3 ( 65164  \xi^{\pm \bar 1}+18896\xi^{\pm \bar 2} + 252 \xi^{\pm \bar 3}  -8\xi^{\pm \bar 4})   \\ \nn
 &&\qquad+c_4 (134192  \xi^{\pm \bar 1} +39280\xi^{\pm \bar 2} + 624 \xi^{\pm \bar 3}  -16\xi^{\pm \bar 4}  )  
 ] \\ \nn
 &&+\, \cO(q^3)  \,, \\ \nn
{\cal F}_{\bG; C^1_E} &=& \nn
q [ c_2 (112 \xi^{\pm \bar 1} + 4 \xi^{\pm \bar 2}) + c_3 (56 \xi^{\pm \bar 1} +2 \xi^{\pm \bar 2})]  \\  \label{E1:S}
&& + \,q^2 [ c_2 (2496 \xi^{\pm \bar 1} + 552 \xi^{\pm \bar 2}) + c_3 (1248 \xi^{\pm \bar 1} +276 \xi^{\pm \bar 2})]  \\ \nn
&& + \,q^3 [ c_2 (26928 \xi^{\pm \bar 1} + 9432  \xi^{\pm \bar 2} + 336 \xi^{\pm \bar 3}) + c_3 (13464 \xi^{\pm \bar 1} + 4716 \xi^{\pm \bar 2}  + 168 \xi^{\pm \bar 3})]  \\ \nn
&&  +\, \cO(q^4)  \,, \\
{\cal F}_{\bG; C^2_E} &=& \nn
q [ c_3 (56 \xi^{\pm \bar 1} +2 \xi^{\pm \bar 2}) + c_4 (112 \xi^{\pm \bar 1} + 4 \xi^{\pm \bar 2})]  \\ \label{E2:S}
&& + \,q^2 [c_3 (1248 \xi^{\pm \bar 1} +276 \xi^{\pm \bar 2}) + c_4 (2496 \xi^{\pm \bar 1} + 552 \xi^{\pm \bar 2}) ]  \\ \nn
&& + \,q^3 [ c_3 (13464 \xi^{\pm \bar 1} + 4716 \xi^{\pm \bar 2}  + 168 \xi^{\pm \bar 3}) + c_4 (26928 \xi^{\pm \bar 1} + 9432  \xi^{\pm \bar 2} + 336 \xi^{\pm \bar 3})]  \\ \nn
&&  +\, \cO(q^4)  \,.
 \eea
Here, the coefficients $c_\alpha$ parametrise the four-form flux $\bG$ as in~\eqref{vU(1)flux-App}. 

 In the main text of the paper we argue that the derivative part of the elliptic genera is given by a formally six-dimensional structure, which manifests itself in terms of the threefolds 
\beq
\pi_i: \mathbb Y^i_3 \to \mathbb B^i_2 \,,\quad i=1,2 \,, 
\eeq
whose two-fold bases are given by
\beq
\mathbb B^i_2 = p^*(C^i)\,.
\eeq
Here we will provide some relevant details of these, for the sample geometry we consider.
Concretely, we have $C^1=C_2$ and $C^2=C_1$ and the respective bases are thus given by
\bea\label{dP2-Model S}
\mathbb B^1_2 &\simeq& dP_2\,, \\ \label{F0-Model S}
\mathbb B^2_2 &\simeq&C^0 \times C_2 \simeq \mathbb F_0  \,.
\eea
Since the self-intersections of $p^*(C^i)$ vanish on $B_3$, the normal bundles $N_{{\mathbb B_2^i}/B_3}$ are trivial and hence so are $N_{{\mathbb Y_3^i}/Y_4}$. This implies that the induced fibrations $\mathbb Y_3^i$ are Calabi-Yau threefolds, once again defined by the polynomials of the form~\eqref{P_MP}. For the geometries under consideration, the restrictions preserve the arithmetic structure of the sections and are described by the classes of the height-pairings by
\beq
{\mathfrak b}^i := b|_{\mathbb B^i_2} = 2 \bar K_{\mathbb B^i_2} \,,
\eeq
where, in the second step, we have used $b= 2 \bar K_{B_3}$ and $N_{\mathbb B_2^i/B_3}=\cO_{\mathbb B_2^i}$. 

As both of the induced fibrations,  $\mathbb Y^1_3$ and $\mathbb Y^2_3$, are torically constructed, they admit a description in terms of an
abelian GLSM, analogous as for $Y_4$ in Table~\ref{tb:Model S}. The most general such descriptions for the bases~\eqref{dP2-Model S} and~\eqref{F0-Model S} can be found, for instance, in \cite{Lee:2018urn} (see Table 4.2, as well as Table 4.1 with $a=0$); for the specific fibrations $\mathbb Y_3^1$ and $\mathbb Y_3^2$ under scrutiny, we set $({\rm x}, {\rm y_1}, {\rm y_2}) = (6, 2, 2)$ in the former case and $({\rm x}, {\rm y})=(4,4)$ in the latter. By determining their Mori cones and intersection rings, we can calculate the BPS invariants, $N^i_{C_{\rm b}}(n,r)$, for the curve classes of the form~\eqref{bfC-AppB} on $\mathbb Y^i_3$, just like it was done for the fourfold $Y_4$. As result we present these invariants via their generating functions of the form,
\be 
{\cal F}^i_{C_{\rm b}} = \sum_{n,r} N^i_{C_{\rm b}}(n, r) q^n \, \xi^r \,, 
\ee
for each of the two geometries and base curves~\eqref{basecurves}. 
For $\mathbb Y^1_3$ we get for the heterotic base curve and for the two E-string curves, respectively:
\begin{align}  
\cF^1_{C^0}&= -2 + \left( 252 + 84 \xi^{\pm 1}\right) q  +\left( 116580  + {65164} \xi^{\pm 1} + 9448 \xi^{ \pm 2} + 84 \xi^{\pm 3} -2 \xi^{\pm 4}\right) q^2 \,\\ \nn
&\quad +\,  \left( 6238536  + {3986964} \xi^{\pm 1} + 965232 \xi^{\pm 2} + 65164 \xi^{\pm 3} +252 \xi^{\pm 4}\right) q^3 + \cO(q^4) \,, \\ 
\cF^1_{C^{i=1,2}_E} &= 
1+ (138 + 56 \xi^{\pm 1} + \xi^{\pm 2}) q + (2358 + 1248 \xi^{\pm 1} + 138 \xi^{\pm 2}) q^2 \\ \nn
&\quad +\,  (23004 + 13464 \xi^{\pm 1} + 2358 \xi^{\pm 2} + 56 \xi^{\pm 3})q^3 + \cO(q^4) \,.
\end{align}
On the other hand, since $\mathbb Y^2_3$ does not contain $(-1)$-curves in its base $\mathbb B_2^2 \simeq \mathbb F_0$,
all we get is invariants for $C_{\rm b}=C^0$:
\begin{align} \nn 
\cF^2_{C^0}&=-2 + \left(288 + 96 \xi^{\pm 1}\right) q 
+ \left(123756  + {69280} \xi^{\pm 1} + 10192 \xi^{\pm 2} + 96 \xi^{\pm 3} - 2 \xi^{\pm 4}\right) q^2 + \cO(q^3) \,. 
\end{align}

\section{Non-Abelian 4d E-String for $B_3=\mathbb F_1 \times \mathbb P^1_{l'}$}    \label{sec:Het-A1} 
 
Here we present some details on the geometry that underlies the four-dimensional E-string model presented in Section~\ref{sec_Estring}.
 The base space of the elliptic fourfold $Y_4$ is  given by $B_3=\mathbb F_1 \times \mathbb P^1_{l'}$, which is related to the base  $dP_2 \times \mathbb P^1_{l'}$ discussed before in Appendix~\ref{app_modelS} by a simple blowdown.
 Hence
   \beq
H^{1,1}(B_3) = {\rm Span}\left< l', f, h\right>\,,
\eeq
with triple intersections
\beq\label{interB3-ModelEstring}
I(B_3) = l' ( f h-h^2) \,
\eeq
and 
 anti-canonical class
\beq\label{Kbar-ModelEstring}
\bar K_{B_3} = 2l' + 3f + 2h  \,. 
\eeq
We construct an elliptic fibration $Y_4$ with base $B_3$ and design an $SU(2)$ gauge symmetry over the divisor $b=h$.
The model is obtained as the resolution of an $SU(2)$ Tate model \cite{Bershadsky:1996nh}. 
For a flat fibration $Y_4$ over $B_3$, one finds two phases that are related via a flop transition. Both phases lead to the same elliptic genus and in the following we present our analysis for just one phase of our choice.  
The GLSM data can be found in Table~\ref{tb:Model A1}.
\begin{table}
\begin{center}
\begin{tabular}{|c ||cccc||cccccc|}
\hline
& $\nu_{x}$ & $\nu_{y}$ & $\nu_{z}$ & $\nu_{e_1}$ & $\nu_{e_0}$ & $\nu_{z_1}$ & $\nu_{z_2}$ & $\nu_{z_3}$ & $\nu_{z_4}$ & $\nu_{z_5}$  \\ \hline \hline
$Z$  &2&3&1&0&0&0&0&0&0&0\\ \hline
$e_1$  &$-1$&$-1$&0&1&$-1$&0&0&0&0&0\\ \hline\hline
$h$  &4&6&0&0&1&0&1&0&0&0\\ \hline
$f$  &6&9&0&0&0&1&1&1&0&0\\ \hline
$l'$ &4&6&0&0&0&0&0&0&1&1\\ \hline
\end{tabular}\end{center}
\caption{GLSM charges of the toric coordinates of the $\mathfrak {su}(2)$-enhanced E-string model on $\mathbb F_1 \times \IP^1_{l'}$.} 
\label{tb:Model A1}
\end{table}
The Mori cone is generated by the following curves,
\beq
\begin{array}{lclrrrrrrrrrrrrr|}
l^{(1)}=&(&1&1&0&-1&1&0&0&0&0&0), \\
l^{(2)}=&(&0&0&-2&0&1&0&1&0&0&0) ,\\
l^{(3)}=&(&-1&0&1&3&-3&0&0&0&0&0) ,\\
l^{(4)}=&(&0&0&-1&0&-1&1&0&1&0&0), \\
l^{(5)}=&(&0&0&-2&0&0&0&0&0&1&1) ,
\end{array}
\eeq
and the intersection polynomial reads
\bea \nn
I(Y_4) &=& 3576 J_1{}^4 + 324 J_1{}^3 J_2 + 18 J_1{}^2 J_2{}^2 + 1236 J_1{}^3 J_3 +
 108 J_1{}^2 J_2 J_3 + 6 J_1 J_2{}^2 J_3  \\ \nn
 && +\, 424 J_1{}^2 J_3{}^2 +
 36 J_1 J_2 J_3{}^2 + 2 J_2{}^2 J_3{}^2 + 144 J_1 J_3{}^3 +
 12 J_2 J_3{}^3 + 48 J_3{}^4 + 188 J_1{}^3 J_4 \\ \nn
 && +\, 18 J_1{}^2 J_2 J_4 +
 68 J_1{}^2 J_3 J_4 + 6 J_1 J_2 J_3 J_4 + 24 J_1 J_3{}^2 J_4 +
 2 J_2 J_3{}^2 J_4 + 8 J_3{}^3 J_4  \\
 && +\, 200 J_1{}^3 J_5 +
 27 J_1{}^2 J_2 J_5 + 3 J_1 J_2{}^2 J_5 + 70 J_1{}^2 J_3 J_5 +
 9 J_1 J_2 J_3 J_5 + J_2{}^2 J_3 J_5  \\ \nn
 && +\, 24 J_1 J_3{}^2 J_5 +
 3 J_2 J_3{}^2 J_5 + 8 J_3{}^3 J_5 + 16 J_1{}^2 J_4 J_5 +
 3 J_1 J_2 J_4 J_5 + 6 J_1 J_3 J_4 J_5  \\ \nn
 && +\, J_2 J_3 J_4 J_5 + 2 J_3{}^2 J_4 J_5 \,,
\eea
where $J_a$ are the Kahler cone generators. 
As a basis of $H^{1,1}(B_3)$, we can pick 
\beq
\begin{array}{lllll}
D_1&=& \nu_{z_1} &=& J_4, \\
D_2&=& \nu_{z_2} &=& J_2,\\
D_3&=& \nu_{z_4} &=& J_5.\\
\end{array}
\eeq
In terms of these,  we take $b= D_2 - D_1$, and furthermore
the zero-section of the elliptic fibration and the exceptional divisor for the $U(1)\subset SU(2)$ subgroup are
\beq
\begin{array}{lllll}
Z&=& -2 J_2 + J_3 - J_4 - 2 J_5, \\
e_1&=& -J_1 + 3 J_3 \,.\\
\end{array}
\eeq
Matter of $U(1)$
 charge $r=1$ arises from an M2-brane wrapping the fibral curve $C^{\rm f}_{r=1} = l^{(1)}$. The latter has
the following  intersection numbers
\be
Z \circ C^{\rm f}_{r=1}  =0 \,, \qquad (-e_1) \circ C^{\rm f}_{r=1}  =1 \,.
\ee  
The $U(1)\subset SU(2)$ flux is then of the form 
\beq
\bG = \pi^{\ast} F \wedge (-e_1) \,,\qquad    F = c_1 D_1 + c_2 D_2 + c_3 D_3 \,.
\eeq

The four-dimensional non-critical E-string in question arises as soliton from a D3-brane that wraps the curve $C_E = l^{(4)}$.
By mirror symmetry we find that the first terms in the expansion of its elliptic genus
\be
Z_{C_E;\bG}(q,\xi) =  - q^{-1/2}  \sum_{n,r} N_{C_E;\bG}(n,r)  q^n \xi^r
\ee
are given by

\bea\label{ellgenEexp1}
q^{1/2}  \, Z_{C_E;\bG}(q,\xi)  &=& \nn
q [ c_1 (16 \xi^{-2}  -160 \xi^{-1} + 768 -2400 \xi + 5616 \xi -10752\xi^3 + 17920 \xi^4 \\ \nn
&&\qquad  -27136\xi^5 + 38400 \xi^6 -51712 \xi^7 + 67072 \xi^8 + \cdots ) \\ \nn
&&\quad + c_3 ( 6\xi^{-2} -40 \xi^{-1} -472\xi + 2042 \xi^2 -4608 \xi^3+ 8192 \xi^4 -12800 \xi^5 \\ \nn
&&\qquad + 18432 \xi^6 -25088\xi^7  + 32768 \xi^8 + \cdots) ]  \\ \nn
&& + q^2 [ c_1(48 \xi^{-4} -640 \xi^{-3} + 3936 \xi^{-2} -15360 \xi^{-1} + 44032 - 101376 \xi \\ \nn
&&\qquad +198816 \xi^2 -345472\xi^3 + 546768 \xi^4 -804864\xi^5  + 1120256 \xi^6 + \cdots ) \\ \nn
&&\quad + c_3 (20 \xi^{-4} -240 \xi^{-3} + 1188 \xi^{-2} -2640\xi^{-1} -11696 \xi + 56156 \xi^2 \\ \nn
&&\qquad -128784 \xi^3 + 229356 \xi^4 - 358400 \xi^5 + 516096  \xi^6 + \cdots)]  \\ \nn
&& + q^3 [ c_1(96 \xi^{-6} -1440 \xi^{-5} + 10272 \xi^{-4} -47040 \xi^{-3} + 158592 \xi^{-2} \\ \nn
&&\qquad - 427872\xi^{-1} + 975360 - 1946784\xi + 3487872 \xi^2 + \cdots ) \\ \nn
&&\quad + c_3 (42 \xi^{-6} -600 \xi^{-5} + 3960 \xi^{-4} -15840 \xi^{-3} + 40854 \xi^{-2} -57432 \xi^{-1} \\ \nn
&&\qquad -154536\xi + 807018 \xi^2 + \cdots)]  \\ \nn
&& + q^4 [ c_1(160 \xi^{-8} -2560\xi^{-7} + 19776 \xi^{-6} -99840 \xi^{-5} + 376480\xi^{-4}  \\ \nn
&&\qquad  -1146240 \xi^{-3} + 2958848 \xi^{-2} -6680192\xi^{-1}+ 13473792 + \cdots ) \\ \nn
&&\quad + c_3 (72 \xi^{-8} -1120\xi^{-7} + 8316 \xi^{-6} -39600\xi^{-5} + 136180 \xi^{-4} \\ \nn 
&&\qquad -353808\xi^{-3} + 671876 \xi^{-2} -740096 \xi^{-1} + \cdots)]  \\ \nn
&&  + \cO(q^5)  \,.\\ \label{E:A1}
\eea


We can interpret the elliptic genus as the derivative of a formal six-dimensional elliptic genus,
possibly augmented by a genuinely four-dimensional piece.
The curve $C_E$ is trivially fibered over the $\mathbb P^1_{l'}$ factor in $B_3$, thereby tracing out the divisor $D_E = h=D_2-D_1$ within $B_3$.
This suggests that we should consider the E-string within the threefold defined by
\be
\mathbb Y_3 = Y_4 |_{\pi^\ast(l')} \,,
\ee
where the divisor $l'= D_3$ cuts out the Hirzebruch surface $\mathbb F_1$ of $B_3$.

The relative BPS invariants for $C_E$ on  $\mathbb Y_3$ can be computed via mirror symmetry as well and be 
packaged into the partition function
\bea \label{E:A1b}  \nn 
Z_{-2,m} (q,\xi)&=&  - q^{-1/2}  \\  \nn
&&-  \left(3 \xi^{-2} -40 \xi^{-1} + 198 +472 \xi -1021 \xi^2 +1536\xi^3 - 2048 \xi^4 +2560\xi^5   \right. \\   \nn 
&&\quad  \left.   -3072 \xi^6 + \cdots \right)q^{1/2} \\  \nn
&& \qquad - ( 5 \xi^{-4} -80 \xi^{-3} + 594 \xi^{-2} -2640\xi^{-1}  +5788 +11696\xi   \\   \nn  
&&   \quad \qquad    -28078 \xi^2 + \cdots )q^{3/2}\\ \nn 
&& \qquad - (7 \xi^{-6} -120 \xi^{-5} +  990 \xi^{-4} -5280\xi^{-3} + 20427 \xi^{-2} + \cdots ) q^{5/2}+ \\ 
&&\qquad - (9 \xi^{-8} -160\xi^{-7} +1386 \xi^{-6} + \cdots ) q^{7/2}  + \cO(q^{9/2}) \,. 
\eea
This can be recognised as the series expansion in $q=e^{2\pi i\tau}$ and $\xi=e^{2\pi iz}$ of the following
meromorphic Jacobi form of weight $w=-2$ and $U(1)$ fugacity index $m=-1$:
\beq
Z_{-2,-1}(\tau,z)\ =\ - \frac{1}{\eta(\tau)^{12}}\sum_{i=2}^4\frac{\theta_i(0,\tau)^{10}}{\theta_i(2z,\tau)^2} \,.
\eeq
As discussed in Section~\ref{sec_4dEstring}, this has the interpretation as the elliptic genus of an instanton string of the six-dimensional
 SCFT with $SU(2)$ gauge symmetry.

\goodbreak
\section{Non-Critical String for $B_3=\mathbb P^3$}\label{app_P3string}

In this Appendix we provide some of the technical details that underlie the example of Section~\ref{sec_P3string}.
It is devoted to a non-critical string obtained by wrapping a D3-brane on the curve $C_{\rm b} = H \cdot H$
on the base $B_3=\mathbb P^3$ of an elliptic fibration.

The model without $U(1)$ gauge symmetry has been discussed in much detail in 
\cite{Klemm:2007in,Haghighat:2015qdq,Cota:2017aal}.
Here we consider an extension of this model in order to support a transversal flux in conjunction with a chiral $U(1)$ gauge symmetry, which allows for a nontrivial elliptic genus. For this we implement the 
gauge symmetry associated with the height-pairing given by $b= 2 \bar K_{B_3} = 8 H$.
The fourfold $Y_4$ we thus consider is characterized by the  Mori vectors
\beq\label{P3mori}
\begin{array}{llrrrrrrrrrr|}
l^{(1)}=&(& -1 ,& -1 ,& 0 ,& 0 ,& 0 ,& 0 ,& 0 ,& 1 ,& 1 ),\\
l^{(2)}=&(&-2 ,& 2 ,& 1 ,& 0 ,& 0 ,& 0 ,& 0 ,& 0 ,& -1), \\
l^{(3)}=&(&0 ,& 0 ,& 0 ,& 1 ,& 1 ,& 1 ,& 1 ,& 0 ,& -4) \,.\\
\end{array}
\eeq
The intersection polynomial, in terms of the dual basis of K\"ahler cone generators, reads
\bea
I(Y_4)&=&1984 J_1{}^4+512 J_1{}^2 J_2{}^2+128 J_2{}^4+28 J_1{}^2 J_3{}^2+8 J_2{}^2 J_3{}^2+256
   J_2{}^3 J_1+3 J_3{}^3 J_1
  \\ 
&&  +\,1024 J_1{}^3 J_2+2 J_3{}^3 J_2 +16 J_3{}^2 J_1 J_2+240
   J_1{}^3 J_3+32 J_2{}^3 J_3+64 J_2{}^2 J_1 J_3+128 J_1{}^2 J_2 J_3\,.
   \nn
\eea
Moreover we define the following curve classes
\bea \label{U-Het}
C_{\rm b}&=&l^{(3)}  \\
\mathbb E_\tau  &=& 3l^{(1)}+2l^{(2)} \nn\\ 
C^{\rm f} &=& l^{(1)} + l^{(2)}  \,,\nn
\eea  
and determine the BPS invariants, 
$N_{{\bf G}; C_{\rm b}}(n,r)$, for the curves of the form
\beq\label{bfC}
{\bf C} = C_{\rm b} + n \mathbb E_\tau + C^{\rm f}_r \,, 
\eeq
with respect to fluxes ${\bf G} \in H^{2,2}_{\rm vert}(Y_4)$.
As a basis of fluxes we pick
\bea 
\bG_{U(1)}&=&   \sigma \wedge \pi^{\ast}(F) \, \qquad {\rm with\ \ }  F = c\, H\,, \nn \\
&=&   \frac{1}{1441}\left[{- 156 \left(4 J_1{}^2+J_1 J_3\right)}+{115 \left(4 J_2{}^2+8 J_1 J_2-2 J_1 J_3+J_2
   J_3\right)}
   \right]\,,\nn
   \\ 
\bG_{H}&=&  \pi^{\ast}(H) \wedge  \pi^{\ast}(H) =   {J_3}{}^2\,,
   \\  
\bG_{0} &=&
  -8 J_3{}^2+\frac{1}{1441} \left[   {316 \left(4 J_1{}^2+J_1 J_3\right)} - {196 \left(4 J_2{}^2+8 J_1 J_2-2 J_1 J_3+J_2
   J_3\right)}
   \right]\,.\nn
\eea
Above, $H$ denotes the hyperplane class  on $\mathbb P^3$ and $c$ a numerical constant. Moreover, as always, $\sigma$ denotes the Shioda map associated with the $U(1)$ symmetry. 
This basis leads to a convenient block-diagonal intersection form on $H^{2,2}_{\rm vert}(Y_4)$ given by
\be
\Omega=\left(
\begin{array}{ccc}
 -8 c^2 & 0 & 0 \\
 0 & 0 & 4 \\
 0 & 4 & 0 \\
\end{array}
\right)\,.
\ee
We now perform a standard mirror symmetry computation to obtain the Gromov-Witten invariants and 
assemble them, for each of the above fluxes, into the following partition function\footnote{We have factored out $1/q^2$ in order to account for the  vacuum energy of the string.}
\be
Z_{\bG; C_{\rm b}}(q,\xi):= -\frac1{q^2}\sum N_{\bG;C_{\rm b}}(n,r) q^n \xi^{r}   \,.
\ee
The fluxes $\bG_{H}$ and $\bG_{0}$ are the fluxes considered in \cite{Haghighat:2015qdq,Cota:2017aal}
and are associated with meromorphic modular forms of weight
$w=-2$ and $w=0$, respectively. These are non-transversal fluxes which do not admit a lift to four-dimensional F-theory.
On the other hand, the transversal flux $\bG_{U(1)}$ leads to
 the $U(1)$-refined elliptic genus we are interested in, with modular weight $w=-1$.

In terms of $\xi^{\pm r}=\xi^r+\xi^{-r}$,  $\xi^{\pm  \bar r}=\xi^r-\xi^{-r}$,
the first few terms of the expansions take the form
\bea
  \frac{1}{c} \, Z_{\bG_{U(1)}; C_{\rm b}}(q,\xi)&=&
-\frac{1152}q \xi^{\pm \bar1}+576  \left(380 \xi^{\pm \bar1}+127 \xi^{\pm \bar2}\right) \\
&& -384 q \left(90633 \xi^{\pm \bar1}+53472 \xi^{\pm \bar2}+7825 \xi^{\pm \bar3}\right)+...\nn
\\
Z_{\bG_H; C_{\rm b}}( (q,\xi)&=&
\frac{20}{q^2}-\frac{1536} q (3+\xi^{\pm 1})+192  \left(4161+2264 \xi^{\pm 1}+343 \xi^{\pm 2}\right)
\\
&& \qquad\qquad \
- 1024q \left(99294+65817 \xi^{\pm 1}+18666 \xi^{\pm 2}+1805 \xi^{\pm 3}\right)+...\nn
\\
 Z_{\bG_{0}; C_{\rm b}}(q,\xi)&=&
-\frac{160}{q^2}+\frac{7680} q (3+\xi^{\pm 1})-768  \left(4179+2248 \xi^{\pm 1}+305 \xi^{\pm 2}\right)
\\
&& \qquad\qquad \
+512 q \left(393396+257535 \xi^{\pm 1}+68112 \xi^{\pm 2}+5405 \xi^{\pm 3}\right)+...\nn
\eea
Matching against Ansaetze of Jacobi forms yields an overdetermined system of equations,
and in this way we find the following closed expression for the weight $w=-2$ flux:
\bea\label{P2Z2}
Z_{\bG_H; C_{\rm b}}(q,\xi)&=&
\frac4{12^4\eta^{48}}\left[
-\frac{5}{72} \left(35 E_4^4
   E_6+37 E_4 E_6^3\right) \phi _{0,1}^4     \nn    \right. \\ \nn
&& \left.  -\frac{1}{432} \left(835 E_4^6 E_6+1252 E_4^3 E_6^3+73 E_6^5\right) \phi _{-2,1}^4\right.
\nn\\   \nn
&&\!\!\!\!\!\!\!\!\!
+\frac{1}{864}
   \left(2173 E_4^7+12406 E_4^4 E_6^2+2701 E_4 E_6^4\right) \phi _{-2,1}^3 \phi _{0,1}
\\
&&\!\!\!\!\!\!\!\!\!
\left.
-\frac{1}{12}
   \left(191 E_4^5 E_6+169 E_4^2 E_6^3\right) \phi _{-2,1}^2 \phi _{0,1}^2   \right. \\ \nn
   && \left. \qquad    +\frac{1}{864} \left(2281
   E_4^6+13342 E_4^3 E_6^2+1657 E_6^4\right) \phi _{-2,1} \phi _{0,1}^3
   \right].\nn
   \eea
Moreover we find for the elliptic genus:
\bea\label{P2Z1}
Z_{\bG_{U(1)}; C_{\rm b}}(q,\xi)&=&  (F \cdot C_{\rm b})       Z_{-1,4}(q,\xi)+   \frac{1}{8} (F \cdot b \cdot H)   \xi \partial_\xi  Z_{-2,4}(q,\xi), \\
\!\!\!\!\!\!\!\!\!\!\!\!\!\!\!\!\!\!\!\!\!\!{\rm where}\qquad\qquad \qquad\qquad\qquad\qquad && \nn\\
Z_{-2,4}(q,\xi) &=& Z_{\bG_H; C_{\rm b}}(q,\xi)\label{ZisZ}
\eea
 and
\bea
Z_{-1,4}(q,\xi)&\!=\!&
\frac1{12^4\eta^{48}}\,\phi _{-1,2}\left[
 64
   E_4^2 E_6 \left(E_4^3-E_6^2\right) \phi _{-2,1} \phi _{0,1}+      \right.  \nn\\
   && \left. \qquad    \frac{1}{6} \left(134 E_4^4 E_6^2+29 E_4 E_6^4-163 E_4^7\right) \phi _{-2,1}^2\right.
\\ 
 &&- \left.
   \frac{1}{6} \left(13 E_4^6+166 E_4^3  E_6^2-179 E_6^4\right) \phi _{0,1}^2\right]   \nn
   \eea
are weak Jacobi forms. 
In (\ref{P2Z1}) we used the fact that $F \cdot C_{\rm b} = c$ and $F \cdot b \cdot H = 8 c$.
Notably, (\ref{ZisZ}) states that the very same function $Z_{-2,4}(q,\xi)$
figures both as generating function $Z_{\bG_H; C_{\rm b}}(q,\xi)$ for the $(-2)$-flux, and in the derivative part of the elliptic genus
(\ref{P2Z1}) associated with the transversal flux $\bG_{U(1)}$. This confirms again our expectations.


\bibliography{papers}
\bibliographystyle{JHEP}

\end{document}